\documentclass[sn-nature]{sn-jnl}% Basic Springer Nature Reference Style/Chemistry Reference Style

\usepackage{graphicx}
\usepackage{multirow}
\usepackage{amsmath,amssymb,amsfonts}
\usepackage{amsthm}
\usepackage{mathrsfs}
\usepackage[title]{appendix}
\usepackage{xcolor}
\usepackage{textcomp}
\usepackage{manyfoot}
\usepackage{booktabs}
\usepackage{algorithm}
\usepackage{algorithmicx}
\usepackage{algpseudocode}
\usepackage{listings}
\usepackage{enumitem}
\usepackage{booktabs}
\usepackage{multirow} 
\usepackage{longtable}  
\usepackage{caption}
\usepackage{makecell}
\usepackage{amsmath}
\usepackage{lineno} 

\theoremstyle{thmstyleone}

\theoremstyle{thmstyletwo}

\theoremstyle{thmstylethree}

\DeclareCaptionLabelSeparator{pipe}{\ \textbar\ }
\begin{document}

% \title[Article Title]{Estimating social segregation as experienced through mobility}
% \title[Article Title]{Urban mobility shapes experiences of social segregation}
\title[Article Title]{Urban transport systems shape experiences of social segregation}

\author[1]{\fnm{Yitao} \sur{Yang}}\email{y.yang@leeds.ac.uk}
\equalcont{These authors contributed equally to this work.}

\author[2]{\fnm{Erjian} \sur{Liu}}\email{17120752@bjtu.edu.cn}
\equalcont{These authors contributed equally to this work.}

\author[2]{\fnm{Bin} \sur{Jia}}\email{bjia@bjtu.edu.cn}

\author*[1]{\fnm{Ed} \sur{Manley}}\email{e.j.manley@leeds.ac.uk}

\affil*[1]{\orgdiv{School of Geography}, \orgname{University of Leeds}, \city{Leeds}, \country{UK}}

\affil[2]{\orgdiv{School of Systems Science}, \orgname{Beijing Jiaotong University}, \city{Beijing}, \postcode{100044}, \country{China}}

\abstract{\noindent\hspace*{1.5em}Mobility is a fundamental feature of human life, and through it our interactions with the world and people around us generate complex and consequential social phenomena. 
Social segregation, one such process, is increasingly acknowledged as a product of one’s entire lived experience rather than mere residential location \cite{de2024PNAS,Cagney2020Annual,Phillips2021Sociological}. 
Increasingly granular sources of data on human mobility have evidenced how segregation persists outside the home, in workplaces, cafes, and on the street \cite{Nilforoshan2023Nature,Cook2024NatureCities,Moro2021NC}. 
Yet there remains only a weak evidential link between the production of social segregation and urban policy \cite{LIAO2025CEUS}.
This study addresses this gap through an assessment of the role of the urban transportation systems in shaping social segregation.
Using city-scale GPS mobility data and a novel probabilistic mobility framework, we establish social interactions at the scale of transportation infrastructure, by rail and bus service segment, individual roads, and city blocks.
The outcomes show how social segregation is more than a single process in space, but varying by time of day, urban design and structure, and service design.
These findings reconceptualize segregation as a product of likely encounters during one's daily mobility practice.
We then extend these findings through exploratory simulations, highlighting how transportation policy to promote sustainable transport may have potentially unforeseen impacts on segregation.
The study underscores that to understand social segregation and achieve positive social change urban policymakers must consider the broadest impacts of their interventions and seek to understand their impact on the daily lived experience of their citizens.

}


\maketitle

\section*{Introduction}\label{Introduction}

 \noindent\hspace*{1.5em}How people interact in their everyday routines---or their lived experience---is increasingly recognized as an important lens for truly understanding social mixing and segregation \cite{Milgram1970Science,Nilforoshan2023Nature}. Our day-to-day social interactions influence our perception and behavior within the city \cite{de2024PNAS,Cagney2020Annual,Phillips2021Sociological,Michael2007MIT}, and several recent studies \cite{Chang2021Nature,Gomez2017NatureHuman,Wang2018PNAS} have begun to unpick the dynamic nature of this experience at increasingly granular scales. Unlike the early studies of residential segregation, new mobility data have shown how segregation unfolds outside of neighborhoods through interactions in shared urban places (e.g., shops, restaurants, and parks) across different times. By going beyond neighborhood-level interactions, we can better understand the transient encounters that reinforce or challenge existing social divides, which ultimately shaping the fabric of urban life in subtle but significant ways \cite{Pappalardo2023Naturecom,Loo2024HSSC,Neira2024SR,LIAO2025CEUS}.

 Recent studies \cite{Cook2024NatureCities,Schläpfer2021nature,Moro2021NC,Athey2021PNAS,Levy2022SA,Abbiasov2024NHB,Xu2025NHB} in computational social science have leveraged large-scale mobility datasets to map daily encounters and rhythms of social contact across diverse urban settings. For instance, research focusing on public transit \cite{Sorath2021JTG,QiLi2022Cities,Lukas2023JTG} have examined how different socioeconomic groups coincide on specific routes or stations using tap-in/tap-out logs, while others have inferred roadside stays \cite{Aiello2025PNAS,Fan2023pnasnexus} or path-crossing events \cite{Nilforoshan2023Nature} in border regions to measure spatial segregation through GPS pings of mobile phone users. Despite this progress, the nature of such data mean that existing studies may overestimate certainty in social interaction events at precise spatial locations. For a fundamental challenge persists: even high-resolution datasets like GPS trajectories suffer from spatiotemporal sparsity, limiting their ability to pinpoint precise encounters \cite{Barreras2024NCS}, nor the spatial context inherent to those interactions. These limitations underscores the need for a paradigm shift toward a more realistic dynamic perspective for measuring segregation, moving beyond conventional deterministic approach that assumes exact knowledge of spatiotemporal overlaps \cite{Xu2019Royal}, failing to account for the inherent uncertainties of human mobility.

 Here we introduce a novel probabilistic framework to quantify the likelihood of transient encounters between individuals as they navigate multimodal transportation systems in the city. By integrating extensive mobile phone data from 11 million users in Beijing, China, we provide a fine-grained, dynamic measure of segregation experienced during daily movements across active, driving, bus and railway journeys. We further propose a multimodal uniformity index to systematically identify spatial inconsistencies in segregation across mobility layers within the same geographic contexts, revealing how urban configurations modulate social divides. Our analysis demonstrates that social segregation is neither static nor universal---it varies across and within geographies, emerging as a dynamic, rhythmic process, shaped by the interplay of daily schedules, transport infrastructure, and urban spatial organization. To further investigate these dynamics and assess potential interventions, we develop an agent-based model of mobility. Through urban-scale simulations, we quantify the distributional effects of current transport policies, highlighting the complex trade-offs policymakers must navigate between efficiency gains, social benefit impacts, and social equity. This study provides a process-oriented understanding of how cities integrate populations through multimodal mobility, paving the way for designing more socially cohesive cities.
 
\section*{Results}\label{Results}

\subsection*{Segregation is dynamic---and uncertain---by definition}\label{Results1}

\begin{figure*}[h]
\centering
\includegraphics[width=1\textwidth]{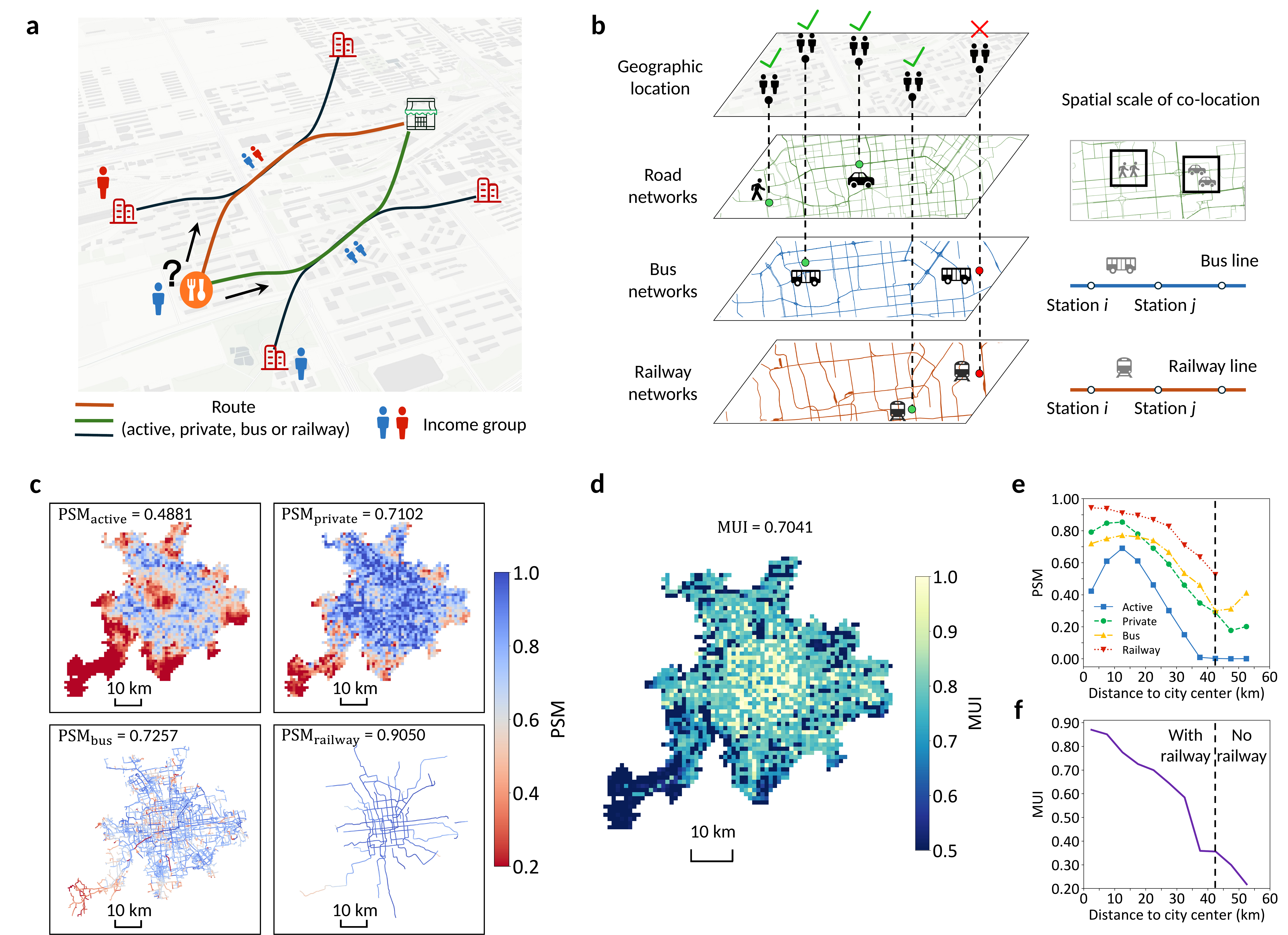}
\caption{\textbf{Probabilistic framework quantifies segregation within multimodal mobility systems}. \textbf{a}, Conceptual overview of the methodology. Individual trajectories from mobile phone data are used to estimate the probability of transient encounters between pairs of individuals as they travel via different transport modes (active, private, bus, railway). \textbf{b}, Definition of mode-specific spatial units used for calculating encounter probabilities. Road-based interactions (active/private modes) are assessed within 1 km $\times$ 1 km grids representing potential roadside proximity. Public transport (bus/railway modes) interactions are assessed along station-to-station segments, representing shared vehicle or platform environments. \textbf{c} Spatial maps displaying the probabilistic segregation measure (\textit{PSM}) calculated for each transport mode across the Beijing metropolitan area. \textit{PSM} quantifies the income diversity within the expected population mix frequenting each spatial unit via a specific mode. Distinct segregation patterns emerge depending on the mode, with mean citywide \textit{PSM} values noted above each map. \textbf{d}, Spatial distribution of multimodal uniformity index (\textit{MUI}), which measures the consistency of \textit{PSM} values across the four transport modes within the same region (1 km $\times$ 1 km grids are shown in the map). \textbf{e} Distance-dependent segregation patterns. Points show \textit{PSM} averages in 5 km concentric zones from the city center. Dashed line marks railway-deprived suburban regions showing increased mixing. \textbf{f}, Average \textit{MUI} versus distance from the city center. The decreasing trend indicates diminishing cross-modal uniformity towards the periphery.}\label{fig1}
\end{figure*}

 \noindent\hspace*{1.5em}To capture the dynamic nature of segregation shaped by multimodal mobility, we develop a probabilistic framework (Fig. \ref{fig1}a) by leveraging high-resolution mobile phone data from 11 million anonymous users in Beijing, China. We first infer socioeconomic status of individuals by geolocating `home' residence to community land value estimated from housing transaction records from LianJia (China’s largest real estate platform), and categorizing individuals into four income quartiles (see the ‘Inferring home locations and workplaces’ and ‘Inferring income quartiles’ in \textit{Methods}). Unlike conventional deterministic segregation measures \cite{Fan2023pnasnexus,Xu2019Royal}, our approach computes the probabilities of individuals encountering each other within urban multilayered spaces using specific transport modes (active, private, bus, or railway). Transient encounters among individuals are defined as simultaneous presence on mode-specific spatial features \textit{s}, defined as 1 km grids for active and private modes (Fig. \ref{fig1}b) and transit segments between stations for public modes (confined to shared vehicle or platform experiences \cite{Joshua2022JTH}). Co-presence is analyzed in 1-hour windows, balancing interaction significance with computational tractability (see Supplementary Note 2.3 for sensitivity analysis of spatiotemporal scales). Aggregating these encountering probabilities across income groups, we define a mode-specific probabilistic measure $PSM_{s,m}$, which quantifies the income segregation within the spatial unit \textit{s} for people traveling through mode \textit{m}. To assess how segregation experiences diverge across different mobility layers within the same geographic context (e.g., a specific region \textit{A}), we further introduce a multimodal uniformity index $MUI_{A}$ by evaluating the consistency of segregation levels across transport modes within the same geographic context (see the ‘Segregation measure’ section of the Methods).

\subsection*{Segregation diverges across mobility layers}\label{Results2}

\noindent\hspace*{1.5em}Urban mobility shapes social segregation, with distinct patterns across transport modes and urban contexts in Beijing metropolitan area (Fig. \ref{fig1}c). Active mobility shows hyper-localized segregation ($PSM_{\mathrm{active}} = 0.4881$), particularly within city centers' specialized functional zones (e.g., office workers in business districts) and the outskirts' homogeneous residential neighborhoods. While private cars exhibit intermediate mixing ($PSM_{\mathrm{private}} = 0.7102$), mainly along arterial roads. Public transport exhibits surprising variation: rail systems achieve the highest levels of integration ($PSM_{\mathrm{railway}} = 0.9050$) except in suburbs, whereas buses maintain considerable segregation ($PSM_{\mathrm{bus}} = 0.7257$) citywide, even in accessible districts. Segregation generally increases towards the periphery (declining \textit{PSM} values; Fig. \ref{fig1}e), yet a paradox emerges: railway-scarce suburbs show pockets of integration driven by shared bus reliance (incremental \textit{PSM} values; Fig. \ref{fig1}e). The distributions of \textit{MUI} (Fig. \ref{fig1}d,f) further expose this spatial complexity, with central areas exhibit consistent inter-modal segregation levels (high \textit{MUI}), while peripheral zones show divergent segregation experiences dependent on travel mode within the same locale, suggesting urban peripheries amplify transport-mediated stratification.

\subsection*{Segregation unfolds with residents' daily rhythms}\label{Results3}

\begin{figure*}[h]
\centering 
\includegraphics[width=1\textwidth]{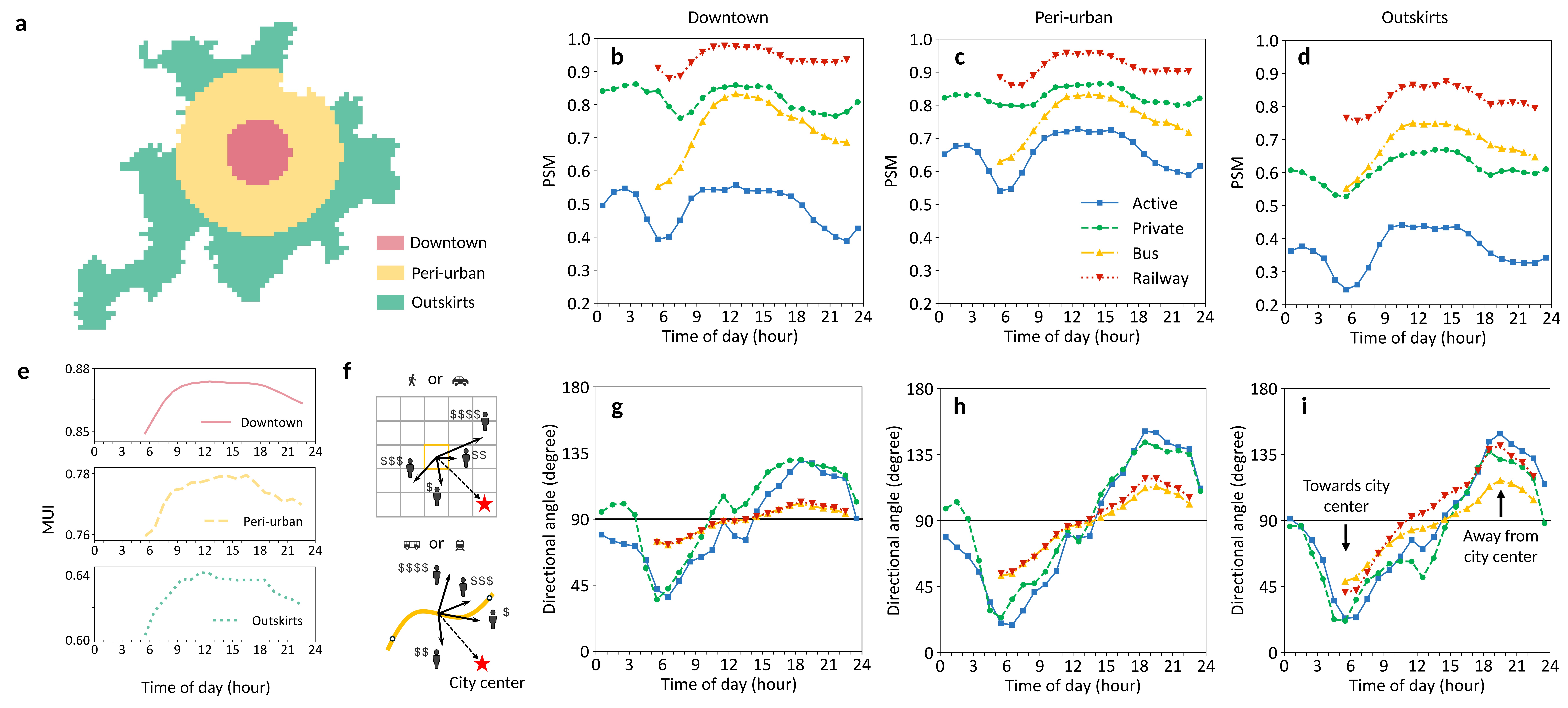}
\caption{\textbf{Daily rhythms drive multimodal segregation dynamics across urban contexts.} \textbf{a}, Beijing metropolitan area partitioned into three zones—downtown, peri-urban, and outskirts—based on equal population distribution. \textbf{b-d}, Hourly variation in \textit{PSM} shown for each transport mode within the three spatial contexts. Points indicate the average \textit{PSM} for a given mode and context during each one-hour interval, revealing a universal midday dip in segregation. \textbf{e}, Temporal fluctuations in the \textit{MUI}. The midday peak indicates the most consistent integration experience across different transport modes, contrasting with lower uniformity during morning and evening commute periods. \textbf{f}, Schematic defining the mobility flow directional angle $\theta_s$ for spatial units (grids or transit segments), calculated relative to the city center (0° towards, 180° away). \textbf{g-i}, Hourly rhythms of the context-specific average directional angle ($\bar{\theta}_\mathcal{C}$). Midday periods exhibit more balanced, multidirectional travel (mean angular dispersion approaching 90°), correlating with enhanced social mixing. In contrast, commuting hours show strongly directional flows (towards center in morning, away in evening) that reinforce socioeconomic sorting.}\label{fig2}
\end{figure*}

\noindent\hspace*{1.5em}Temporal analysis across Beijing’s downtown, peri-urban, and outskirts zones (Fig. \ref{fig2}a) reveals how daily routines shape segregation. A midday mixing window (e.g., 11:00–13:00) emerges across all transport modes, with synchronized peaks in \textit{PSM} and \textit{MUI} (Fig. \ref{fig2}b–e), driven by radially balanced mobility flows (mean angular dispersion 90°, Fig. \ref{fig2}g-i; see the ‘Mobility flow directionality’ section of the Methods). This midday integration is fueled by diverse activities, temporarily increasing inter-group contact. In contrast, morning (06:00–08:00) and evening (post-18:00) commutes intensify socioeconomic sorting, as income-stratified flows converge on downtown workplaces and disperse to peripheral residences. Notably, bus systems serve as key social integrators in transport-scarce peripheral areas, as evidenced by the relative increase in \textit{PSM} values from downtown to outskirts (Fig. \ref{fig2}b-d). These findings challenge conventional assumption that extensive public transit ensures equity \cite{Karen2012TP}, showing segregation dynamically tracks daily rhythms and infrastructural limits, informing targeted urban cohesion strategies.

\subsection*{Transport infrastructure shapes segregation dynamics}\label{Results4}

\begin{figure*}[h]
\centering
\includegraphics[width=1\textwidth]{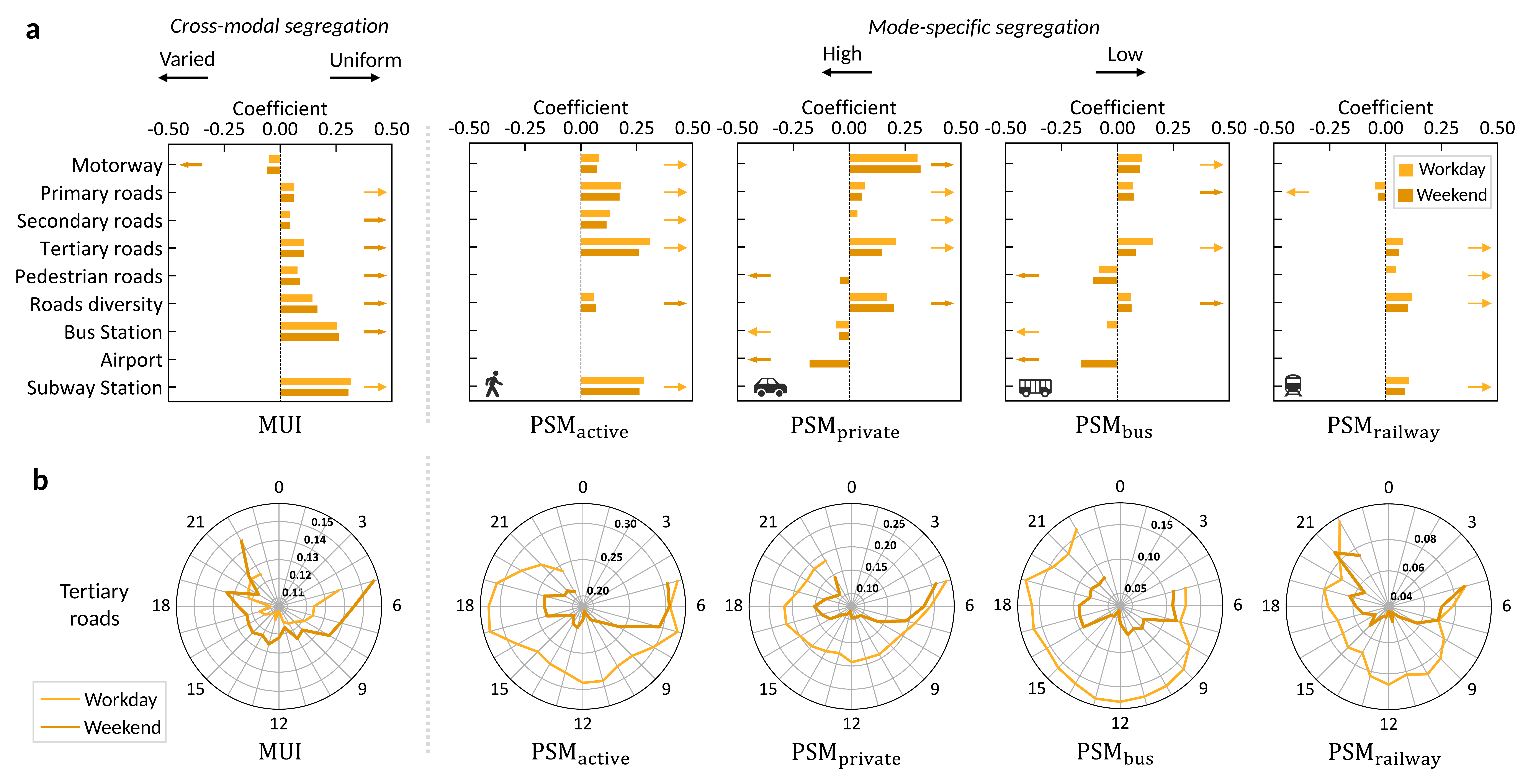}
\caption{\textbf{OLS regression models explain segregation.} \textbf{a}, Contribution of transport infrastructure features on segregation estimated using the daily granularity model. The leftmost column of the graph shows the feature names. Each feature is associated with one or two bars, representing the significant coefficients ($\beta$) on workdays and weekends respectively. A positive coefficient ($\beta>0$, rightward arrow) indicates the feature promotes mixing (for \textit{PSM}) or cross-modal uniformity (for \textit{MUI}), while a negative coefficient ($\beta<0$, leftward arrow) indicates increased segregation or more varied segregation patterns among modes. Bold arrow denotes a feature with more pronounced weekend effects, while non-bold arrow marks stronger workday effects. \textbf{b}, Variation in the influence of selected feature (Tertiary roads) throughout the day estimated using the hourly granularity model. In the compass plot, each point's angular position represents a specific hour of the day (0:00 to 23:00 arranged radially), and the associated numeric label indicates the coefficient's value for that hour and day type.}\label{fig3}
\end{figure*}

\noindent\hspace*{1.5em}To unravel how the urban built environment influences segregation, we use OLS regression models linking grid-level transport infrastructure features to mode-specific segregation (\textit{PSM}) and cross-modal uniformity (\textit{MUI}) (see the ‘OLS regression models’ section of the Methods; Supplementary Note 3). Generally, denser roads and transit stations enhance income mixing within the modes they primarily serve (Fig. \ref{fig3}a; Supplementary Table 2). For instance, motorways significantly boost mixing among private vehicle users ($\beta_{\text{PSM,Private}}^{***} > 0$), by improving accessibility for car owners across income brackets in those areas, while tertiary roads ($\beta_{\text{PSM,Active}}^{***} > 0$) and subway stations ($\beta_{\text{PSM,Active/Railway}}^{***} > 0$) benefit active and rail users. However, the impact on \textit{MUI} reveals a divergence: infrastructure supporting diverse travel enhances more equitable segregation patterns between modes (e.g., $\beta_{\text{MUI}}^{***} > 0$ for roads diversity); whereas motorways decrease uniformity ($\beta_{\text{MUI}}^{**} < 0$) by selectively boosting private transport mixing, an outcome largely due to their exclusive infrastructure that bars other modes, thus exacerbating cross-modal segregation disparities. These variances motivate further work to understand the lived experience of mixing in each modal context (i.e. how bus-based mixing differs from road-based mixing).

The hourly regression coefficients reveal infrastructure's dynamic influence, linked to daily activity patterns (Fig. \ref{fig3}b; Supplementary Figs. 18-22). Taking tertiary roads as an example, their mixing-enhancement effect peaks during workday commute times (8:00/17:00) for active/private modes and strengthens during daytime hours (9:00–19:00) for public transport, reflecting work-related travel needs. Conversely, their contribution to cross-modal uniformity (\textit{MUI}) is most pronounced on weekends, fostering more equitable segregation outcomes during leisure travel.

\subsection*{Model simulation for policy interventions}\label{Results5}

\begin{figure*}[h]
\centering 
\includegraphics[width=1\textwidth]{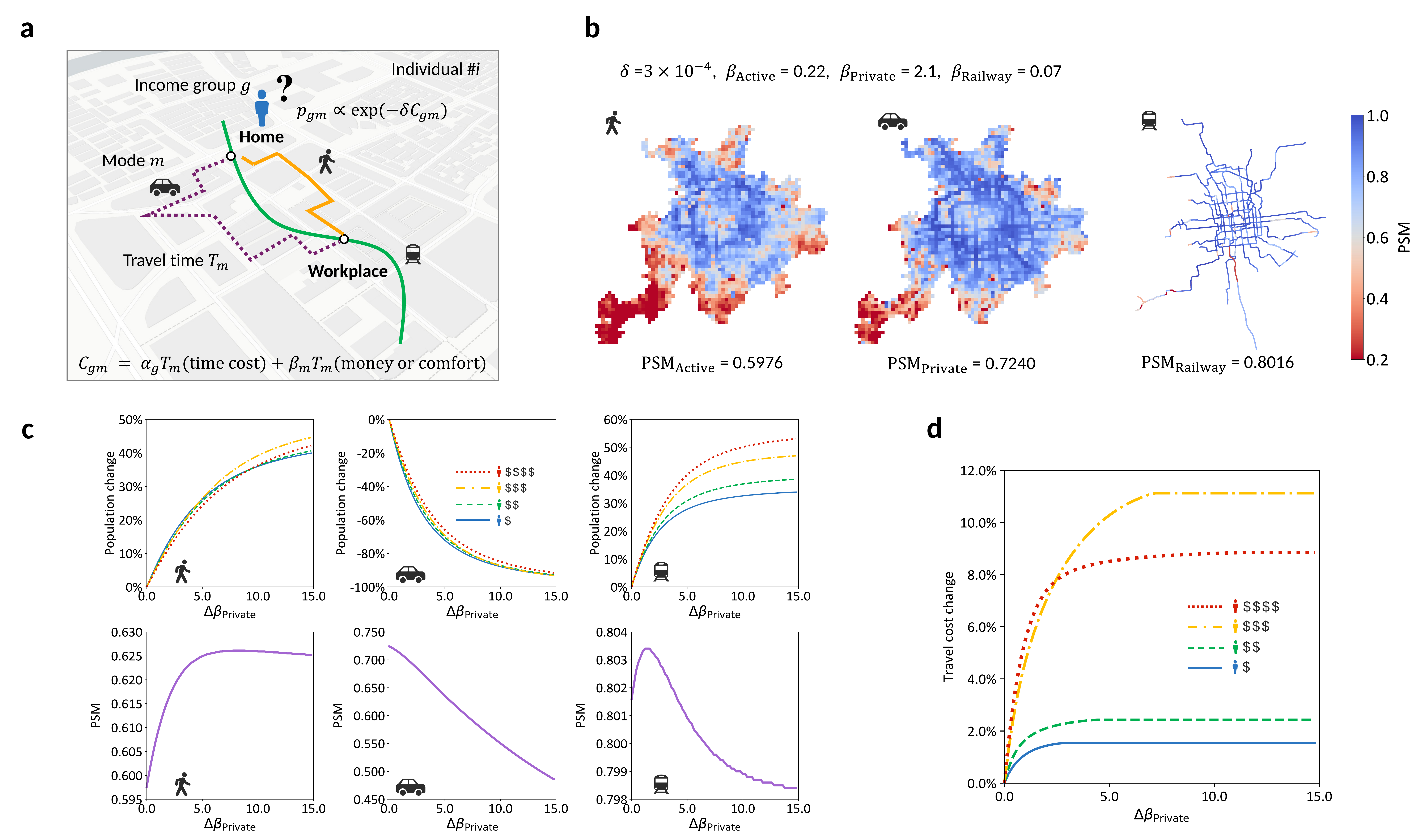}
\caption{\textbf{Individual mobility model simulates distributional effects of transport policy.} \textbf{a}, Schematic of the agent-based mobility model. Individuals choose between active, private car, or railway modes probabilistically to minimize perceived travel cost, which incorporates travel time ($T_m$), income-specific value of time ($\alpha_g$), and mode-specific costs ($\beta_m$). \textbf{b}, Spatial distribution of model-predicted \textit{PSM} for active, private, and railway travel during the morning commuting hour. The optimal model parameters calibrated using empirical data are displayed above the maps. \textbf{c}, Simulated impact of a uniform citywide increase in private car cost ($\Delta\beta_{\text{private}}$). Top row: Proportional change in mode usage relative to baseline by income groups. Bottom row: Change in overall citywide segregation measure (\textit{PSM}) for each mode. \textbf{d}, Impact of the uniform private car policy on average travel costs relative to baseline by income group, showing a generally progressive burden.}\label{fig4}
\end{figure*}

\noindent\hspace*{1.5em}To examine how income-differentiated travel preferences shape segregation and evaluate policy interventions, we develop an agent-based model of mobility grounded in discrete choice theory \cite{benakiva1985discrete}. This model simulates morning peak home-to-work commutes for distinct income groups, assuming individuals probabilistically choose between active, private car, or railway travel to minimize perceived cost (Fig. \ref{fig4}a; see the ‘Individual mobility model’ section of the Methods). The model is calibrated by matching its predicted mode-specific segregation (\textit{PSM}) with empirical data, demonstrating excellent performance in capturing observed commuting patterns (Fig. \ref{fig4}b; Supplementary Note 4.2).

Using the calibrated model, we simulate three policy archetypes (private car control \cite{Creutzig2020NCC,Cramton2018Nature}, public transport subsidies \cite{Martin1989Science}, active travel promotion \cite{Ding2024LancetPH,timmons2024active}) to assess their distributional effects across income groups and urban space. First, policies controlling private car use reveal complex outcomes sensitive to spatial design. A uniform citywide cost increase $\Delta\beta_{\text{private}}$ (akin to fuel taxes \cite{Ross2017NatureEnergy} or parking fees \cite{Donald1997plan}) most reduces lower-income car usage and shifts higher-income groups towards alternatives (Fig. \ref{fig4}c), increasing segregation among remaining drivers (\textit{PSM} decreases) while showing complex effects on active mode and railway segregation (initial mixing, then re-segregation) (Fig. \ref{fig4}c). This policy yields a progressive outcome: although travel costs rise for the society, the proportional burden increase is higher for wealthier groups (Fig. \ref{fig4}d), before a threshold is reached where other modes become more attractive. In contrast, downtown-targeted policy (mimicking congestion pricing \cite{Cramton2018Nature}) impacts fewer lower-income individuals (less commuting downtown), paradoxically reducing car users' segregation citywide while increasing it for active/railway modes due to the specific commuter groups affected (Supplementary Note 4.3). Second, public transport subsidies offer larger travel cost reductions for lower-income group but heighten segregation within rail and among remaining drivers, as lower-income groups shift disproportionately. Yet, overall impacts appear muted even with significant subsidies, suggesting diminishing returns from further fare-based interventions alone (Supplementary Note 4.4). Third, promoting active travel (through infrastructure or safety improvements \cite{timmons2024active}) also delivers progressive benefits and reduces overall active/private segregation, but creates stark spatial contrasts: fostering mixing downtown while increasing segregation in suburbs for active travel, with inverse spatial effects for other modes (Supplementary Note 4.5). Overall, transport policies generate complex, spatially-dependent distributional outcomes, revealing critical trade-offs between efficiency and social equity that demand integrated planning.

\section*{Discussion}\label{Discussion}

\noindent\hspace*{1.5em}Social segregation emerges a persistent challenge in rapidly urbanizing societies. Multimodal transportation systems hold promise for mitigating segregation by facilitating interactions among diverse populations, especially in sprawling, car-centric cities \cite{Pappalardo2023Naturecom,Laura2023EnvirB}. To examine their actual effectiveness, we proposed a dynamic approach to quantify how segregation unfolds through daily transient encounters between individuals as they travel by different modes. Departing from conventional segregation measures reliant on precise spatiotemporal overlaps, our metrics (\textit{PSM} and \textit{MUI}) reconceptualize segregation as a dynamic, relational phenomenon, shifting focus from where people live or go, towards the likelihood of whom they encounter during their daily journeys under multimodal infrastructural and temporal conditions. This interaction-based perspective allows us to assess segregation not as a fixed spatial or spatiotemporal pattern, but as an evolving social experience embedded within fabrics of urban life. Through this revised lens discrete experiences of segregation can occur in the same space and time.

Our findings challenge the longstanding paradigm \cite{Eagle2006Personal,Chetty2022Nature1,Chetty2022Nature2} that physical proximity or shared infrastructure ensure social integration. We demonstrate that the mixing potential of mobility infrastructure is contingent on when, where, and how diverse populations use it. High-capacity rail systems achieve integration only within narrow temporal windows (e.g., midday), while localized bus networks paradoxically sustain segregation through fragmented routes and income-stratified travel schedules, as our data confirm. This resonates strongly with time-geography principles \cite{Nigel1981Progress}, as interaction opportunities are constrained by societal rhythms. Crucially, our simulations reveal that policy interventions aiming to influence mobility yield asymmetric impacts on segregation and travel costs across income groups and neighborhoods. These findings compel infrastructure-centric urban design \cite{MichaelScience2008,GrahamMarvin2001} to move beyond simple expansion and instead embrace strategies that coordinate infrastructure, policy, and operations with urban rhythms, a perspective central to the concept of chrono-urbanism \cite{moreno2021}.

While the probabilistic framework provides a powerful, transferable lens for analyzing dynamic segregation, our empirical findings are derived solely from Beijing. Therefore, the direct generalizability of the specific quantitative results to other global cities requires caution. Further comparative studies across diverse urban settings can test the universality of observed patterns, and provide further evidence of the role of urban and transport design in shaping social processes.

%TC:ignore

\section*{Methods}\label{Methods}
\subsection*{Datasets}\label{Methods1}
\bmhead{Mobile phone data} The anonymous mobile phone dataset come from a telecommunications service provider in China. The dataset was collected over one-month period (June) in 2023, and consists of anonymized records of GPS locations (“pings”) from users that have signed up to provide data through explicit consent agreements outlined in the telecoms company’s privacy notice. The mobile phone users were informed about how their data will be used and stored under the regulation of China’s Personal Information Protection Law (PIPL). In alignment with China's data protection and cybersecurity regulations, all data are thoroughly anonymized to remove personally identifiable information and processed to ensure privacy. This work was exempt from the ethical review by the Business, Environment, Social Sciences Faculty Research Ethics Committee in the University of Leeds (Reference: {BESS+ FREC - 2024 1663-2108}). The raw data consist of 11,018,253 users (covering 50\% population) and 8,386,564,428 pings in Beijing, China. Each ping consists of a de-identified user ID, latitude, longitude and timestamp. The mean number of raw pings associated with a user in a month is 1008 and the median number of pings is 872. To ensure data reliability, we removed duplicate pings to ensure accuracy. We filtered out users with fewer than 300 pings to eliminate noise. The filtered dataset consists of 7,562,482 users and 4,815,969,539 pings. The population representativeness of the dataset is validated in Supplementary Note 1.2.

\bmhead{LianJia housing} The LianJia housing dataset \cite{LianJia}, sourced from China’s largest real-time property transaction platform, offers a detailed repository of residential property information in Beijing. As of June 2023, this dataset encompasses 9,501 communities, covering 97.3\% of Beijing’s residential market. This dataset provides granular metrics including average transaction prices (RMB/m²), architectural typologies, household counts, and precise geographic coordinates. As China’s residential neighborhoods are predominantly planned, gated communities with homogeneous pricing and shared amenities, the dataset captures near-complete market dynamics, enabling robust socioeconomic inference.

\bmhead{Transport facilities} We collected a dataset of 64,805 transport facilities in Beijing, China, using Application Programming Interface (API) of Amap \cite{Amap}, a leading mapping and navigation service provider in China. Each record includes name, address, latitude, longitude and category. The categories of transport facilities encompassed in this dataset include: Airport, Train Station, Port, Intercity Bus Station, Subway Station, Bus Station, Parking Lot, Toll Station, and Highway Service.

\bmhead{Transport networks} We retrieved the urban road networks from OpenStreetMap \cite{OpenStreetMap}, including both driving roads and pedestrian pathways. Urban road networks in Beijing consist of 307,978 junctions and 437,337 segments. We used Amap API \cite{Amap} to obtain public transport networks, encompassing bus and railway routes. The API generates 4,238 directional bus routes (15,876 stations and 44,370 segments) and 118 directional railway routes (432 stations and 1,028 segments) in Beijing. Each record includes the route names, station names, station coordinates (latitude and longitude), average travel duration between stations, and route operational hours.

\bmhead{Geolife} The Geolife GPS Trajectory dataset \cite{geolife2009} released by Microsoft Research Asia  comprises 17,621 trajectories collected from 182 volunteer users between April 2007 and August 2012. It includes raw GPS points (latitude, longitude, timestamp) and labels for transport modes. We used a processed subset of this dataset, totaling 4,425 trips with unambiguous labels (1,819 active, 881 private, 1,725 public) for training and validating our travel mode inference model (Supplementary Note 1.4) and validating route generation accuracy (Supplementary Note 1.5).

\bmhead{MemDA} The MemDA (Memory-Augmented Deep Architecture) dataset \cite{Cai2023MemDA} is an open-sourced dataset providing traffic speeds on major road segments in Beijing over a 75-day period (May 12 - July 25, 2022). This dataset serves as an independent ground truth to validate the reliability of our inferred mobility patterns, particularly the travel mode inference and route generation processes (Supplementary Note 1.6).

\subsection*{Inferring home locations and workplaces}\label{Methods2}

 \noindent\hspace*{1.5em}We first detected significant stays during individuals' trips through a spatiotemporal clustering approach \cite{Hariharan2004}. These stays were defined by high-density clusters of trajectory points within a 50-meter radius and a minimum of 10 points, subsequently linked to nearby POIs within 100 meters for contextual validation. Stays shorter than 15 minutes or longer than 24 hours were excluded to ensure focus on meaningful activities. Home locations were determined by analyzing stays during nighttime hours (21:00–6:00) across multiple days, selecting the location with the highest cumulative duration and at least 25 visits over the 30-day period, situated in residential areas as confirmed by Amap POIs. Workplaces were identified from stays during typical working hours (9:00–17:00) on workdays, requiring a minimum of 4 visits per 5 workdays, cross-referenced with commercial POIs. This method reliably assigned home/work locations to 604,7368 individuals. The inference reliability was validated by comparing home-based trip frequencies with China’s National Population Census in 2020 \cite{NBSCensus2021}, adjusted for population growth (see Supplementary Note 1.1 and 1.2).

\subsection*{Inferring income quartiles}\label{Methods3}

\noindent\hspace*{1.5em} To estimate socioeconomic status, we leveraged China's distinct residential structure—characterized by uniformly priced, gated communities with shared amenities—which enables housing expenditure to serve as robust proxies for income stratification. Individuals' geolocated homes were systematically matched to property transaction data from LianJia, China's largest real estate platform (\textit{n} = 9,501 Beijing communities). We constructed Voronoi polygons around each community to delineate localized socioeconomic zones, ensuring spatial contiguity while minimizing edge effects through nearest-neighbor allocation. Individuals within a polygon were assigned corresponding community’s average transaction price as a proxy for socioeconomic status. This method capitalizes on the inherent homogeneity of Chinese residential communities where housing prices strongly correlate with residents' economic capacity. Validation confirms 80\% of inferred home locations resided within 250 meters (90\% within 500 meters) of matched communities, ensuring spatial reliability. Individuals were categorized into four income quartiles based on 25th, 50th and 75th percentiles of community price distributions. Spatial analysis reproduces the distinct residential segregation patterns in Beijing, reinforcing the validity of our approach (see Supplementary Note 1.3).

\subsection*{Segregation measure}\label{Methods4}
\noindent\hspace*{1.5em} To quantify segregation experiences shaped by multimodal mobility, we developed two entropy-based measures grounded in probabilistic co-location patterns. The foundation is the calculation of the likelihood that individuals are simultaneously present at specific locations along their journeys. This involves first estimating the probability of individuals choosing specific transport modes—active (walking/cycling), private (car/taxi), bus, or railway—for each trip using a machine learning model pre-trained on Geolife GPS trajectory dataset (see Supplementary Note 1.4). The most probable route for each potential mode is then generated using the Amap navigation API, incorporating real-time traffic, transit schedules, and original GPS waypoints to enhance realism (see Supplementary Note 1.5). The likelihood of any two individuals encountering each other via the same mode \textit{m} within a specific spatial unit (\textit{s}) is calculated as the product of the probabilities that their routes intersect within that unit and their estimated travel times overlap. The reliability of these co-location likelihood estimates was confirmed via cross-validation using MemDA traffic speed data (see Supplementary Note 1.6). Aggregating these encounter probabilities across all individuals belonging to the four inferred income quartiles (\textit{q}), we estimated the expected population mix in each spatial unit $pop_{qsm} = \sum_{i \in q} p_{ism}$, where $p_{ism}$ is the probability that individual \textit{i} from group \textit{q} occupies spatial unit \textit{s} via mode \textit{m} during a specific time window. Our first measure, the mode-specific Probabilistic Segregation Measure (\textit{PSM}), quantifies the income group diversity within this expected population mix using a normalized entropy metric \cite{Shannon2001}:

 \begin{equation}
PSM_{s,m} = -\frac{1}{\log(4)} \sum_{q} \tau_{qsm} \log\big(\tau_{qsm}\big),
\label{eq:PSM_{s,m}}
\end{equation}

\noindent where $\tau_{qsm} = pop_{qsm}/\sum_{q} pop_{qsm}$. $PSM_{s,m}$ spans 0 (complete segregation, single-group dominance) to 1 (full mixing, equal group representation).

Our second measure, the Multimodal Uniformity Index (\textit{MUI}), evaluates the regional consistency of these segregation experiences across the different transport modes within a specific geographic area \textit{A}. It is calculated by computing the normalized entropy \cite{Shannon2001} of the mode-averaged \textit{PSM} values within that region:

 \begin{equation}
MUI_{A} = -\frac{1}{\log(4)} \sum_{m} r_{A,m} \log\big(r_{A,m}\big),
\label{eq:MUI_{A}}
\end{equation}

\noindent where $r_{A,m}=PSM_{A,m}/\sum_{m} PSM_{A,m}$. Here, $PSM_{A,m}$ represents the average segregation level across all the mode-specific spatial units located at region \textit{A}. A higher $MUI_{A}$ value (close to 1) indicates greater uniformity in segregation (or mixing) experiences across four modes within region \textit{A}. Conversely, a lower value (closer to 0) signals more divergent experiences. Together, these measures capture how distinct mobility networks—each with unique spatial granularity and interaction dynamics—collectively shape socioeconomic exposure in cities.

\subsection*{Mobility flow directionality}\label{Methods5}

\noindent\hspace*{1.5em}We adopt a vector-based approach \cite{zhao2024nc} to capture the spatial orientation of population movements across urban contexts. The city center is first defined as the point of highest population density within the Beijing metropolitan area. For each spatial unit (1 km × 1 km grid for active and private modes; transit segment for bus and railway modes), mobility flow vector for each groups is calculated based on trip origin-destination pairs recorded over hourly intervals. The vector magnitude represents the total number of trips weighted by the probabilities of individuals from the specific group being present in that spatial unit during the given hour. Next, the group-specific vectors within each spatial unit \textit{s} are summed into a composite mobility flow vector $\vec{V}_s$. The directional angle of this composite vector is then computed as the angle between the vector and the reference line connecting the unit’s centroid to the city center: $\theta_s = \arccos\left( \frac{\vec{V}_s \cdot \vec{e}_c}{\lVert \vec{V}_s \rVert} \right)$, where $\vec{e}_c$ is the unit vector pointing from \textit{s} to city center. Angles measured in radians are converted to degrees on a 0°–180° scale (0°: strictly toward center, 180°: strictly outward). For each urban context \(\mathcal{C}\) (downtown, peri-urban, or outskirts), the mean directional angle \(\bar{\theta}_\mathcal{C}\) in a given hour is computed as the arithmetic average of directional angles from all spatial units within the context: $\bar{\theta}_\mathcal{C} = \frac{1}{N} \sum_{s \in \mathcal{C}} \theta_s$, where \textit{N} denotes the number of spatial units.

\subsection*{OLS regression models}\label{Methods6}

\noindent\hspace*{1.5em}We employ ordinary least squares (OLS) regression model \cite{Greene2018} to quantify how transport infrastructure influences spatiotemporal segregation patterns. These models operate at the 1 km × 1 km grid level, linking grid-specific segregation measures to transport-related explanatory variables $\{T_i\}$. Specifically, either the probabilistic segregation measure for a given mode \textit{m} ($PSM_{A,t,m}$) or the multimodal uniformity index ($MUI_{A,t}$) within grid \textit{A} at time \textit{t} (collectively denoted $M_t$) is modeled as a function of these transport infrastructure features. The features $\{T_i\}$ encompass characteristics such as the lengths of different road types, road type diversity, and counts of transport facilities. A detailed list and description of these grid-level variables are provided in Supplementary Table 1. To capture fine-grained temporal dynamics, we fitted separate models for each hour of the day, distinguishing between workdays and weekends (see Supplementary Note 3 for further details on the model specification).

\subsection*{Individual mobility model}\label{Methods7}

\noindent\hspace*{1.5em}An agent-based model of individual mobility is developed to simulate how commuters' travel mode choices influence segregation at the city scale. For a trip (given origin and destination), an individual $i$ from income group $g \in \mathcal{G}$ ($|\mathcal{G}|=4$) probabilistically chooses a mode $m \in \mathcal{M} = \{\text{active, private, railway}\}$ to minimize perceived travel cost $\mathcal{C}_{gm}$ using a multinomial logit formulation \cite{mcfadden1974}

\begin{equation}
p_{gm} = \frac{\exp(-\delta C_{gm})}{\sum_{m' \in \mathcal{M}} \exp(-\delta C_{gm'})},
\label{eq:p_{gm}}
\end{equation}

\noindent where $\delta$ denotes the sensitivity parameter. Perceived cost $\mathcal{C}_{gm}$ combines income-specific value of time ($\alpha_g$) and other mode-specific costs ($\beta_m$; additional monetary costs or inconvenience) applied to travel time $T_m$ (computed as the duration of the shortest path $\textbf{R}_{m}$ traveling via mode \textit{m} within urban transport networks)

\begin{equation}
\mathcal{C}_{gm} = \alpha_g T_m + \beta_m T_m.
\label{eq:mathcal{C}_{gm}}
\end{equation}

We use the model to simulate travel preferences of individuals during the morning peak hour (9:00–10:00 AM) as they commute from their homes to workplaces. The model outputs, including individual probabilities $p_{gm}$ and shortest paths $\textbf{R}_{m}$, are inputs for calculating the mode-specific segregation measure (\textit{PSM}). Model parameters are calibrated using empirical data: Relative time values ($\alpha_g$) are fixed based on group incomes, while the sensitivity ($\delta$) and mode-specific costs ($\beta_m$) are estimated by minimizing the difference between predicted and observed \textit{PSM} during the peak hour. Detailed calibration procedures and performance validation are provided in Supplementary Note 4.2.

We utilize the calibrated model to evaluate the distributional effects of three common transport policy archetypes. These policies are simulated by systematically modifying the relevant calibrated mode-specific cost parameters ($\beta_m$) to reflect their impact on perceived travel costs. For each policy scenario, we recalculate individual mode choices and analyze the resulting changes in mode-specific \textit{PSM}, mode shares, and average travel costs across income groups to assess impacts on social benefits and equity. Detailed simulation setups and results are provided in Supplementary Notes 4.3–4.5. 

\section*{Data availability}\label{Data availability}

 \noindent\hspace*{1.5em}The source data supporting the findings of this study, including segregation measures across space and time, OLS regression model results, and policy simulation outcomes, are available online (\url{https://github.com/UrbanMobility-y/multimodal-segregation/tree/main/Source%20data}). The raw mobile phone mobility data are not publicly available to preserve individual privacy and user confidentiality. The datasets of transport facilities and public transport networks are commercially available and may be requested for research use (\url{https://lbs.amap.com/}). Road networks were from OpenStreetMap (\url{https://www.openstreetmap.org/}). LianJia housing transaction data (\url{https://bj.lianjia.com/xiaoqu/}), China National Population Census 2020 data (\url{https://www.stats.gov.cn/english/PressRelease/202105/t20210510_1817185.html}), Geolife trajectory data (\url{https://www.microsoft.com/en-us/download/details.aspx?id=52367}) and MemDA traffic speed data (\url{https://github.com/deepkashiwa20/Urban_Concept_Drift}) are available online.

\section*{Code availability}\label{Code availability}

 \noindent\hspace*{1.5em}All analysis was conducted using Python. Code is publicly available at GitHub (\url{https://github.com/UrbanMobility-y/multimodal-segregation}).

\section*{Acknowledgments}

\noindent\hspace*{1.5em}We thank our funders who supported this research - the Leverhulme Trust via Philip Leverhulme Prize funding (Ref: PLP-2022-262) and the National Natural Science Foundation of China (Refs: 72288101, 72242102).

\section*{Author contributions}

\noindent\hspace*{1.5em}Y. Y. and E. M. conceived the study, overall research design, and developed the probabilistic framework. Y. Y. and E. L. performed data processing, implemented the software, ran the formal analyses and simulations, prepared the figures, and drafted the manuscript. B. J. curated the mobility data and provided transport-systems expertise in result interpretation. E. M. supervised the project, provided critical feedback, and reviewed and edited the manuscript. All authors discussed the results and approved the final version of the manuscript.

\section*{Competing interests}

The authors declare no competing interests.

\bibliography{sn-article}

%TC:endignore

\end{document}

\title[Article Title]{Supplementary Information for `Urban transport systems shape experiences of social segregation'}

\author[1]{\fnm{Yitao} \sur{Yang}}\email{y.yang@leeds.ac.uk}
\equalcont{These authors contributed equally to this work.}
\author[2]{\fnm{Erjian} \sur{Liu}}\email{17120752@bjtu.edu.cn}
\equalcont{These authors contributed equally to this work.}
\author[2]{\fnm{Bin} \sur{Jia}}\email{bjia@bjtu.edu.cn}
\author*[1]{\fnm{Ed} \sur{Manley}}\email{e.j.manley@leeds.ac.uk}
\affil*[1]{\orgdiv{School of Geography}, \orgname{University of Leeds}, \city{Leeds}, \country{UK}}
\affil[2]{\orgdiv{School of Systems Science}, \orgname{Beijing Jiaotong University}, \city{Beijing}, \postcode{100044}, \country{China}}

\maketitle

\tableofcontents  
\clearpage 
\listoffigures  
\listoftables 
\clearpage

\section*{Supplementary Information}\label{Supplnotes}
\section{Mobility data treatment}\label{Supplnote1}
\subsection{Home and workplace identification}\label{Supplnote1.1}
\noindent\hspace*{1.5em}In this study, we utilize an anonymized mobile phone dataset provided by a telecommunications company in China, covering one month (June 2023) of GPS “pings” from users who gave explicit consent. The dataset, which complies with China’s Personal Information Protection Law, includes de-identified user IDs, latitudes, longitudes, and timestamps, ensuring privacy and preventing re-identification attempts. After removing duplicates and excluding users with fewer than 300 pings, our final dataset comprises 7.56 million users and roughly 4.82 billion pings.

To ensure robust inference of home and workplace locations from trajectory data, we implement a multi-stage methodology with rigorous validation. For each individual, we first detect significant stays using the DBSCAN algorithm \cite{ester1996density}.  We set the tuning parameters carefully for spatial distance of 50 meters and minimum 10 points to identify high-density clusters of trajectory points, representing visited places or stays. Once clusters are formed, we assign each cluster to the nearest Point of Interest (POI) within a predefined radius of 100 meters, ensuring that each significant stay is contextually anchored to a known venue. Clusters that are too small (fewer than 10 points) or that do not correspond to any recognized venue are discarded to minimize noise and improve data reliability. Next, we refine these significant stays by applying temporal filters to capture meaningful activities. Specifically, we exclude any stays with durations of less than 15 minutes, as such brief stops are unlikely to represent significant activities, and we also filter out stays exceeding 24 hours, which may indicate data errors. We have extracted 241 million trips for 7.56 million users. The trip characteristics in Beijing metropolitan area are shown in Supplementary Fig. \ref{FigS1}.

\begin{figure}[h]
\centering
\includegraphics[width=1\textwidth]{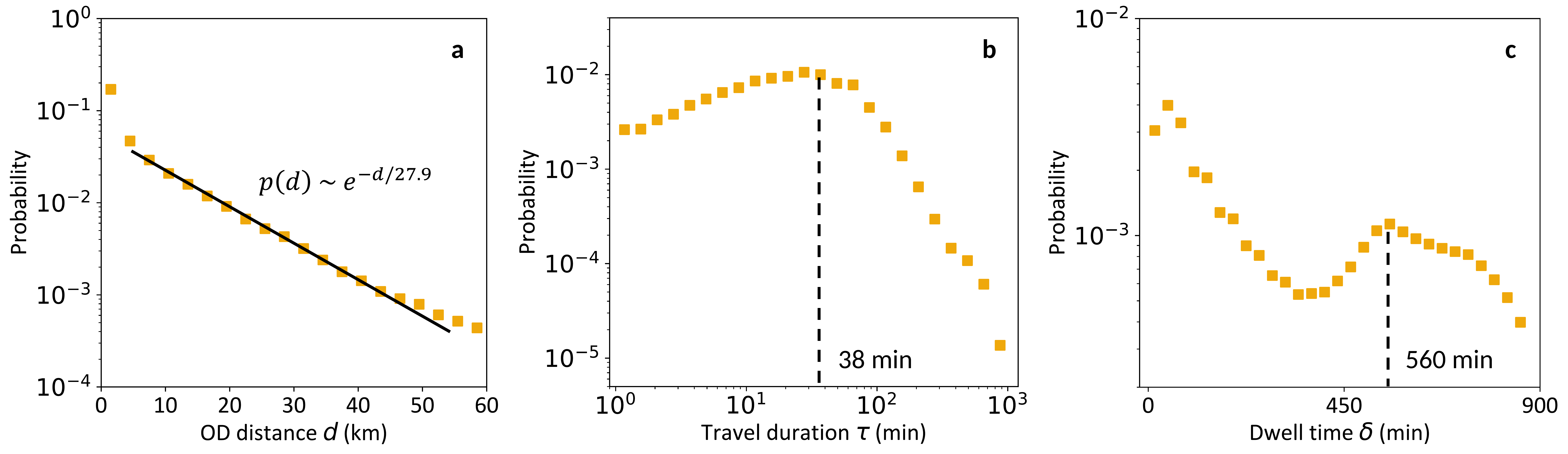}
\caption[Trip characteristics in Beijing metropolitan area.]{\textbf{Trip characteristics in Beijing metropolitan area.} \textbf{a} Distribution of trip distance \textit{d}, which follows an exponential decay \(p(d) \sim e^{-d/27.9}\). \textbf{b} Distribution of travel duration $\tau$, with most trips lasting around 38 minutes. \textbf{c} Distribution of dwell time at stays $\delta$, capturing that residents tend to remain at home for around 560 minutes.}\label{FigS1}
\end{figure}

To infer home locations, we analyze the refined stays by examining both their temporal patterns and visit frequencies. We focus on locations visited most frequently during nighttime hours (i.e., 21:00 to 6:00) across multiple days. The location with the highest cumulative duration during these nighttime periods is designated as the likely home location for an individual, provided that it meets a minimum threshold of 25 visits over the 30-day observation period. Additionally, we validate these candidates by comparing weekend visit durations, under the assumption that true home locations typically exhibit higher activity during weekends. Amap residential POIs are leveraged to ensure that the identified candidate is situated within a residential area. If multiple locations meet these criteria, the candidate with the longest total nighttime duration is selected as the home location. For workplace detection, we apply a similar approach by focusing on significant stays during typical working hours (i.e., 9:00 to 17:00) on workdays. We identify locations that exhibit both high frequency and long cumulative durations of stays during these hours, setting a threshold of at least 4 visits per 5 workdays to qualify as potential workplace candidates. Amap commercial POIs are used to validate these candidates, ensuring that the identified location is consistent with common workplace settings. Among the candidates, the location with the longest total working-hour duration is selected as the workplace for each individual. We successfully infer home locations and workplaces of 6.05 million individuals, ensuring that only users with robust and consistent activity patterns are included in the final dataset. Unidentified users, for whom home or workplace cannot be reliably determined, are excluded to maintain the accuracy and reliability of our analysis.

\begin{figure}[h]
\centering
\includegraphics[width=1\textwidth]{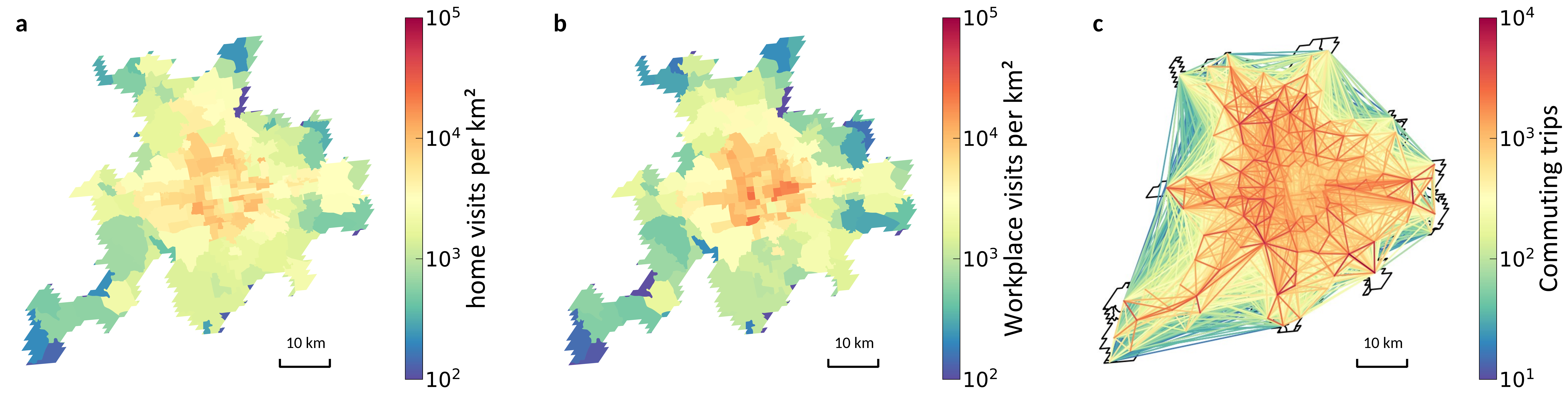}
\caption[Jobs-housing structure of Beijing metropolitan area.]{\textbf{Jobs-housing structure of Beijing metropolitan area.} \textbf{a} Spatial clusters of home locations. Individuals' home-based trips are aggregated to township administrative boundaries, and the mobility counts are normalized by jurisdictional area (km²). \textbf{b} Spatial clusters of workplaces. \textbf{c} Spatial distribution of commuting flows.}\label{FigS2}
\end{figure}

\begin{figure}[h]
\centering
\includegraphics[width=1\textwidth]{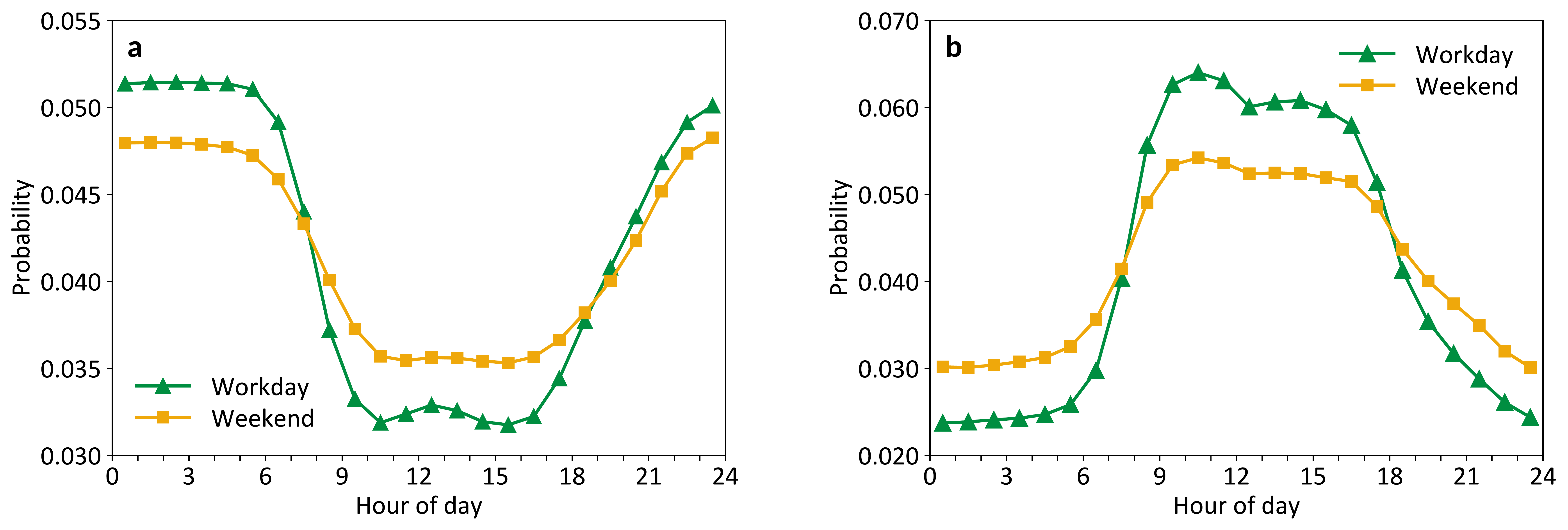}
\caption[Daily patterns  of time spent at homes and workplaces.]{\textbf{Daily patterns  of time spent at homes and workplaces.} \textbf{a} Home presence probabilities across hours of day. \textbf{b} Workplace presence probabilities across hours of day.}\label{FigS3}
\end{figure}

Distributions of home locations and workplaces of individuals (Supplementary Fig. \ref{FigS2}) suggest that both residential and employment densities are higher in the city center, while commuting flows tend to be more localized within nearby neighborhoods. Daily movements between homes and workplaces show predictable time patterns. People's time spent at home and workplaces follows opposite daily rhythms---home presence typically peaks overnight particularly on weekends, whereas workplace presence peaks during daytime hours and on workdays (Supplementary Fig. \ref{FigS3}).

\subsection{Population representativeness}\label{Supplnote1.2}

\noindent\hspace*{1.5em}We validate the population representativeness of mobile phone data through cross-validation with China's Seventh National Population Census \cite{NBSCensus2021}. This nationwide census, conducted by the National Bureau of Statistics (NBS), provides comprehensive demographic data across 41,636 township-level administrative units encompassing all 31 provincial divisions. To address temporal discrepancies between the decennial census (2020) and our mobility dataset (2023), we incorporate annual population growth estimates (0.87\% average increase) derived from Beijing Municipal Statistical Yearbooks (2020-2023) \cite{beijingstats2020,beijingstats2021,beijingstats2022,beijingstats2023}. The population distribution in Beijing's metropolitan region is visually summarized in Supplementary Fig. \ref{FigS4}a. Our validation strategy involves examining the correlation between census-recorded resident populations and home-based trip frequencies derived from mobility data at matched township units. This approach is grounded on the inherent stability of residential behavior \cite{PierrePNAS2014}, hypothesizing that home-based trip frequencies constitute reliable proxies for static population distributions. Across all township units, we compute Pearson’s correlation coefficient (\textit{r}) between the two datasets. A robust coefficient of \textit{r} = 0.8426 (***\textit{p} $<$ 0.001; 95\% CI [0.7943, 0.8804]) indicates a statistically significant positive association, explaining 71.0\% of shared variance (\textit{r}$^2$ = 0.710) (Supplementary Fig. \ref{FigS4}b). This strong correspondence confirms the capacity of mobile phone data to reliably approximate population distribution patterns at fine spatial scales.

\begin{figure}[h]
\centering
\includegraphics[width=1\textwidth]{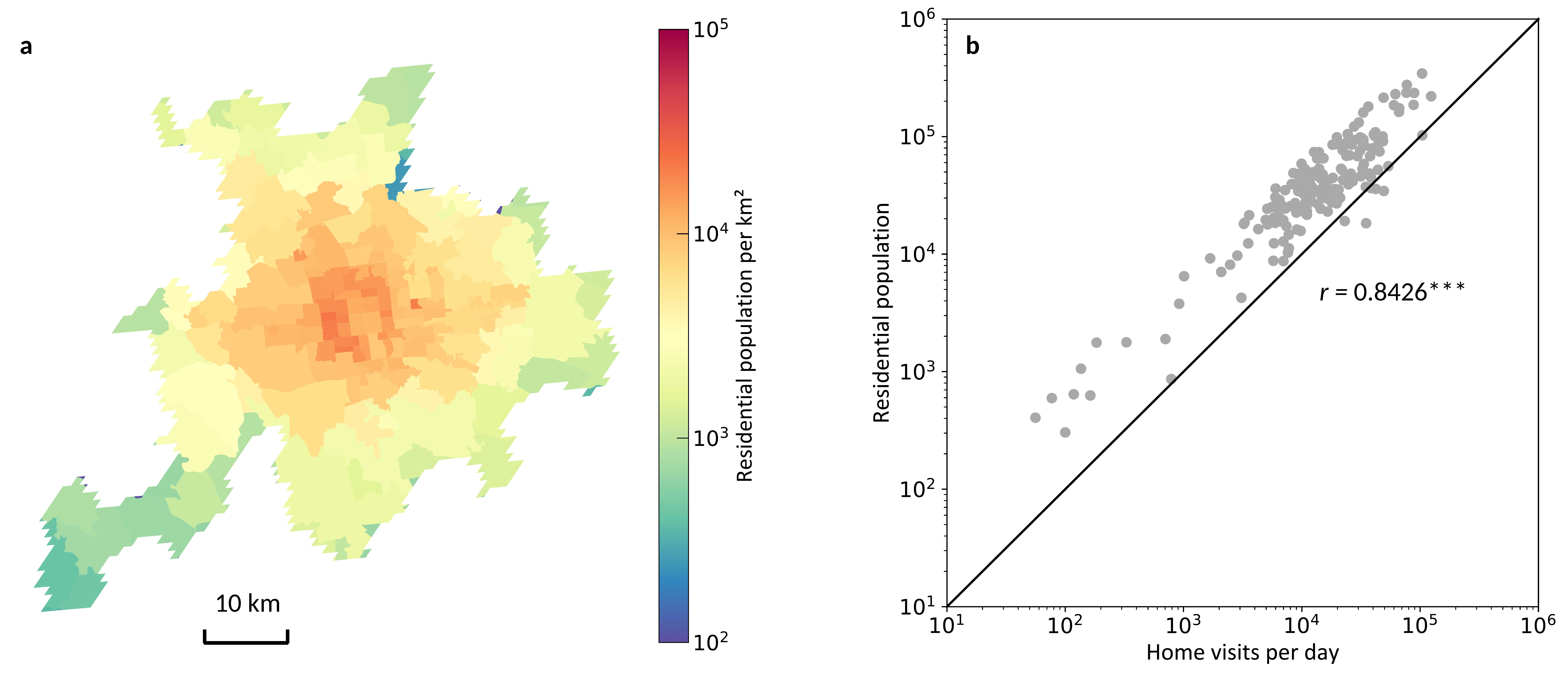}
\caption[Population representativeness validation.]{\textbf{Population representativeness validation.} \textbf{a} Census population distribution in Beijing metropolitan region. \textbf{b} Scatter plot illustrating correlation between census population and home-based trip frequency indicated by Pearson’s correlation coefficient \textit{r}, with the diagonal line providing the reference. Point represents a township-level administrative unit. *** indicates statistical significance\textit{p} $<$ 0.001.}\label{FigS4}
\end{figure}

\subsection{Socioeconomic status inference}\label{Supplnote1.3}

\noindent\hspace*{1.5em}To infer the income levels of individuals, we establish a connection between the individuals' inferred home locations and the LianJia property database---China's largest real-time property transaction platform covering 97.3\% of residential markets. This database provides detailed, geotagged records of residential communities, including information such as community name, average transaction price (in RMB per square meter), architectural type (high-rise towers, slab complexes, bungalows), number of households and buildings in the community, and exact location (precise latitude and longitude coordinates). It is important to note that housing in China is typically organized into well-defined residential communities. Unlike many Western settings where neighborhoods might comprise a mix of varied housing styles and unplanned developments, these communities are generally gated, uniformly managed, and offer shared amenities such as green spaces and retail facilities. This structured arrangement not only standardizes property types within a community but also results in more homogenous pricing and quality measures across the board. Therefore, platforms like LianJia can efficiently capture a near-complete snapshot of the housing market, thereby serving as a reliable proxy for inferring the socioeconomic status of residents based on their home locations.

In the Beijing metropolitan area, the database lists 9,501 communities with price data as of June 2023, as illustrated in Supplementary Fig. \ref{FigS5}a. We perform a spatial query to match the individual's home coordinates (latitude and longitude) with communities listed in the LianJia database. Specifically, we construct Voronoi polygon around each community, creating non-overlapping zones where all points within a polygon are geographically closer to its central community than to others (Supplementary Fig. \ref{FigS5}b). These Voronoi polygons effectively capture localized market conditions, as residents within the same polygon are likely to experience similar socioeconomic environments. For every individual, we identify the polygon containing the inferred home location and assign the associated community transaction price as an approximate measure of that individual’s income level. To ensure reliability, we compute the geographic distance between each individual's home and the matched community. Our analysis reveals that 80\% of home locations are within 250 meters of a community, and 90\% are within 500 meters (Supplementary Fig. \ref{FigS5}c). These findings confirm that the majority of individuals reside in close proximity to the communities used in our analysis, thereby demonstrating the robustness of income inference.

\begin{figure}[h]
\centering
\includegraphics[width=1\textwidth]{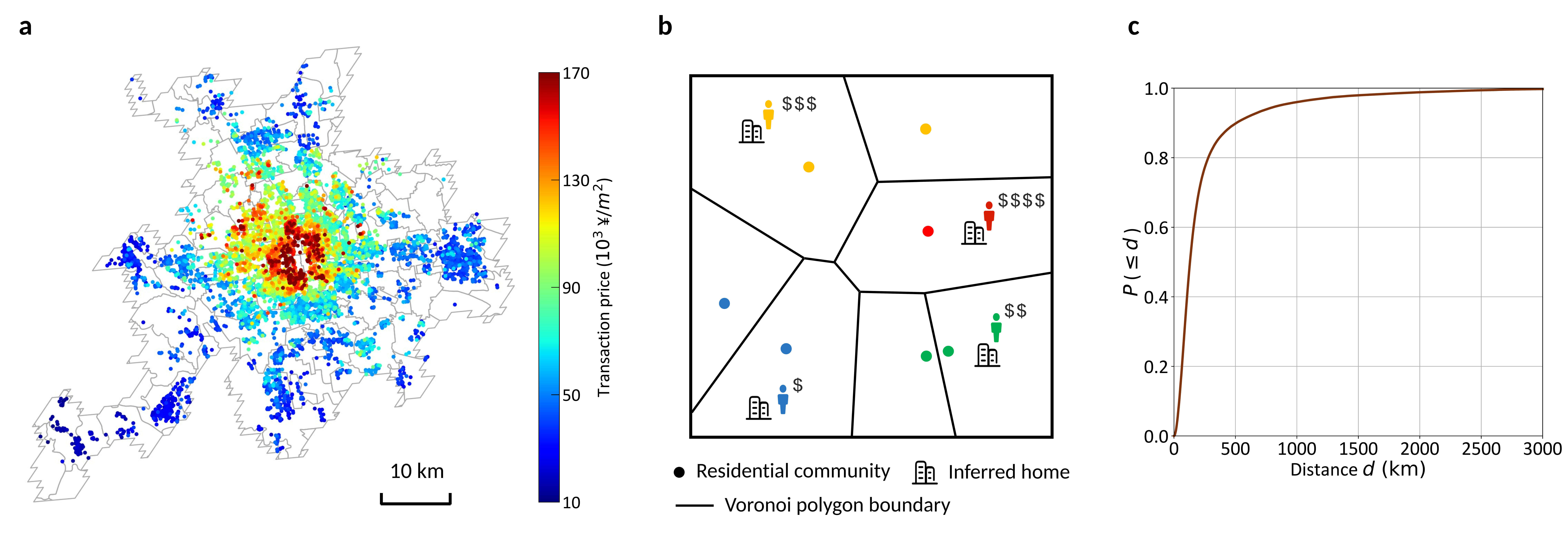}
\caption[Spatial matching analysis based on LianJia property data.]{\textbf{Spatial matching analysis based on LianJia property data.} \textbf{a} Geospatial visualization of 9,501 communities listed on the LianJia platform, with color coding indicating average transaction prices as of June 2023. \textbf{b} Construction of Voronoi polygons around each community. Individual's home location within a polygon is assigned the transaction price of the corresponding community, establishing localized socioeconomic proxies. \textbf{c} Distance distribution between residences and matched communities.}\label{FigS5}
\end{figure}

Individuals are divided into four equal groups based on the 25th, 50th, and 75th percentiles of the inferred income levels derived from the matched community transaction prices. Each quartile corresponds to a distinct income group. For instance, the first quartile, which contains individuals with property transaction prices at or below the 25th percentile, is assumed to represent the lower-income group. The second quartile (between the 25th and 50th percentiles) represents the lower-middle-income group, the third quartile (between the 50th and 75th percentiles) represents the upper-middle-income group, and the fourth quartile (above the 75th percentile) represents the higher-income group. The spatial distributions of home and workplace locations for these four income groups are shown in Supplementary Fig. \ref{FigS6}.

\begin{figure}[h]
\centering
\includegraphics[width=1\textwidth]{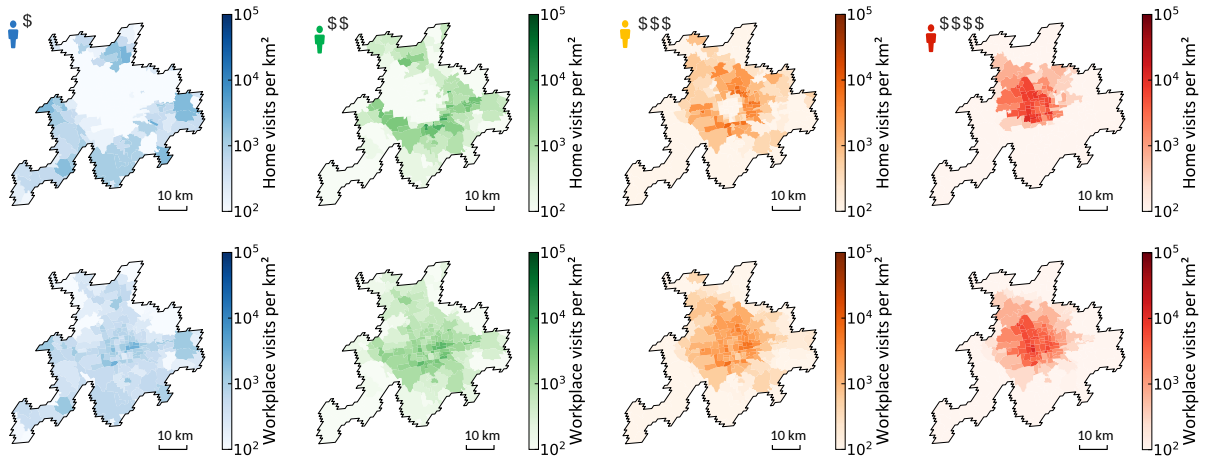}
\caption[Spatial distributions of residential locations and workplaces for four income groups.]{\textbf{Spatial distributions of residential locations and workplaces for four income groups.} Top row: Residential distributions reveal strong income stratification, with higher-income groups concentrated within city core area, transitioning to lower-income groups in peripheral districts. Bottom row: Workplace distributions display more evenly spatial patterns, maintaining partial concentration in central business districts across all income groups.}\label{FigS6}
\end{figure}

\subsection{Travel mode choices}\label{Supplnote1.4}

\noindent\hspace*{1.5em}For each individual trip, we compute the probabilities that an individual travels through particular transport mode (active, private or public) using a pre-trained random forest model calibrated on the publicly available Geolife dataset \cite{geolife2009}. The Geolife dataset, collected by Microsoft Research Asia, comprises GPS trajectories of over 180 users in a range of cities, primarily in Beijing, China, over several years. We segment each trajectory into multiple contiguous trips, defined by a minimum dwell time of 15 minutes between trips. Each trip is labeled with ground-truth transport modes (one or more) used, including walking, cycling, car, taxi, bus, railway. We consolidate similar modes (e.g., walking and cycling into "active"; car and taxi into "private"; bus, railway into "public") to align with our defined categories. Trips involving a combination of private and public modes are excluded from the analysis to ensure unambiguous mode classification. Only trips exclusively involving active modes are labeled as "active"; all other trips are categorized as either "private" or "private" based on their dominant mode. In Beijing metropolitan area, this process yields 1,819 active trips, 881 private trips, and 1,725 public trips.

To train the random forest model, we extract five features relevant to transport mode identification, including: 'Route length' (total distance traveled during the trip), 'OD distance' (Euclidean distance between origin and destination), 'O\_pubstation\_dist' (distance from the origin to the nearest public transportation station obtained from Amap POIs), 'D\_pubstation\_dist' (distance from the destination to the nearest public transportation station), and 'Travel time' (duration of the trip). The model is trained using a subset of the Geolife data (80\% for training, 20\% for validation) and optimized for classification accuracy. To mitigate potential sample imbalance, the model is configured to automatically adjust the weights assigned to each class based on their prevalence in the data, ensuring that classes with fewer samples are given more importance during training. The contribution of each feature to the model's predictions is shown in Supplementary Fig. \ref{FigS7}. Rather than assigning a single, definitive mode for each trip, the model generates probabilistic mode assignments, reflecting pre-trip decision uncertainty. For instance, a particular trip might be assigned probabilities of 0.1 for "active", 0.6 for "private", and 0.3 for "public". This suggests that while private mode is the most probable, there's still a non-negligible chance of choosing public transport. Such probability vectors capture travelers' latent preference influenced by contextual factors prior to a trip.

\begin{figure}[h]
\centering
\includegraphics[width=1\textwidth]{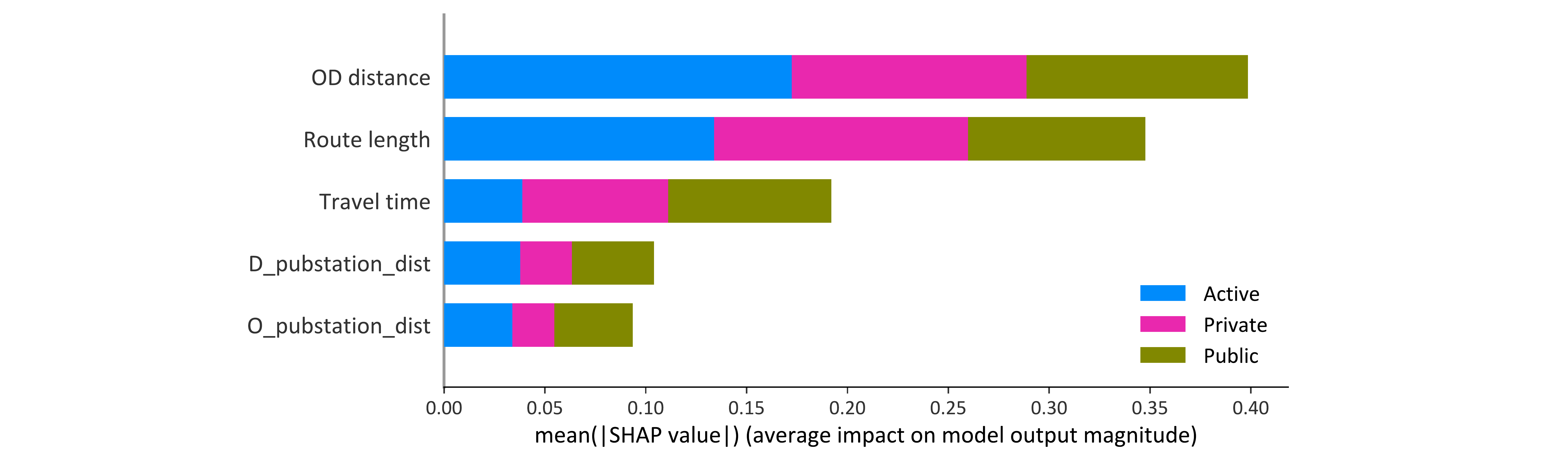}
\caption[Overall feature importance based on SHAP (SHapley Additive exPlanations) values for the travel mode inference model.]{\textbf{Overall feature importance based on SHAP (SHapley Additive exPlanations) values for the travel mode inference model.} Features are listed on the vertical axis, ordered from most to least important. The horizontal axis represents the mean absolute SHAP value for each feature.  A longer bar indicates a greater overall impact of that feature on the model's predictions across the entire dataset.}\label{FigS7}
\end{figure}

\begin{figure}[h]
\centering
\includegraphics[width=1\textwidth]{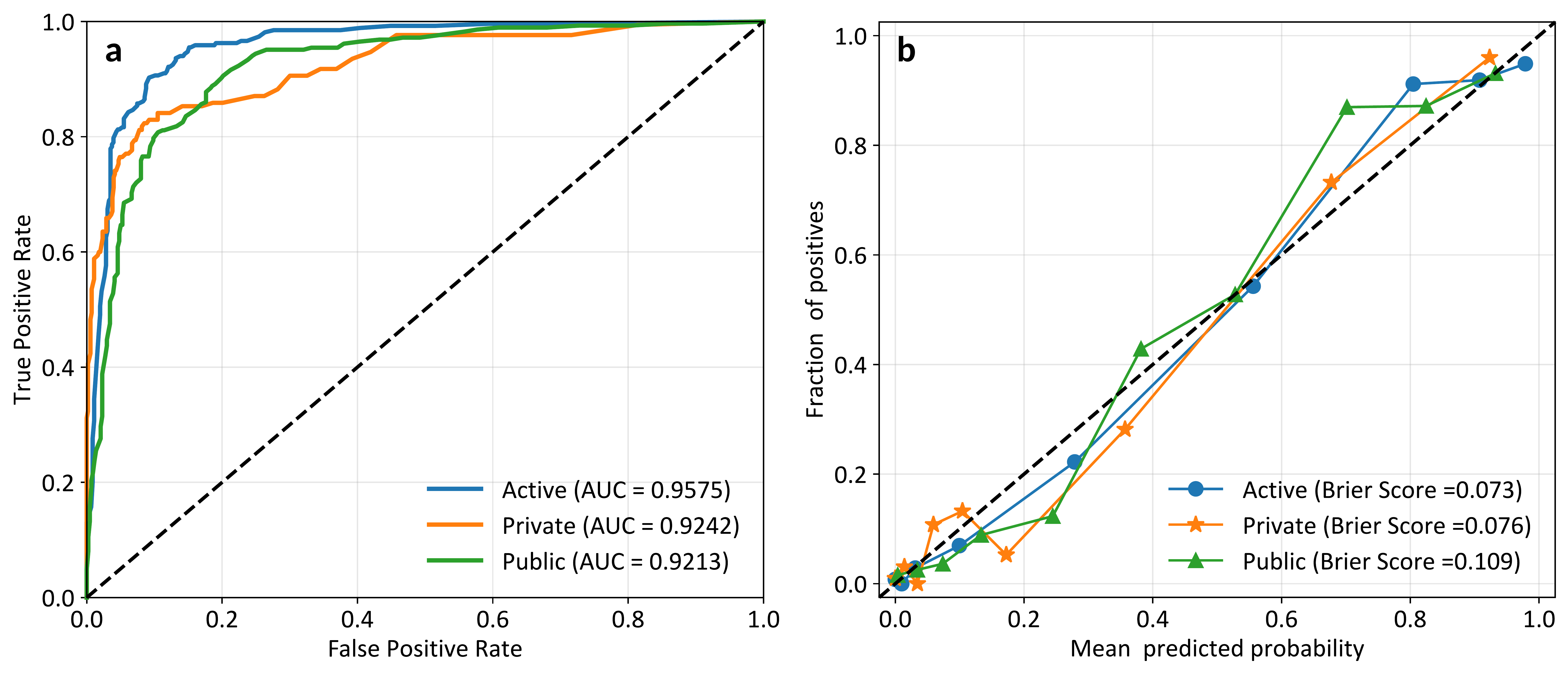}
\caption[Model performance evaluation.]{\textbf{Model performance evaluation.} \textbf{a} Receiver Operating Characteristic (ROC) curves. Each curve plots the True Positive Rate against the False Positive Rate at various threshold settings. The Area Under the Curve (AUC) for each mode is indicated in the legend. High AUC values (close to 1) demonstrate the model's excellent ability to distinguish between each transport mode and the others. \textbf{b} Calibration curves assess the reliability of the predicted probabilities by plotting the observed fraction of positives against the predicted probabilities. Ideally, the calibration curves should closely follow the diagonal (dashed line), indicating well-calibrated probabilities where predicted probabilities align with actual event frequencies. Lower Brier scores (close to 0) indicate better calibration.}\label{FigS8}
\end{figure}

The model's performance is evaluated through metrics appropriate for both classification and probability estimation. Specifically, Receiver Operating Characteristic (ROC) curves \cite{hanley1982meaning} (Supplementary Fig. \ref{FigS8}a), which assess the model's ability to discriminate between classes, yield high Area Under the Curve (AUC) scores: 0.9575 for active mode, 0.9242 for private mode, and 0.9213 for public mode, indicating strong discriminatory capacity across all modes. Furthermore, the calibration of the probability estimates is assessed using the Brier score \cite{brier1950verification}, which measures the mean squared difference between predicted probabilities and actual outcomes. The Brier scores are also favorable (Supplementary Fig. \ref{FigS8}b): 0.073 for active mode, 0.076 for private mode, and 0.109 for public mode, demonstrating well-calibrated probability predictions.

This pre-trained random forest model is then applied to our mobile phone dataset.  For each trip in this dataset, we calculate the same five features: 'Route length', 'OD distance', 'O\_pubstation\_dist', 'D\_pubstation\_dist', and 'Travel time'.  By inputting these features into the trained model, we estimate the probability distribution across active, private, and public transport modes for each trip. These probability estimates are then used for further mobility analysis.

\subsection{Travel route generation}\label{Supplnote1.5}

\noindent\hspace*{1.5em}For each trip of an individual, we generate a most probable travel route for active, private, and public modes respectively using Amap navigation API \cite{Amap}, a sophisticated service renowned for its routing capabilities in China. The Amap API is configured to generate routes by considering a range of input parameters including the trip origin and destination coordinates (latitude and longitude), waypoints, departure time and desired travel modes. Critically, the API computes the most time-efficient route for the specified mode, dynamically factoring in real-time traffic conditions, public transit schedules, estimated costs, and general traveler preferences as modeled within its algorithms. To enhance the realism of these generated routes, we incorporate all GPS trajectory points from each original trip record as intermediate waypoints when querying the API. This strategy allows the navigation system to compute routes that more accurately capture potential deviations, detours, and individual preferences that may have influenced the observed travel behavior. The resulting output from the Amap API delivers comprehensive navigation information. For active and private modes, this includes a breakdown of route details by road segment, specifying the roads to be taken and the estimated travel duration for each segment, accounting for real-time traffic where applicable. For public mode routes, the API details the specific transit lines to utilize, the sequence of stations, the estimated travel time between stations, and any necessary transfer points. The use of Amap API allows for privacy-preserving travel planning by inferring potential routes without directly accessing sensitive location data from the individual’s mobile device.

To validate the accuracy of route generation process, we leverage the high-resolution GPS trajectories provided in the Geolife dataset as ground truth. Notably, the majority of these trajectories (91.5\%) are recorded at a dense sampling rate, typically every 1–5 seconds, ensuring a detailed and accurate representation of travel paths. For each trip, we generate a route between its origin and destination using the Amap API corresponding to the actual travel mode recorded in Geolife data. To quantitatively assess the spatial similarity between the generated route and the actual GPS trajectory, we create a buffer around both the generated route $B_{generated}$ and the original GPS trajectory $B_{real}$, and calculate the Jaccard index, representing the ratio of the intersection area to the union area of the two buffers

\begin{equation}
J = \frac{Area(B_{generated} \cap B_{real})}{Area(B_{generated} \cup B_{real})}
\label{eq:J}
\end{equation}

\begin{figure}[h]
\centering
\includegraphics[width=1\textwidth]{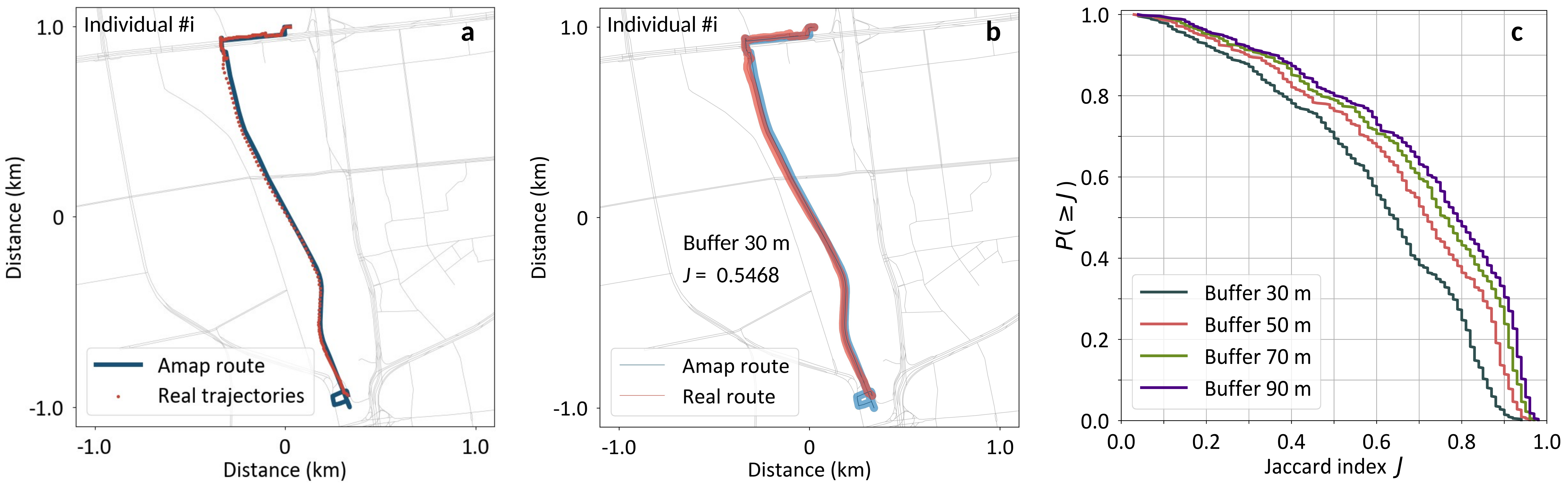}
\caption[Validation of Amap route generation against Geolife GPS trajectories.]{\textbf{Validation of Amap route generation against Geolife GPS trajectories.} \textbf{a} An example of an Amap-generated route overlaid with the corresponding real GPS trajectory from Geolife for a single trip. \textbf{b} Buffered representation (30-meter buffer) of the generated and real routes. \textbf{c} Cumulative distribution of Jaccard Index \textit{J} across 4,425 mode-labeled trips from the Geolife dataset, shown for different buffer distances (30m, 50m, 70m, and 90m).}\label{FigS9}
\end{figure}

\noindent\hspace*{0em}This index provides a measure of overlap, with higher values (close to 1) indicating greater agreement between the generated route and the real-world trajectory. For 4,425 mode-labeled trips in Geolife dataset, we test a range of buffer distances—30 meters, 50 meters, 70 meters, and 90 meters—around both the Amap-generated routes and the corresponding real GPS trajectories. We observe that even under a stringent 30-meter buffer, over 70\% of generated routes achieve a Jaccard index greater than 0.5 (Supplementary Fig. \ref{FigS9}). This threshold of 0.5 signifies a substantial level of overlap, suggesting that the generated routes closely align with the real-world trajectories in a majority of cases. When examining the results across different modes, we observe no significant differences in performance. This validation demonstrates the Amap API’s effectiveness in generating realistic routes.

\subsection{Cross-data validation}\label{Supplnote1.6}

\noindent\hspace*{1.5em}To further validate the reliability of travel mode inference and route generation processes, we leverage an additional independent dataset to compare the consistency of mobility flow distributions in urban spaces. We use the open-sourced MemDA data \cite{Cai2023MemDA}, which comprise traffic speeds from major roads in Beijing collected in a period of 75 days (from May 12, 2022, to July 25, 2022). In urban road traffic, the average speed of a road segment is typically negatively correlated with traffic volume under most normal conditions, as higher vehicle densities tend to reduce speeds due to congestion. We process the traffic speed data by aggregating measurements into hourly intervals for each major road segment over the 75-day period. The average speed per segment per hour is computed as the mean of all recorded speeds within that time window, providing a proxy for potential traffic condition. For the mobile phone dataset, we calculate the expected traffic volume for each road segment in a given hourly interval by summing the estimated private mode probabilities of all trips whose Amap-generated routes include that segment during that time period. For example, a trip with a 0.7 probability of private mode contributes 0.7 vehicle units to the traffic volume of each segment along its route. This probabilistic aggregation reflects the uncertainty in mode choice predictions and provides a robust estimate of traffic flow. To compare the two datasets, we normalize the MemDA-derived average speeds and the mobile phone-derived expected traffic volumes for matching road segments and hourly intervals. The spatial distributions of these normalized values during the morning peak hour (9:00-10:00) is visualized in Supplementary Fig. \ref{FigS10}, revealing a notable spatial consistency, which is supported by a significant negative Pearson correlation coefficient of \textit{r} = -0.3191 (\textit{p} $<$ 0.001; 95\% CI [-0.3504, -0.2871]) (Supplementary Fig. \ref{FigS11}a). We also observe similarly significant negative correlations during the midday (13:00-14:00, \textit{r} = -0.3521; 95\% CI [-0.3826, -0.3209]) and evening peak (17:00-18:00, \textit{r} = -0.3513; 95\% CI [-0.3819, -0.3201]) hours (Supplementary Fig. \ref{FigS11}bc). This temporal and spatial consistency across the MemDA and mobile phone datasets underscores the method reliability in capturing road traffic patterns.

\begin{figure}[h]
\centering
\includegraphics[width=1\textwidth]{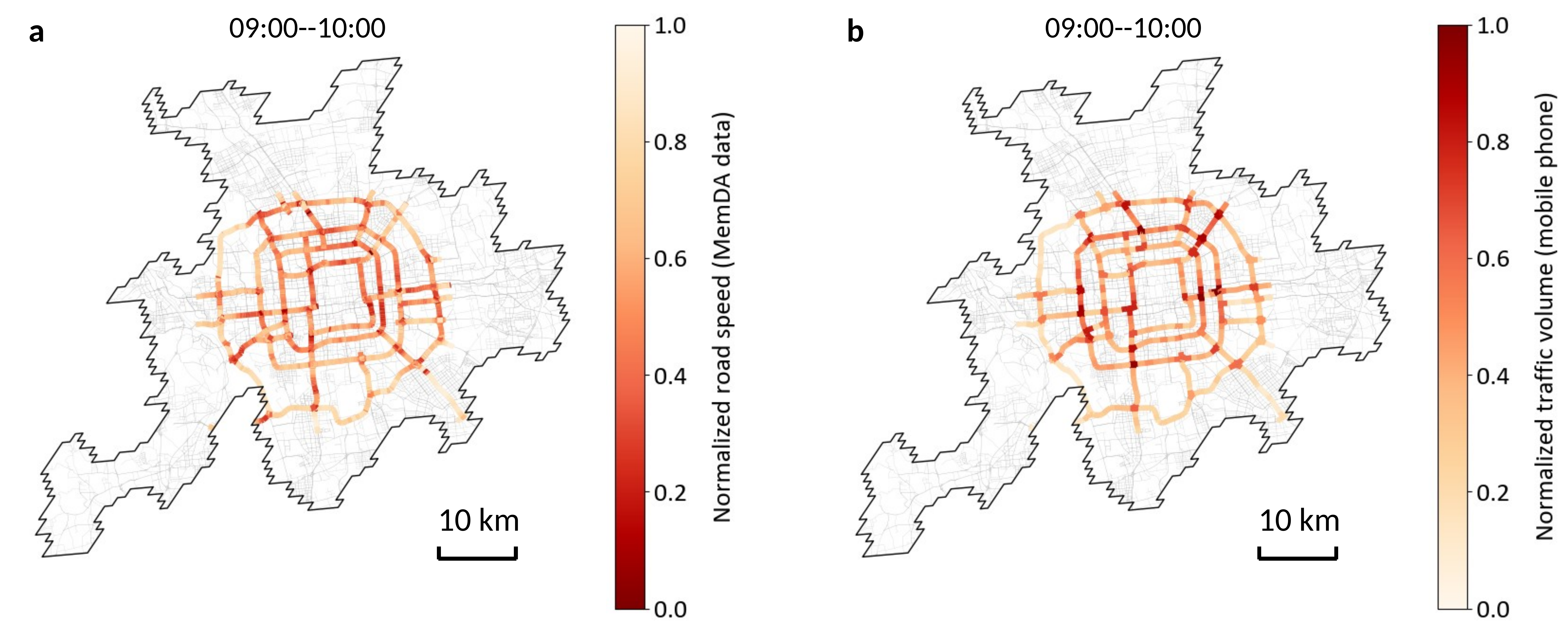}
\caption[Spatial consistency of mobility flow distributions across datasets in Beijing metropolitan area.]{\textbf{Spatial consistency of mobility flow distributions across datasets in Beijing metropolitan area.} \textbf{a}, Spatial distribution of normalized road average speed from MemDA data during the morning peak hour (9:00-10:00). \textbf{b}, Spatial distribution of normalized expected road traffic volumes inferred from mobile phone data for private mode during the morning peak hour.}\label{FigS10}
\end{figure}

\begin{figure}[h]
\centering
\includegraphics[width=1\textwidth]{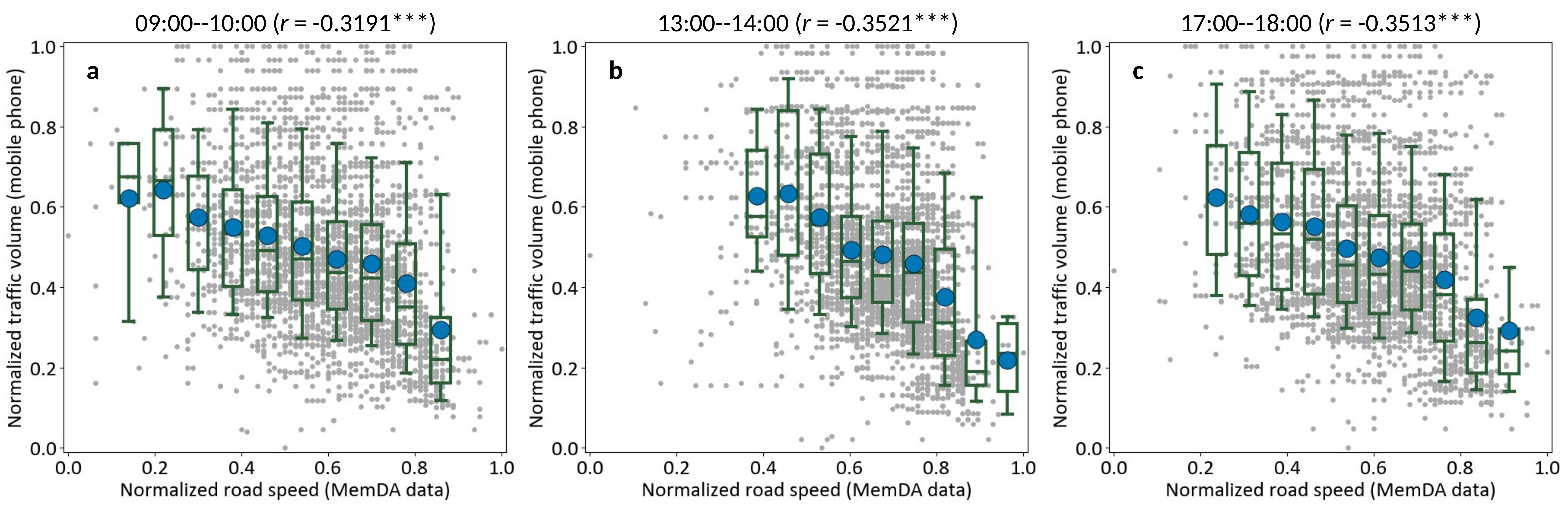}
\caption[Cross-data validation performance.]{\textbf{Cross-data validation performance.} Correlation between normalized average traffic speeds from MemDA data and normalized expected traffic volumes inferred from mobile phone data for private modes during the morning peak (9:00-10:00; panel \textbf{a}), midday (13:00-14:00; panel \textbf{b}) and evening peak (17:00-18:00; panel \textbf{c}) hours. Pearson correlation coefficient of \textit{r} is marked on the title, with *** indicating statistical significance \textit{p} $<$ 0.001. In all panels, grey points represent road segments, plotted according to their values from the two datasets being compared. Boxplots are grouped by bins of values from the reference dataset on the x-axis, and show the distribution of the corresponding values from mobile phone data on the y-axis within each bin. Blue points represent the average value within each bin, summarizing the overall trend.}\label{FigS11}
\end{figure}

\section{Measuring segregation in interconnected urban spaces}\label{Supplnote2}

\noindent\hspace*{1.5em}We develop two measures to quantify the segregation experiences through multimodal mobility at the city scale. The first one is mode-specific segregation measure (\textit{PSM}), which is designed to capture the extent of segregation experienced by individuals when using a particular transport mode. The second one is the multimodal uniformity index (\textit{MUI}), which builds upon \textit{PSM} to assess the consistency of segregation across different travel modes within a geographical region.

\subsection{Probabilistic segregation measure}\label{Supplnote2.1}

\noindent\hspace*{1.5em}Mode-specific probabilistic segregation measure (\textit{PSM}) is calculated based on the probabilities of individual paths crossing in multilayered urban spaces traveling via a specific mode. After data fusion processing, we have estimated the probabilities that an individual travels through different transport modes (active, private or public) (Supplementary Section \ref{Supplnote1.4}), and generated a most probable travel route corresponding to each mode (Supplementary Section \ref{Supplnote1.5}). For road transportation (active and private), the generated route specifies the road segments to be taken and the estimated duration. For public transportation (bus and railway), the generated route provides the station-by-station trajectories within transit systems, including the specific sequence of stations, along with the estimated travel times between them. For a given departure time, the geographical location at any moment along the route of a specific mode can be determined, as shown in Supplementary Fig. \ref{FigS12}a.

For simplicity, we assume that individuals experience segregation from others traveling via the same mode through transient encounters in time and space. These co-locations represent moments when two or more individuals are present in the same spatial unit simultaneously while using the same mode. To capture these encounters, the urban space is partitioned into mode-specific spatial units, reflecting the distinct ways individuals perceive and interact with their surroundings. The spatial scales for active and private modes is defined as 1 km × 1 km grids. For active (or private) mode, two individuals' routes are mapped onto the 1 km × 1 km grids over time. A co-location occurs when they occupy the same grid within a specific time frame, indicating a potential encounter. For bus and railway mode, the spatial units are the transit segments between stations. In this context, co-locations occur within the confined spaces of transit vehicles or at stations, capturing the shared experience inherent to public transit. These spatial units are analyzed within 1-hour time frames, a temporal resolution chosen to balance computational feasibility with the need to capture significant social interactions.

To calculate \textit{PSM} for each spatial unit for a mode, and 1-hour time frame, the expected number of individuals from each income group is computed by summing the product of their mode-choice probability and an indicator of their presence in that spatial unit. Mathematically, for a spatial unit \textit{s}, time frame \textit{t}, and income group \textit{q}, the expected population can be expressed as:

\begin{equation}
E_{s,t,q} = \sum_{i \in q} p_{i,m} \cdot I_{i,s,t}
\label{eq:E_{s,t,q}}
\end{equation}

\noindent\hspace*{0em}where \( p_{i,m} \) is the probability that individual \( i \) travel via mode \( m \),  \( I_{i,s,t} \) is an indicator variable equal to 1 if individual \( i \) is present in spatial unit \( s \) during time frame \( t \) using mode \( m \), and 0 otherwise. Each individual contributes a fractional value—between 0 and 1—to the expected population, rather than a whole unit of 1, reflecting their partial likelihood of being present. This expected value accounts for the inherent uncertainty in individual travel behavior, providing a more dynamic and realistic estimate of population distribution. For example, if an individual from group \textit{q} has a 0.9 probability of choosing to drive and their driving route passes through spatial unit \( s \) between 8 and 9 AM, the contribution to \( E_{s,t,q} \) for driving would be \( 0.9 \times 1 = 0.9 \). If the route does not intersect that unit, the contribution becomes \( 0.9 \times 0 = 0 \). A schematic illustration of the calculations is shown in Supplementary Fig. \ref{FigS12}bc. After computing the expected population \( E_{s,t,q} \) for each of the four income groups, the \textit{PSM} is derived using entropy metric to quantify the diversity of income group representation, calculated as:

\begin{equation}
PSM_{s,t,m} = -\frac{1}{\log(4)} \sum_{q=1}^{4} \tau_{s,t,q} \cdot \log(\tau_{s,t,q})
\label{eq:PSM_{s,t,m}}
\end{equation}

\noindent\hspace*{0em}where \( \tau_{s,t,q} = \frac{E_{s,t,q}}{\sum_{q=1}^{4} E_{s,t,q}} \) denotes the proportion of the total expected population in spatial unit \( s \), time frame \( t \), and mode \( m \) that belongs to income group \( g \). The normalization factor \( \frac{1}{\log(4)} \) scales the entropy to a range between 0 and 1, since there are four income groups. A \( PSM_{s,t,m} \) value of 0 indicates complete segregation (i.e., only one income group is present), while a value of 1 indicates maximum diversity (i.e., all four income groups are equally represented).

\begin{figure}[h]
\centering
\includegraphics[width=1\textwidth]{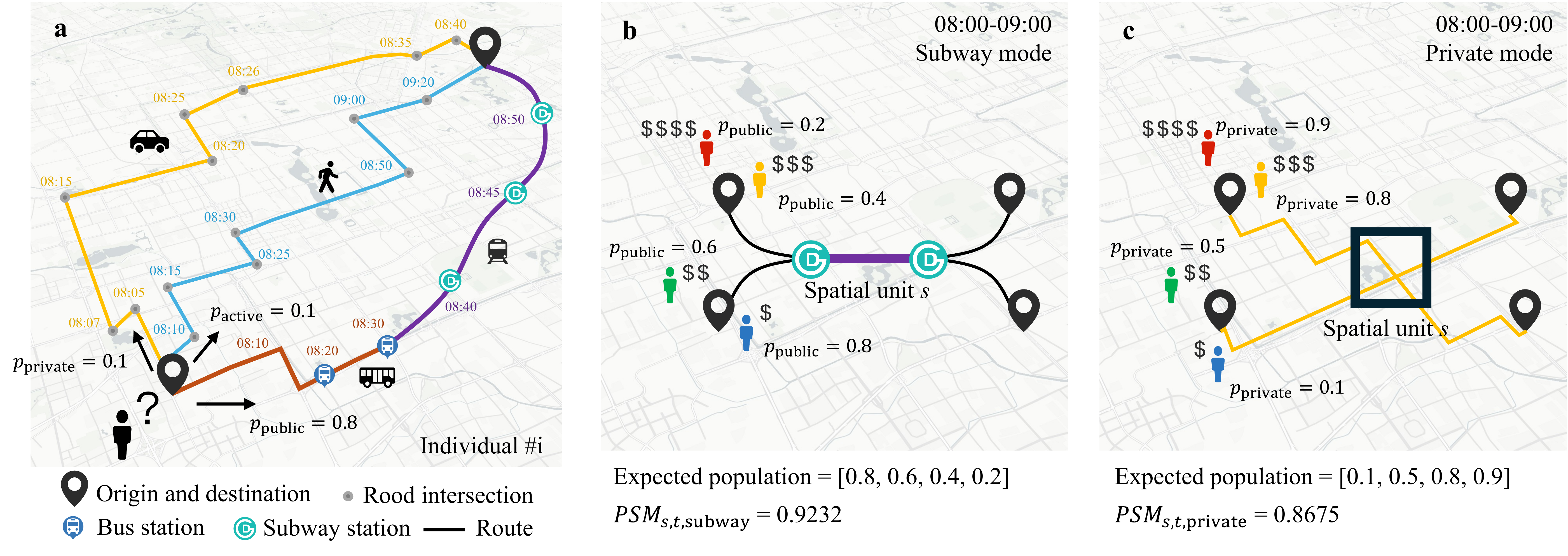}
\caption[Illustration of calculation of mode-specific segregation measure (\textit{PSM}).]{\textbf{Illustration of calculation of mode-specific segregation measure (\textit{PSM}).} \textbf{a} Individual mode choice and time-stamped paths for active, private, and public modes. \textbf{b} Example of PSM calculation for a single spatial unit (transit segment between two stations) for a railway line. Assuming four individuals from four income groups are co-located at this segment between 8–9 AM, their contributions to the expected population \( E_{s,t,q} \) equal their mode probabilities as marked in the panel. \textbf{c} Example of PSM calculation for a single spatial unit (1 km × 1 km grid) for individuals driving.}\label{FigS12}
\end{figure}

\subsection{Multimodal uniformity index}\label{Supplnote2.2}

\noindent\hspace*{1.5em}The multimodal uniformity index (\textit{MUI}) is introduced to evaluate how consistently or diversely segregation is experienced when individuals travel through different travel modes within the same geographical region. For a specific region \textit{A} and time frame \textit{t}, we first calculate the average segregation level \( PSM_{A,t,m} \) for a given mode \textit{m} across all relevant spatial units \textit{s} in region \textit{A} during period \textit{t}:

\begin{equation}
PSM_{A,t,m} = \frac{1}{|S_A|} \sum_{s \in S_A} PSM_{s,t,m}
\label{eq:PSM_{A,t,m}}
\end{equation}

\begin{figure}[h]
\centering
\includegraphics[width=1\textwidth]{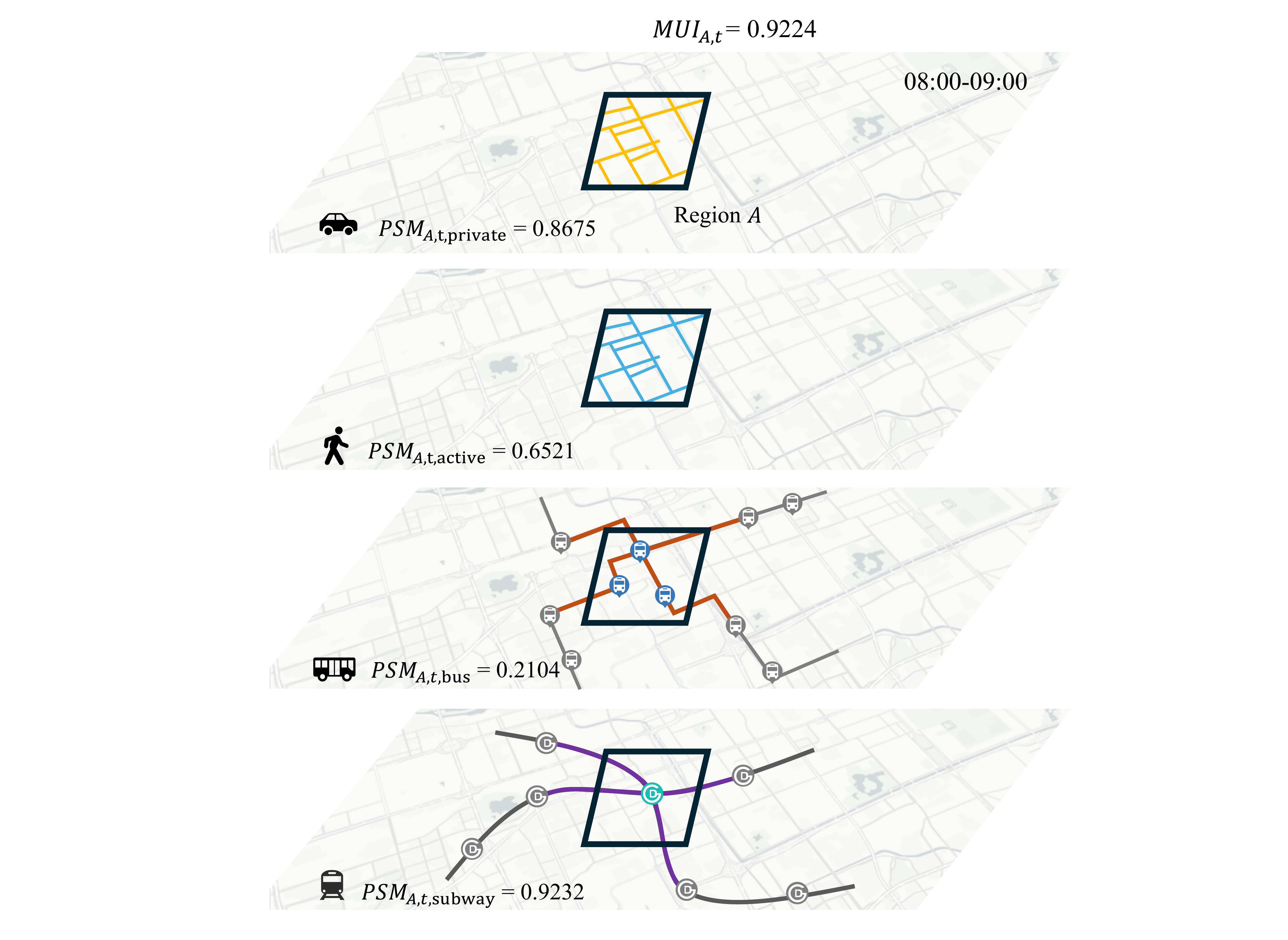}
\caption[Illustration of calculation of multimodal uniformity index (\textit{MUI}).]{\textbf{Illustration of calculation of multimodal uniformity index (\textit{MUI}).} Region-level segregation level \( PSM_{A,t,m} \) for each mode \( m \) (active, private, bus, railway) during time frame \( t \) is first calculated by aggregating all relevant spatial units in region \( A \). \( MUI_{A,t} \) is then computed by normalizing these \( PSM_{A,t,m} \) values into proportions and applying the entropy formula, indicating the uniformity of segregation experiences across modes in region \( A \) during time \( t \).}\label{FigS13}\end{figure}

\noindent\hspace*{0em}where \( S_A \) is the set of spatial units (grids for active and private modes, transit segments for bus and railway modes) associated with region \( A \), and \( |S_A| \) is the number of such units. The spatial units in \( S_A \) are defined based on the characteristics of each travel mode. For active and private modes, a spatial unit \textit{s} (1 km × 1 km grid) is included in \( S_A \) if any portion of it overlaps with the region, ensuring comprehensive coverage of movement patterns. For public transportation, specifically bus and railway, a spatial unit (transit segment) is included in \( S_A \) if at least one of its endpoint stations is located within this region, reflecting the service provision within the defined geographical area (Supplementary Fig. \ref{FigS13}). With \( PSM_{A,t,m} \) computed for each mode—active, private, bus, and railway—we then normalize these values into proportions \( r_{A,t,m} = PSM_{A,t,m} / \sum_{m} PSM_{A,t,m} \), where the sum is taken over the four modes. The \textit{MUI} for region \textit{A} at time \textit{t}, denoted \( MUI_{A,t} \), is then calculated using the entropy formula:

\begin{equation}
MUI_{A,t} = -\frac{1}{\log(4)} \sum_{m} r_{A,t,m} \log(r_{A,t,m}),
\label{eq:MUI_{A,t}}
\end{equation}

An \( MUI_{A,t} \) value close to 1 indicates high uniformity, meaning segregation (or mixing) experiences are similar across all modes in region \textit{A} during time frame \textit{t}, while a value near 0 suggests diverse experiences, with significant variation in segregation levels between modes at that time. This index thus provides a time-specific measure of how evenly transportation modes contribute to social exposure patterns within a region.

\subsection{Sensitivity analysis of spatiotemporal scales}\label{Supplnote2.3}

\noindent\hspace*{1.5em}In this study, the primary results are presented using spatiotemporal scales that balance computational efficiency with the ability to capture meaningful social interactions and segregation patterns, specifically a temporal resolution of 1 hour and spatial scales of 1 km × 1 km grids for active and private modes, and transit segments for public transportation modes, as outlined in Supplementary Sections \ref{Supplnote2.1} and \ref{Supplnote2.2}. To evaluate the robustness of the mode-specific segregation measure (\textit{PSM}) and the multimodal uniformity index (\textit{MUI}) to variations in these scales, we conduct a sensitivity analysis by systematically testing alternative temporal and spatial resolutions to evaluate their impact.

For temporal scales, we examine window sizes ranging from 3 to 60 minutes (3, 5, 10, 20, 30, and 60 minutes), recognizing that shorter intervals might capture more transient interactions while longer windows could aggregate mobility patterns. Spatial scales for active and private modes are tested at 250 m, 500 m, 1 km, and 2 km grid resolutions, spanning a spectrum from highly localized interactions to coarser regional patterns, while keeping the transit segment definition unchanged for public transportation (bus and railway) due to its inherent station-based structure.

\begin{figure}[h]
\centering
\includegraphics[width=1\textwidth]{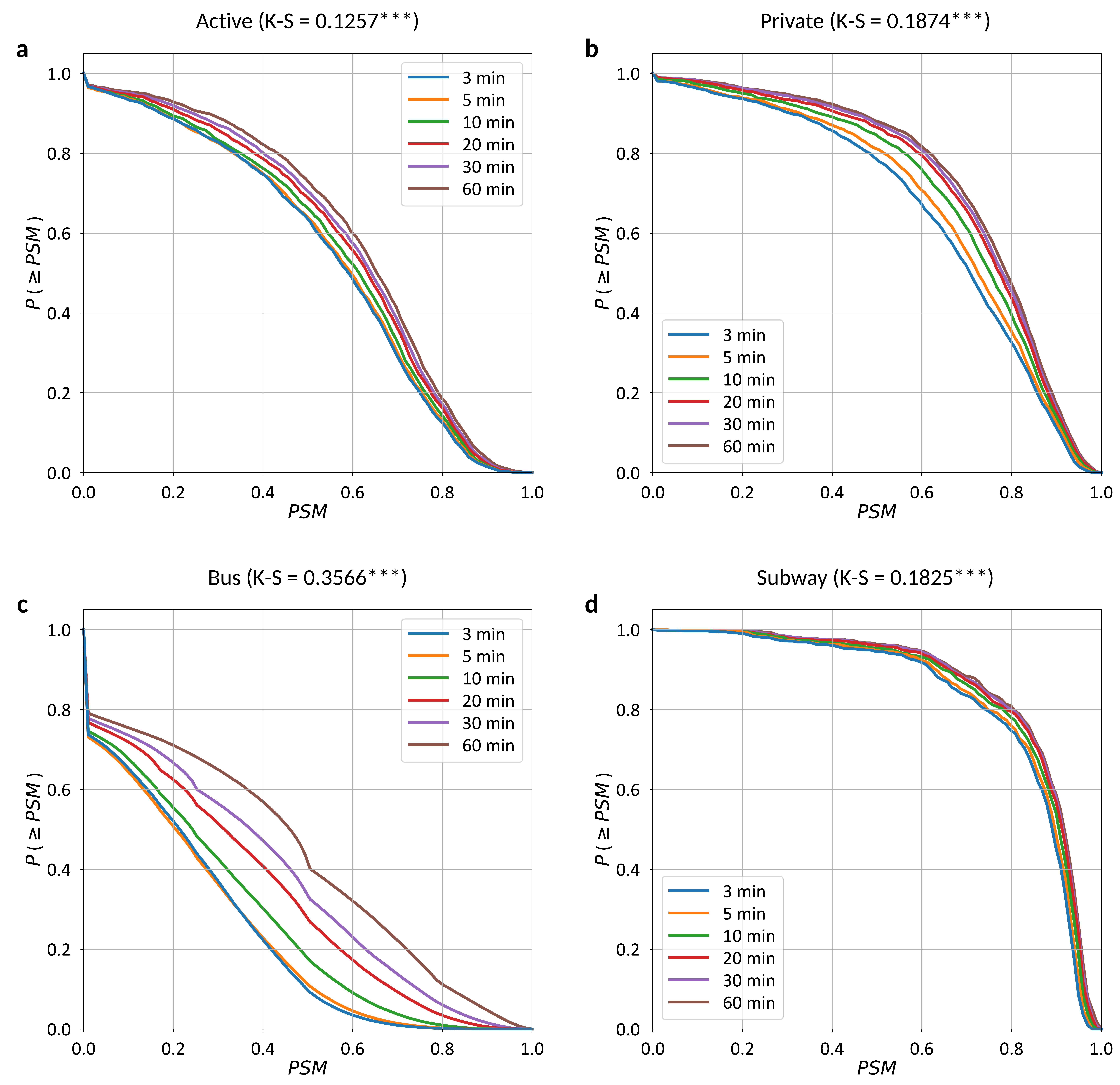}
\caption[Cumulative distributions of \textit{PSM} for four travel modes across different temporal scales under fixed 1 km × 1 km grids.]{\textbf{Cumulative distributions of \textit{PSM} for four travel modes across different temporal scales under fixed 1 km × 1 km grids.} The Kolmogorov-Smirnov (K-S) statistics (*** indicates p-value \( < 0.001 \)) are provided to quantify the distributional divergence between the 3-minute and 60-minute temporal scales, shown in the title.}\label{FigS14}\end{figure}

Figure \ref{FigS14} presents the cumulative distributions of \textit{PSM} for each of the four travel modes across the range of temporal scales. As the temporal scale increases, the cumulative distribution curves generally shift upwards across all modes, indicating a tendency towards higher \textit{PSM} (greater entropy; reduced segregation) with longer periods. This effect is most pronounced for the bus mode, as evidenced by the Kolmogorov-Smirnov (K-S) statistic \cite{kolmogorov1933sulla} comparing 3-minute and 60-minute windows: bus mode exhibits the largest distributional divergence (K-S = 0.3566, \( p < 0.001 \)), followed by private (K-S = 0.1874), railway (K-S = 0.1825), and active modes (K-S = 0.1257; all \( p < 0.001 \)). The heightened sensitivity of bus systems likely stems from their variable ridership patterns and frequent stops, which amplify transient co-location noise in short time frames. In contrast, railway, active, and private modes exhibit more stable distributions, with relatively smaller K-S distances confirming that their segregation patterns are governed by persistent route structures rather than momentary interactions. 

\begin{figure}[h]
\centering
\includegraphics[width=1\textwidth]{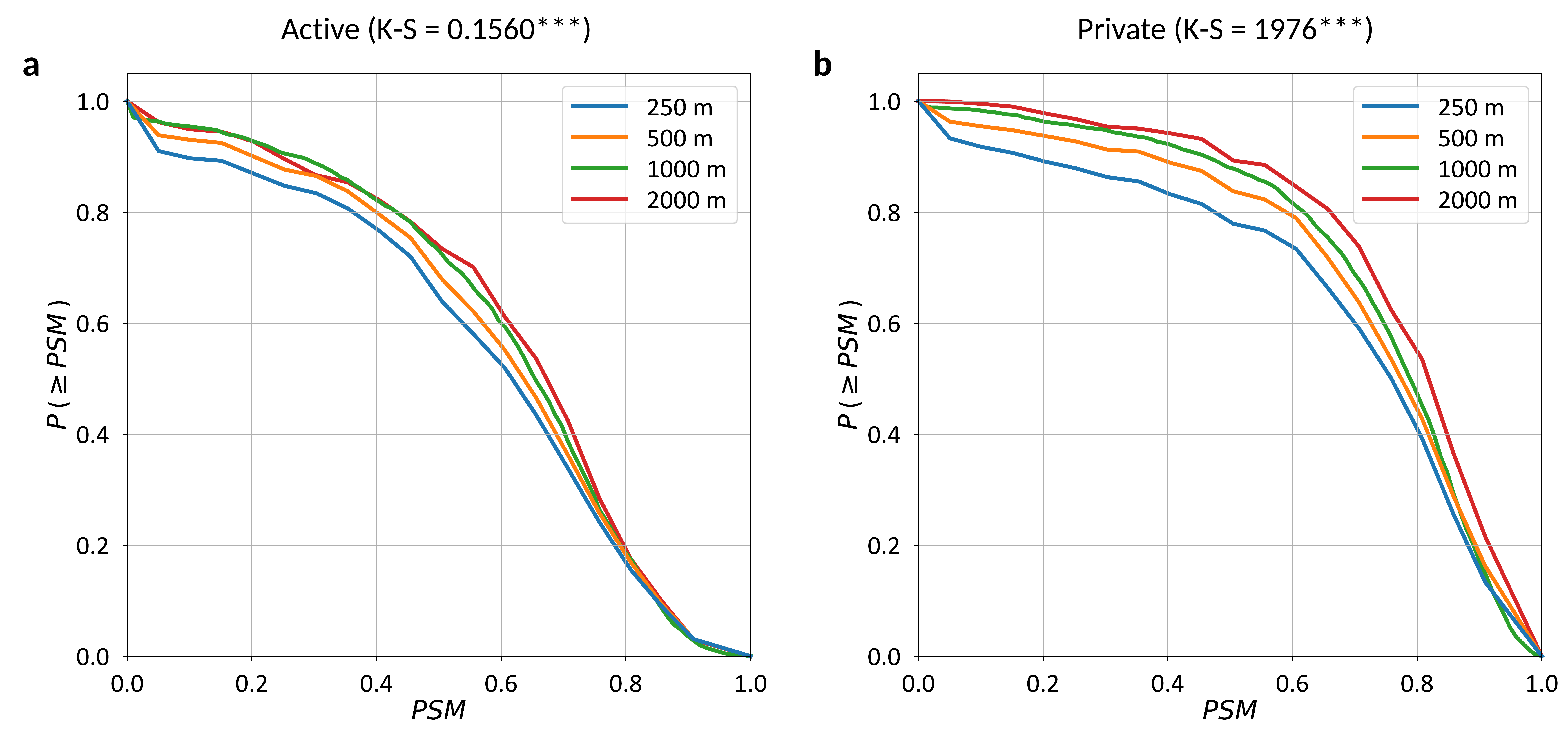}
\caption[Cumulative distributions of \textit{PSM} for active and private  modes across spatial scales under a fixed 60-minute temporal scale.]{\textbf{Cumulative distributions of \textit{PSM} for active (panel a) and private (panel b) modes across spatial scales under a fixed 60-minute temporal scale.} The Kolmogorov-Smirnov (K-S) statistics (*** indicates p-value \( < 0.001 \)) are provided to quantify the distributional divergence between the 250 m and 2 km grid scales shown in the title.}\label{FigS15}\end{figure}

Figure \ref{FigS15} presents spatial scale sensitivity by comparing cumulative \textit{PSM} distributions for active and private modes across grid resolutions ranging from 250 m to 2 km. Larger spatial scales produce upward-shifted distribution curves for both modes, reflecting higher \textit{PSM} values at coarser resolutions. The K-S statistic quantifies this divergence, with private modes exhibiting greater sensitivity (K-S = 0.1976 between 250 m and 2 km grids, \( p < 0.001 \)) compared to active modes (K-S = 0.1560, \( p < 0.001 \)). The relatively low magnitude of the K-S statistics suggests that \textit{PSM} distributions for both active and private modes remain reasonably stable across the tested range of spatial scales.

\begin{figure}[h]
\centering
\includegraphics[width=1\textwidth]{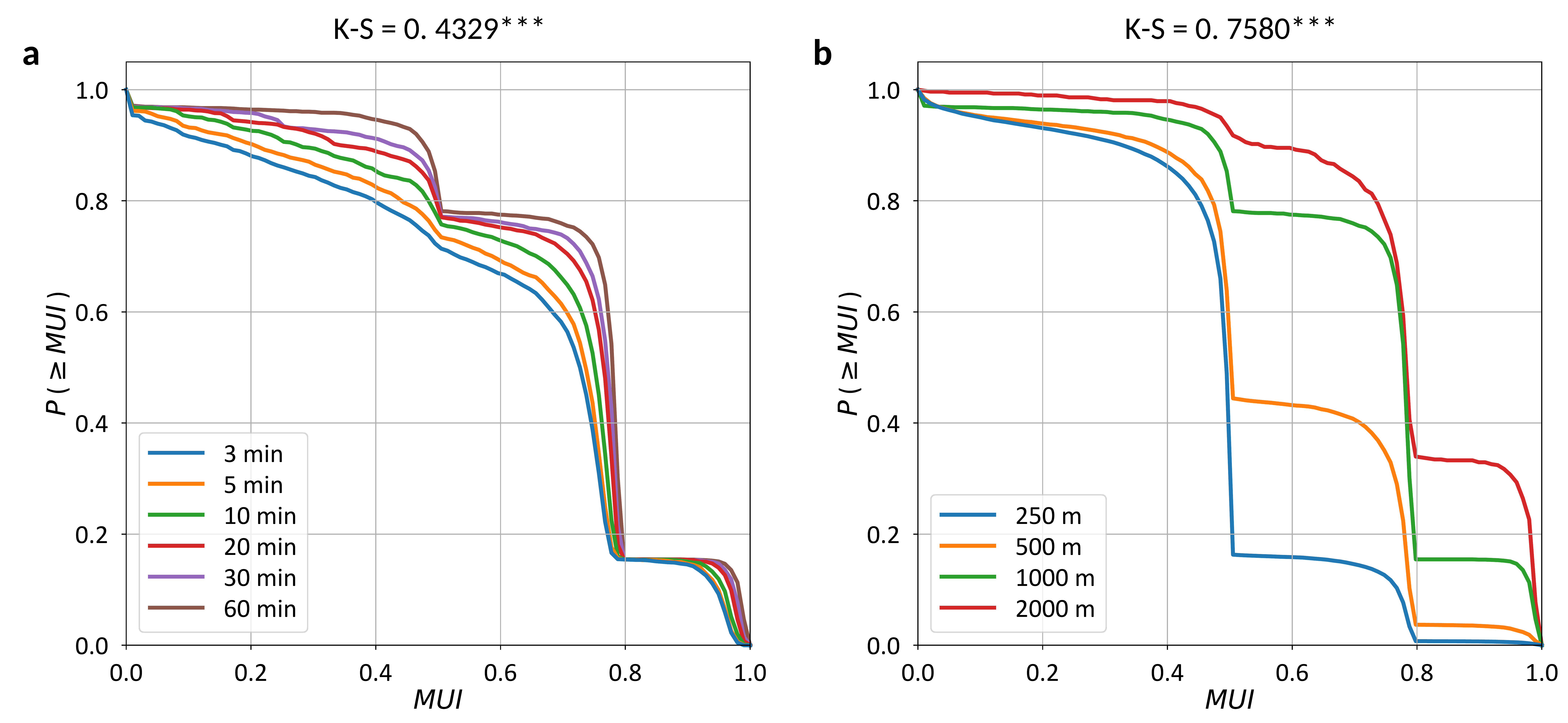}
\caption[Cumulative distributions of \textit{MUI} for urban grids across different spatiotemporal scales.]{\textbf{Cumulative distributions of \textit{MUI} for urban grids across different spatiotemporal scales.} \textbf{a} Temporal sensitivity analysis with each curve representing a temporal resolution (3, 5, 10, 20, 30, and 60 minutes) under fixed 1 km × 1 km grids. \textbf{b} Spatial sensitivity analysis with each curve corresponding to a spatial grid size (250 m, 500 m, 1 km, and 2 km) under a fixed 60-minute temporal scale. In both panels, the x-axis represents \textit{MUI} values (0 to 1), and the y-axis shows the cumulative probability (0 to 1).The Kolmogorov-Smirnov (K-S) statistic (*** indicates p-value \( < 0.001 \)) are provided to quantify the distributional divergence between the 3-minute and 60-minute temporal scales (panel \textbf{a}), and between 250 m and 2 km grid scales (panel \textbf{b}), shown in the title.}\label{FigS16}\end{figure}

Figure \ref{FigS16} presents the sensitivity of the multimodal uniformity index (\textit{MUI}) to spatiotemporal variations. In Supplementary Fig. \ref{FigS16}a, the spatial scale is fixed at 1 km × 1 km grids, and the cumulative \textit{MUI} distributions are compared across temporal scales from 3 to 60 minutes. The K-S statistic between the 3-minute and 60-minute windows is 0.4329 (\( p < 0.001 \)), indicating a notable shift in uniformity as temporal resolution changes. In Supplementary Fig. \ref{FigS16}b, the temporal scale is fixed at 60 minutes, and the spatial scale is varied from 250 m to 2 km, yielding a much larger K-S statistic of 0.7580 (\( p < 0.001 \)). A key observation in Supplementary Fig. \ref{FigS16} is the presence of two significant phase transitions in the \textit{MUI} distribution, occurring at approximately 0.5 and 0.8. These transitions correspond to distinct grid characteristics related to transit availability. The phase transition at \( MUI \approx 0.5 \) captures grids that lack both bus and railway stations, leaving only active and private modes as contributors to the \textit{MUI} calculation. In the entropy-based \textit{MUI} formula, normalized by \( \log(4) \) for four modes, if only two modes are present with equal segregation levels, the \textit{MUI} approximates \( \frac{\log(2)}{\log(4)} = 0.5 \). The second transition at \( MUI \approx 0.8 \) corresponds to grids without railway stations but potentially retaining bus stations, meaning three modes (active, private, and bus) contribute. For three modes with roughly equal segregation levels, the \textit{MUI} approximates \( \frac{\log(3)}{\log(4)} \approx 0.7925 \), which is close to 0.8. These phase transitions reflect discrete drops in the number of travel modes available in a grid, directly impacting the uniformity of segregation experiences. At finer spatial resolutions, the increased number of grids amplifies the effect of these absences, leading to a greater shift in the \textit{MUI} distribution and explaining the pronounced K-S statistic in Supplementary Fig. \ref{FigS16}b.

\begin{figure}[h]
\centering
\includegraphics[width=1\textwidth]{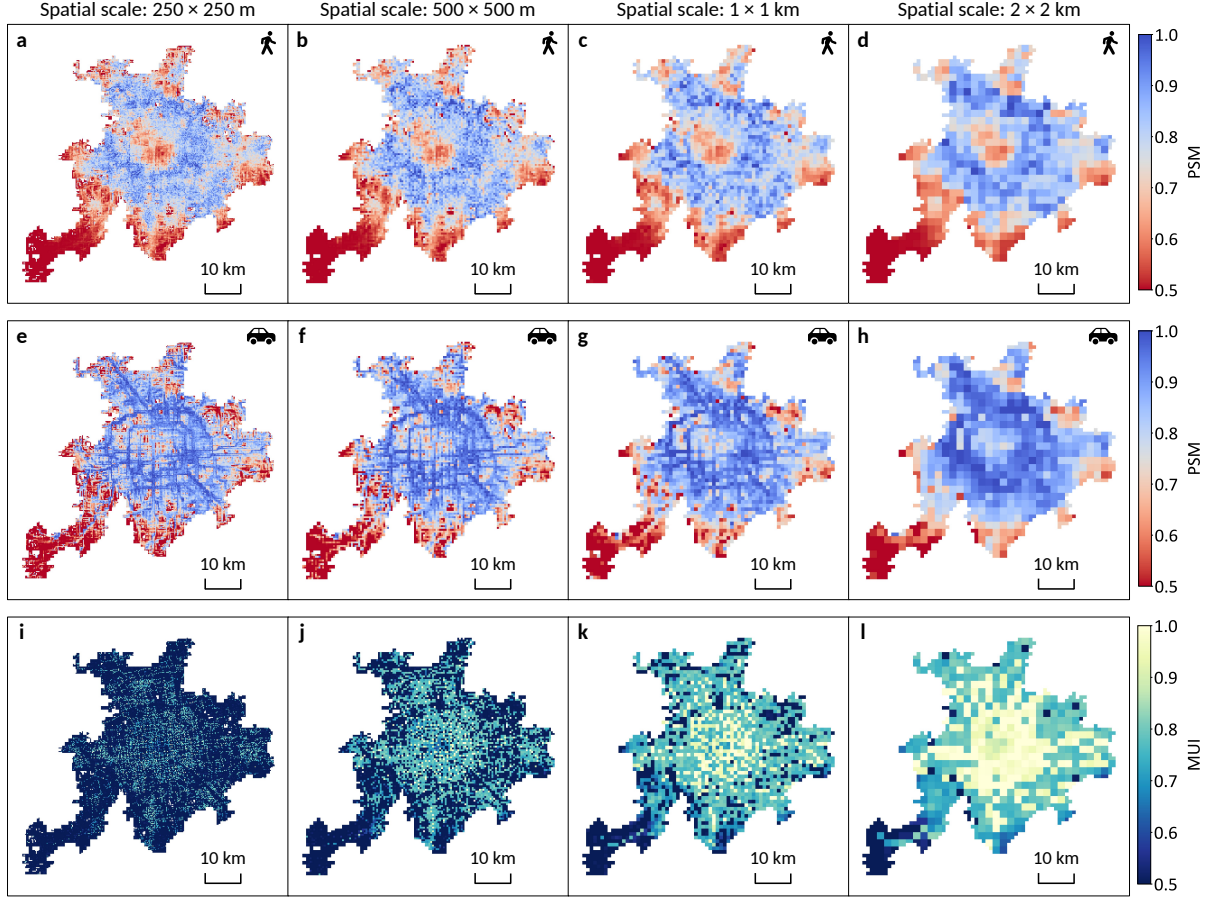}
\caption[Illustration of \textit{PSM} and \textit{MUI} distributions at different spatial scales in the Beijing metropolitan area.]{\textbf{Illustration of \textit{PSM} and \textit{MUI} distributions at different spatial scales in the Beijing metropolitan area.}}\label{FigS17}\end{figure}

Despite the substantial influence of spatial scale on segregation metric distributions (e.g., \textit{PSM} and \textit{MUI}), universal patterns are captured across different spatial resolutions (Supplementary Fig. \ref{FigS17}). Specifically, the spatial distributions of \textit{PSM} and \textit{MUI} in the Beijing metropolitan area reveal consistent trends, such as higher segregation in peripheral regions and greater uniformity in central urban cores, regardless of whether the grid size is 250 m or 2 km. These findings suggest that while the absolute values of the metrics may shift with scale, the underlying spatial organization of segregation and mode-specific interactions exhibits robust, scale-invariant characteristics that reflect the city's socioeconomic and infrastructural layout.

\section{OLS models explaining spatiotemporal patterns of segregation}\label{Supplnote3}

\noindent\hspace*{1.5em}We use ordinary least squares (OLS) regression models to explain how transport infrastructure influence segregation patterns observed through the probabilistic segregation measure (\textit{PSM}) and the multimodal uniformity index (\textit{MUI}). The form of the OLS regression model focusing on these factors is:

\begin{equation}
M_t = \beta_0 + \sum_{i} \beta_{T_i} T_i + \epsilon_t,
\label{eq:M_t_transport}
\end{equation}

where:

\begin{itemize}[leftmargin=1.5em]
    \item $M_t$ is the dependent variable, representing either the observed regional-level segregation metric ($PSM_{A,t,m}$, Eq.~\ref{eq:PSM_{A,t,m}}) for a specific travel mode $m$ (active/private/bus/railway) or the $MUI_{A,t}$ (Eq.~\ref{eq:MUI_{A,t}}) across four modes within region $A$ at a specific period $t$. The analysis is conducted at a spatial scale of 1\,km $\times$ 1\,km grids, which define the regions $A$.
    \item $\{T_i\}$ denotes the set of transport infrastructure-related explanatory variables. These include lengths of different road types provided by OpenStreetMap (Motorway, Primary, Secondary, Tertiary, Pedestrian roads), road diversity (calculated using the entropy of road types within a grid), and counts of transport facilities provided by Amap (e.g., Bus Stations, Subway Stations, Airports), reflecting the transportation and built environment features of the grids. Grid-level statistics for these variables are detailed in Supplementary Table~\ref{tab:explanatory_variables}.
    \item $\beta_0$ is the intercept of the regression model, representing the baseline level of $M_t$ when all transport infrastructure explanatory variables are zero.
    \item $\beta_{T_i}$ are the regression coefficients corresponding to the transport infrastructure variables, quantifying their individual contributions to $M_t$.
    \item $\epsilon_t$ is the error term, assumed to be normally distributed with a mean of zero and constant variance, capturing unexplained variation in the model at time $t$.
\end{itemize}

\vspace{1\baselineskip}
To capture temporal heterogeneity in segregation dynamics related to transport infrastructure, we design two model variants with distinct temporal granularity. \textbf{(1) Daily granularity model:} Separate models are estimated for workdays and weekends. For a specific day type \textit{d} (workday or weekend), the dependent variable $M_t$ for each grid $A$ is calculated as the average value of the segregation metric ($PSM_{A,t',m}$ or $MUI_{A,t'}$) across all hourly periods $t'$ falling within that day. \textbf{(2) Hourly granularity model:} Separate models are estimated for each specific hour of the day, differentiated by day type (workday vs. weekend). For a specific hour \textit{h} (e.g., 8:00--9:00 AM) and day type \textit{d} (workday or weekend), the dependent variable $M_t$ for each grid $A$ is the observed value of the segregation metric ($PSM_{A,t',m}$ or $MUI_{A,t'}$) specifically for that hour \textit{h} on days of type \textit{d}.

% ---------------------Summary of explanatory variables------------------------
\small
\begin{longtable}{ccccc} % Changed from cccccc to ccccc
\caption[Summary of explanatory variables and grid-level statistics.]{\textbf{Summary of explanatory variables and grid-level statistics.}} \\ % Added title
\toprule
% Removed "Variable group &" from the header
Primary category & Subcategory & Sum value & Median value & Max value \\
\midrule
\endfirsthead
\caption{Summary of explanatory variables and grid-level statistics.} \\ % Added title
\toprule
% Removed "Variable group &" from the header
Primary category & Subcategory & Sum value & Median value & Max value \\
\midrule
\endhead
\bottomrule
% Changed multicolumn from 6 to 5
\multicolumn{5}{r}{Continued on next page} \\
\endfoot
\bottomrule
\endlastfoot
% main data - Removed the first column data ("Infrastructure &") from each row
Transport Facility & Airport & 9 & 1 & 6 \\
Transport Facility & Train Station & 334 & 1 & 102 \\
Transport Facility & Port & 63 & 1 & 7 \\
Transport Facility & Intercity Bus Station & 40 & 1 & 3 \\
Transport Facility & Subway Station & 1489 & 4 & 17 \\
Transport Facility & Bus Station & 6098 & 3 & 17 \\
Transport Facility & Parking Lot & 56628 & 13 & 278 \\
Transport Facility & Toll Station & 125 & 2 & 6 \\
Transport Facility & Highway Service & 19 & 2 & 2 \\
Transport networks (km) & Motorway & 2242.997 & 0 & 12.949 \\
Transport networks (km) & Primary roads & 1770.722 & 0 & 8.366 \\
Transport networks (km) & Secondary roads & 2222.137 & 0.3885 & 6.451 \\
Transport networks (km) & Tertiary roads & 4199.528 & 1.5375 & 11.193 \\
Transport networks (km) & Pedestrian roads & 6867.383 & 2.5405 & 18.555 \\
Roads diversity & Roads diversity & & 0.6276 & 0.9909
\label{tab:explanatory_variables}
\end{longtable}
\normalsize

% --------------------------------PSM and MUI comparison table - Multirow Headers------------------------------
\footnotesize  % Reduce font size to fit page
\begin{longtable}{lcccccccccc} % l: left-align (Variable Name), c: center-align (coefficients)
\caption[Regression coefficients explaining probabilistic segregation measures (\textit{PSM}) and multimodal uniformity index (\textit{MUI}).]{\textbf{Regression coefficients explaining probabilistic segregation measures (\textit{PSM}) and multimodal uniformity index (\textit{MUI}).} Columns 2-9 represent \textit{PSM} models ($PSM_{A,t,m}$); Columns 10-11 represent MUI models ($MUI_{A,t}$). Significance levels: * $p < 0.1$, ** $p < 0.05$, *** $p < 0.001$. R$^2$ is the coefficient of determination, and MSE is the Mean Squared Error. Observations represent the number of grids included. Only significant variables are shown.} \label{tab:psm_mui_infrastructure_multirow} \\
\toprule
% --- Top Level Header: Model Type ---
& \multicolumn{8}{c}{Models for $PSM_{A,t,m}$} & \multicolumn{2}{c}{Models for $MUI_{A,t}$} \\
\cmidrule(lr){2-9} \cmidrule(lr){10-11} % Rule under Model Type
% --- Second Level Header: Mode ---
% --- Third Level Header: Time Period ---
% Using multirow for the last two columns to span rows 2 & 3
Variable & \multicolumn{2}{c}{Active Mode} & \multicolumn{2}{c}{Private Mode} & \multicolumn{2}{c}{Bus Mode} & \multicolumn{2}{c}{Railway Mode} & \multirow{2}{*}{\makecell{Workday}} & \multirow{2}{*}{\makecell{Weekend}} \\ % Start multirow here
\cmidrule(lr){2-3} \cmidrule(lr){4-5} \cmidrule(lr){6-7} \cmidrule(lr){8-9} % Rules only under modes
% --- Row 3 Placeholders for Modes / Empty for multirow cols ---
& \makecell{Workday} & \makecell{Weekend} & \makecell{Workday} & \makecell{Weekend} & \makecell{Workday} & \makecell{Weekend} & \makecell{Workday} & \makecell{Weekend} & & \\ % Time periods for modes, empty placeholders for multirow
\midrule
\endfirsthead

% --- Repeating Headers for Subsequent Pages ---
\caption[]{Regression coefficients explaining probabilistic segregation measures (PSM) grouped by mode and mobility uniformity index (MUI). Columns 2-9 represent PSM models ($PSM_{A,t,m}$); Columns 10-11 represent MUI models ($MUI_{A,t}$). Significance levels: * $p < 0.1$, ** $p < 0.05$, *** $p < 0.001$. R$^2$ is the coefficient of determination, and MSE is the Mean Squared Error. Observations represent the number of grids included. Only significant variables are shown. (Continued)}\\
\toprule
% --- Top Level Header: Model Type ---
& \multicolumn{8}{c}{Models for $PSM_{A,t,m}$} & \multicolumn{2}{c}{Models for $MUI_{A,t}$} \\
\cmidrule(lr){2-9} \cmidrule(lr){10-11} % Rule under Model Type
% --- Second Level Header: Mode ---
% --- Third Level Header: Time Period ---
% Using multirow for the last two columns to span rows 2 & 3
Variable & \multicolumn{2}{c}{Active Mode} & \multicolumn{2}{c}{Private Mode} & \multicolumn{2}{c}{Bus Mode} & \multicolumn{2}{c}{Railway Mode} & \multirow{2}{*}{\makecell{Workday}} & \multirow{2}{*}{\makecell{Weekend}} \\ % Start multirow here
\cmidrule(lr){2-3} \cmidrule(lr){4-5} \cmidrule(lr){6-7} \cmidrule(lr){8-9} % Rules only under modes
% --- Row 3 Placeholders for Modes / Empty for multirow cols ---
& \makecell{Workday} & \makecell{Weekend} & \makecell{Workday} & \makecell{Weekend} & \makecell{Workday} & \makecell{Weekend} & \makecell{Workday} & \makecell{Weekend} & & \\ % Time periods for modes, empty placeholders for multirow
\midrule
\endhead

\bottomrule
\multicolumn{11}{r}{Continued on next page} \\ % Adjusted colspan
\endfoot

\bottomrule
\endlastfoot
% --- Data Section ---
Motorway         & 0.083$^{***}$ & 0.071$^{**}$  & 0.306$^{***}$ & 0.319$^{***}$ & 0.11$^{***}$  & 0.1$^{***}$   & -0.006        & -0.002        & -0.048$^{**}$  & -0.058$^{**}$  \\
Primary roads    & 0.178$^{***}$ & 0.173$^{***}$ & 0.069$^{**}$  & 0.058$^{**}$  & 0.07$^{***}$  & 0.074$^{***}$ & -0.047$^{**}$ & -0.036$^{*}$  & 0.062$^{***}$  & 0.06$^{**}$    \\
Secondary roads  & 0.13$^{***}$  & 0.115$^{***}$ & 0.037$^{*}$   & 0.017         & 0.012         & 0.003         & -0.02         & -0.012        & 0.046$^{***}$  & 0.046$^{***}$  \\
Tertiary roads   & 0.309$^{***}$ & 0.258$^{***}$ & 0.211$^{***}$ & 0.148$^{***}$ & 0.157$^{***}$ & 0.082$^{***}$ & 0.079$^{***}$ & 0.058$^{**}$  & 0.107$^{***}$  & 0.108$^{***}$  \\
Pedestrian roads & -0.028        & -0.036        & -0.016        & -0.041$^{*}$  & -0.081$^{***}$& -0.109$^{***}$& 0.048$^{*}$   & 0.03          & 0.078$^{***}$  & 0.089$^{***}$  \\
Roads diversity  & 0.059$^{**}$  & 0.069$^{***}$ & 0.17$^{***}$  & 0.2$^{***}$   & 0.062$^{***}$ & 0.063$^{***}$ & 0.119$^{***}$ & 0.101$^{***}$ & 0.145$^{***}$  & 0.167$^{***}$  \\
Bus Station      & -0.011        & 0.032         & -0.058$^{**}$ & -0.044$^{*}$  & -0.045$^{*}$  & -0.035        & 0.037         & 0.036         & 0.253$^{***}$  & 0.262$^{***}$  \\
Airport          & -0.078        & -0.141        & -0.084        & -0.177$^{**}$ & -0.092        & -0.163$^{*}$  & -0.005        & -0.005        & 0.043          & 0.053          \\
Subway Station   & 0.283$^{***}$ & 0.262$^{***}$ & 0.039         & 0.018         & -0.003        & -0.041        & 0.104$^{***}$ & 0.087$^{***}$ & 0.316$^{***}$  & 0.306$^{***}$  \\
\midrule % Rule separating variables from summary stats
Observations     & 2116          & 2112          & 2114          & 2108          & 1687          & 1684          & 334           & 334           & 2116           & 2115           \\
R$^2$            & 0.229         & 0.227         & 0.219         & 0.249         & 0.081         & 0.067         & 0.15          & 0.129         & 0.423          & 0.412          \\
MSE              & 0.03          & 0.028         & 0.028         & 0.024         & 0.018         & 0.016         & 0.007         & 0.005         & 0.015          & 0.017          \\
\end{longtable}
\normalsize % Restore normal font size

Prior to finalizing each OLS model (for each temporal granularity and dependent variable), a systematic feature selection process is employed. Begin with the set of candidate transport infrastructure explanatory variables listed in Supplementary Table~\ref{tab:explanatory_variables}. The Variance Inflation Factor (\textit{VIF}) is calculated for each explanatory variable. If high multicollinearity is detected ($VIF > 10$) for a variable, this variable is removed. Variables contributing most significantly to multicollinearity are iteratively removed until all remaining variables have acceptable \textit{VIF}s ($VIF \leq 10$), ensuring the robustness and validity of the model estimates. The statistical results focusing on the significant transport infrastructure variables for the daily granularity models explaining both \textit{PSM} and \textit{MUI} are presented in Table~\ref{tab:psm_mui_infrastructure_multirow}. The results detailing the hourly variations in the influence of these transport variables on the mode-specific \textit{PSM} are shown in Figs.~\ref{FigS18}--\ref{FigS21}, and those for the \textit{MUI} are presented in Supplementary Supplementary Fig.~\ref{FigS22}.

\begin{figure}[h]
\centering
\includegraphics[width=1\textwidth]{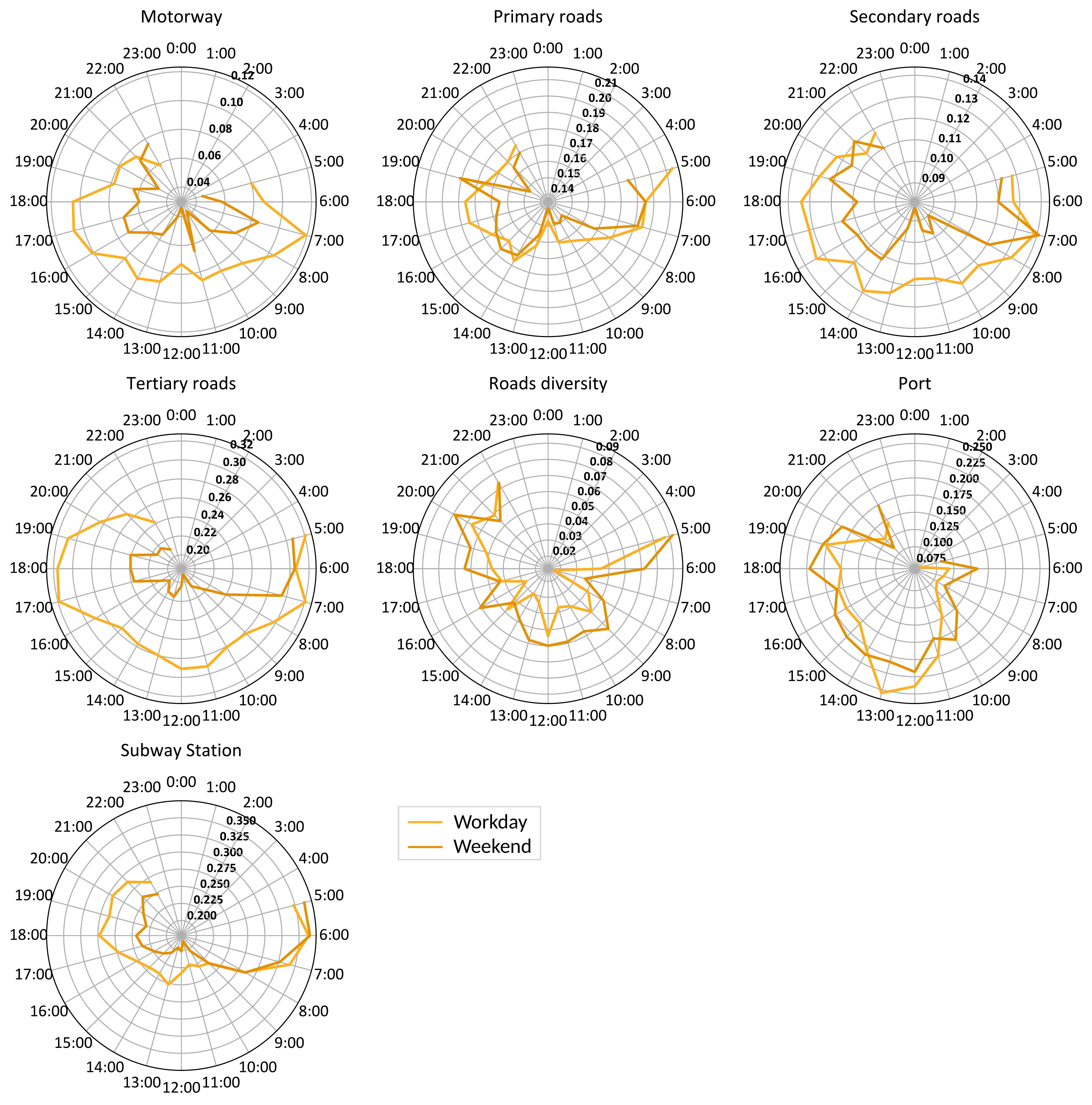}
\caption[Hourly \textit{PSM} dynamics for active mode across workdays and weekends.]{\textbf{Hourly \textit{PSM} dynamics for active mode across workdays and weekends.} This figure displays a series of compass plots, where each panel visualizes the variation of the regression coefficients (numeric label next to each circle) throughout hours of day for a specific explanatory variable (named in the title). Only statistically significant variables are displayed.}\label{FigS18}\end{figure}

\begin{figure}[h]
\centering
\includegraphics[width=1\textwidth]{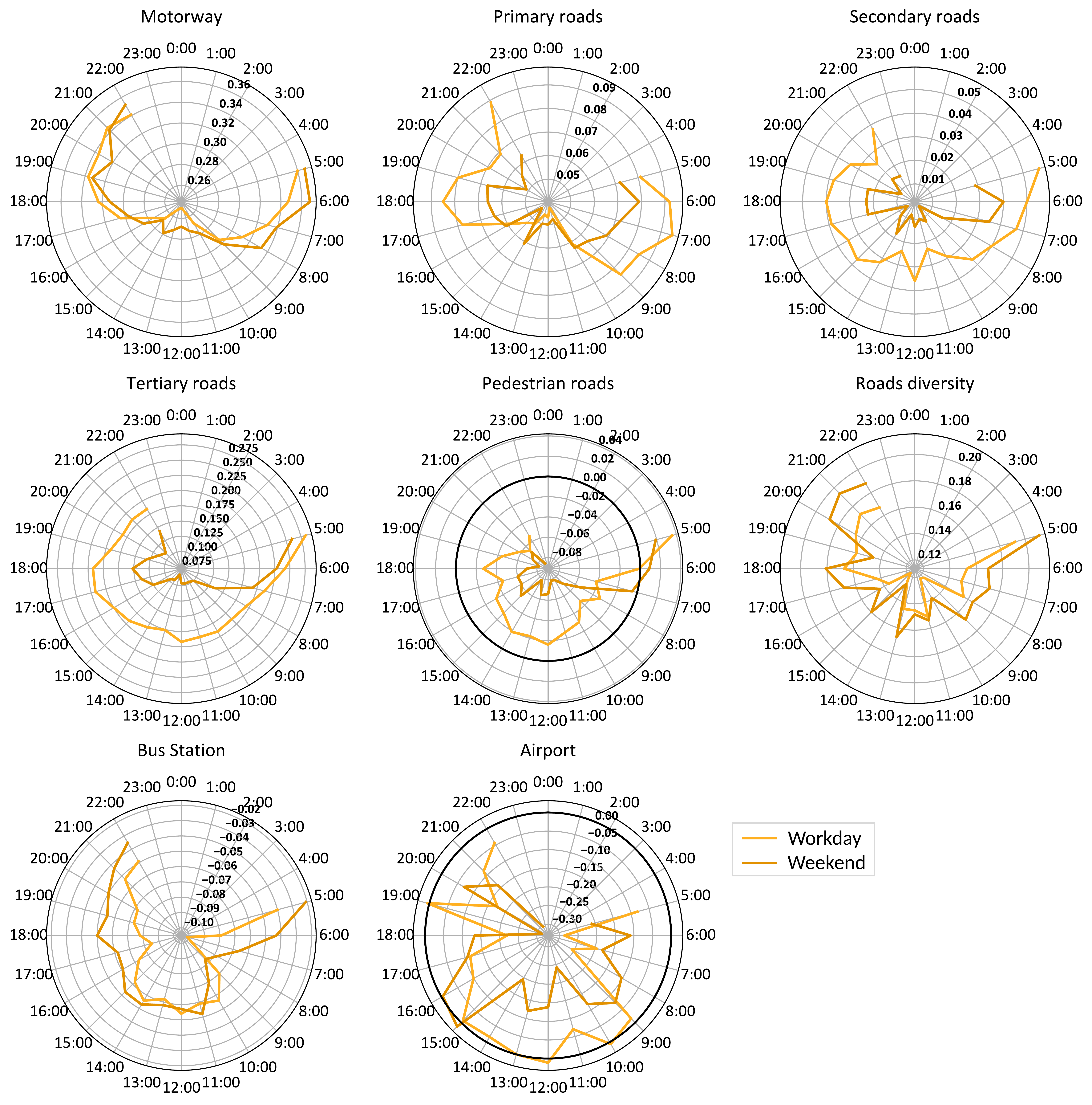}
\caption[Hourly \textit{PSM} dynamics for private mode across workdays and weekends.]{\textbf{Hourly \textit{PSM} dynamics for private mode across workdays and weekends.} This figure displays a series of compass plots, where each panel visualizes the variation of the regression coefficients (numeric label next to each circle) throughout hours of day for a specific explanatory variable (named in the title). Only statistically significant variables are displayed.}\label{FigS19}\end{figure}

\begin{figure}[h]
\centering
\includegraphics[width=1\textwidth]{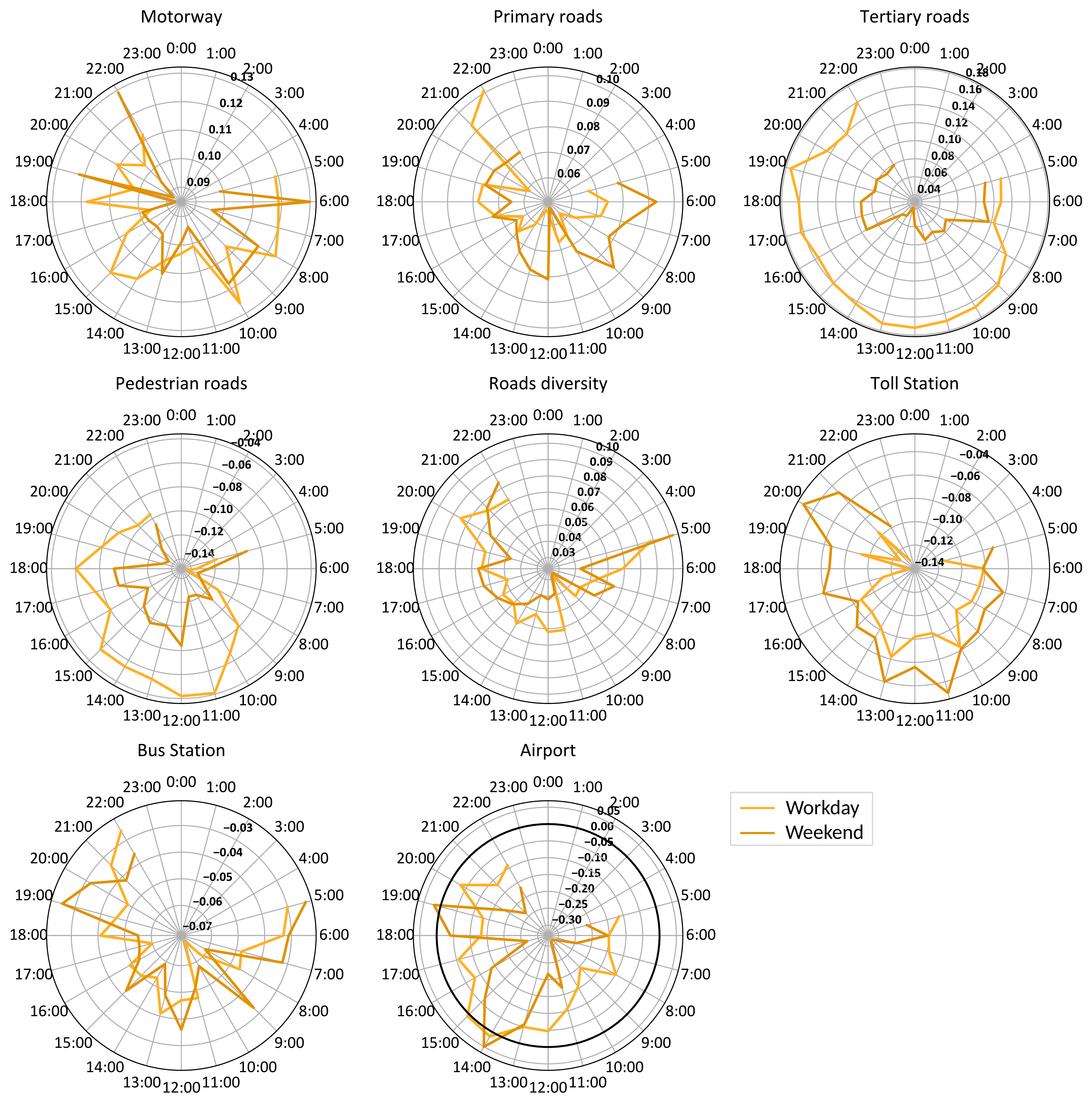}
\caption[Hourly \textit{PSM} dynamics for bus mode across workdays and weekends.]{\textbf{Hourly \textit{PSM} dynamics for bus mode across workdays and weekends.} This figure displays a series of compass plots, where each panel visualizes the variation of the regression coefficients (numeric label next to each circle) throughout hours of day for a specific explanatory variable (named in the title). Only statistically significant variables are displayed.}\label{FigS20}\end{figure}

\begin{figure}[h]
\centering
\includegraphics[width=1\textwidth]{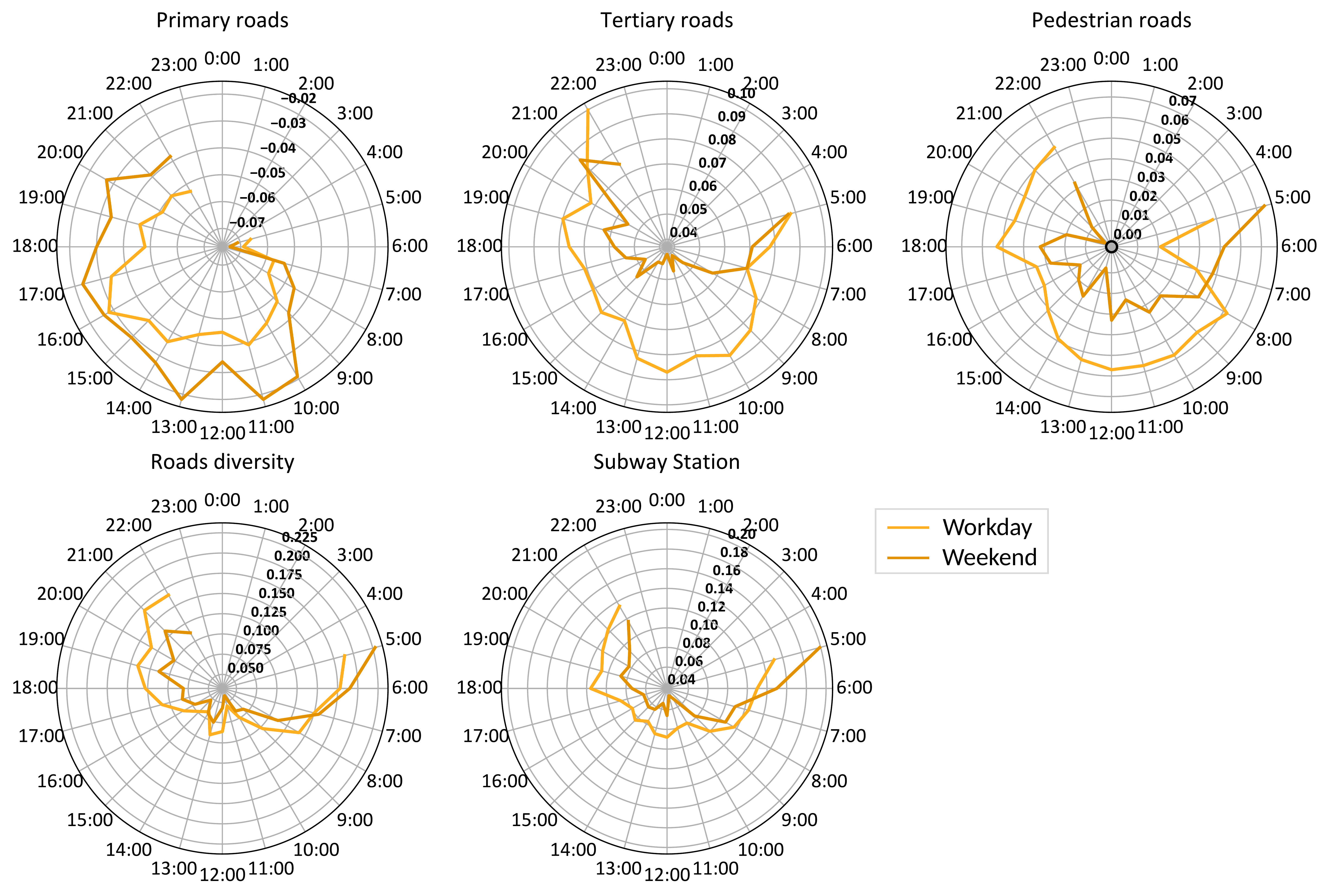}
\caption[Hourly \textit{PSM} dynamics for railway mode across workdays and weekends.]{\textbf{Hourly \textit{PSM} dynamics for railway mode across workdays and weekends.} This figure displays a series of compass plots, where each panel visualizes the variation of the regression coefficients (numeric label next to each circle) throughout hours of day for a specific explanatory variable (named in the title). Only statistically significant variables are displayed.}\label{FigS21}\end{figure}

\begin{figure}[h]
\centering
\includegraphics[width=1\textwidth]{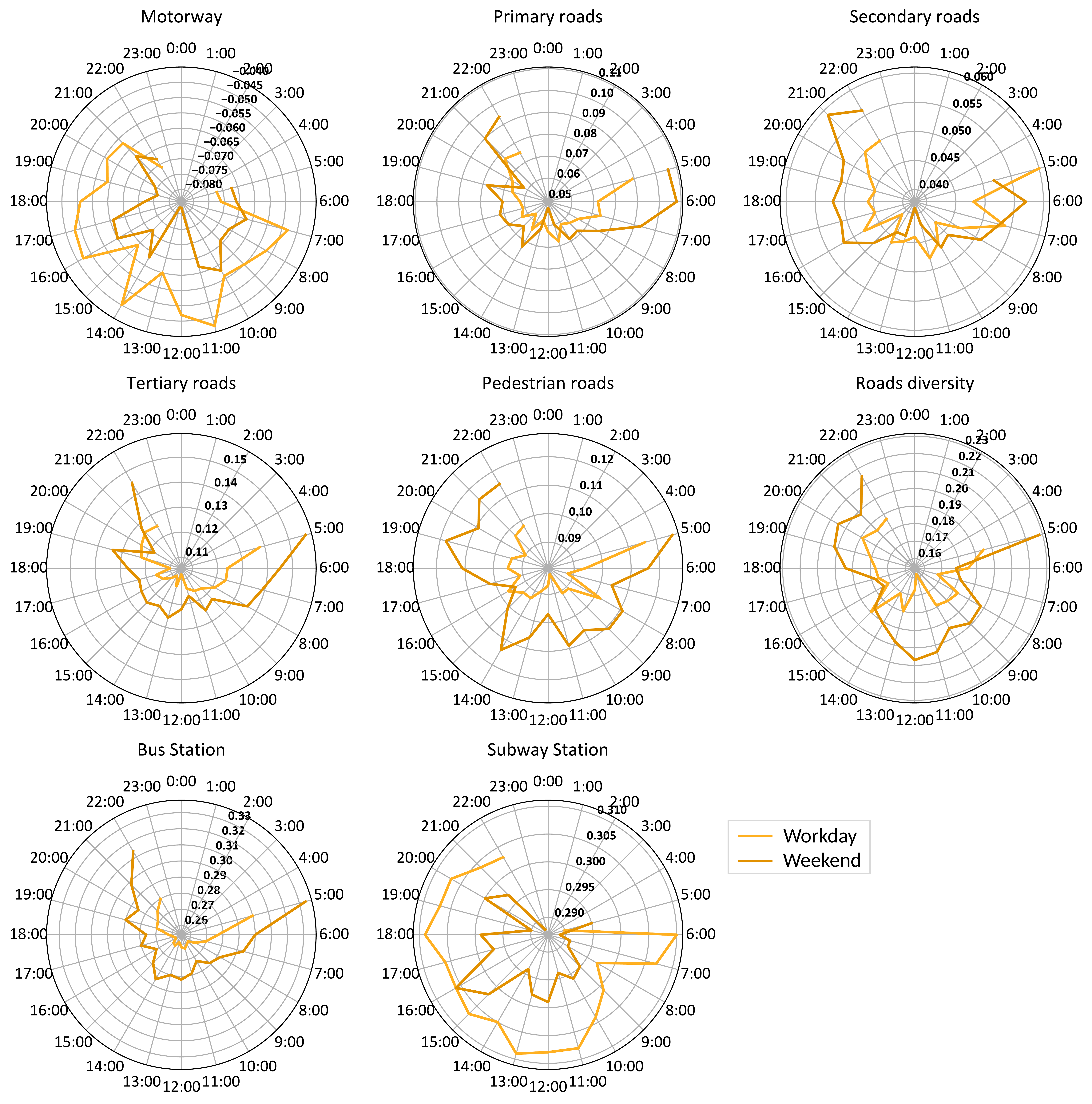}
\caption[Hourly \textit{MUI} dynamics across workdays and weekends.]{\textbf{Hourly \textit{MUI} dynamics across workdays and weekends.} This figure displays a series of compass plots, where each panel visualizes the variation of the regression coefficients (numeric label next to each circle) throughout hours of day for a specific explanatory variable (named in the title). Only statistically significant variables are displayed.}\label{FigS22}\end{figure}

\section{Agent-based model of individual mobility}\label{Supplnote4}

\noindent\hspace*{1.5em}This section provides detailed information about the agent-based mobility model used to simulate travel mode choices and their impact on urban segregation. The model is grounded in discrete choice theory \cite{benakiva1985discrete} and simulates the behavior of individuals commuting from home to work during morning peak hours (9:00-10:00 AM).

\subsection{Model specification}\label{Supplnote4.1}

\noindent\hspace*{1.5em}Model considers a population of individuals, each belonging to a specific income group $g \in \mathcal{G}$ (where $\mathcal{G}$ represents the set of defined income groups, $|\mathcal{G}|=4$ in this study). Each individual $i$ needs to make a trip from their home location to their workplace location. For this commute, they choose a travel mode $m$ from a set of available modes $\mathcal{M}$. For model simplification, we focus on three key modes: active, private, and railway. The choice is probabilistic and assumes individuals aim to minimize their perceived travel cost $\mathcal{C}_{gm}$ associated with each mode $m$ for their income group $g$. The probability $p_{gm}$ that an individual from group $g$ chooses mode $m$ is given by the multinomial logit formula:

\begin{equation}
p_{gm} = \frac{\exp(-\delta \mathcal{C}_{gm})}{\sum_{m' \in \mathcal{M}} \exp(-\delta \mathcal{C}_{gm'})},
\label{eq:supp_p_gm}
\end{equation}

\noindent\hspace*{0em}where $\delta$ is the sensitivity parameter, reflecting how strongly cost differences influence mode choice. A higher $\delta$ indicates greater sensitivity to cost variations. The perceived travel cost $\mathcal{C}_{gm}$ for an individual of income group $g$ using mode $m$ is defined as:

\begin{equation}
\mathcal{C}_{gm} = \alpha_g T_m + \beta_m T_m = (\alpha_g + \beta_m) T_m,
\label{eq:supp_mathcal_C_gm}
\end{equation}

\noindent\hspace*{0em}where:
\begin{itemize}
    \item $T_m$ is the estimated travel time for mode $m$. This is calculated as the duration of the shortest path $\textbf{R}_{m}$ from an individual's home to workplace using the real-world transport networks in Beijing metropolitan area. For active or private mode, shortest path is generated on the road networks. For railway mode, the shortest path calculation leverages the unified graph to determine the optimal door-to-door journey time, which includes estimated walking time from home to the nearest origin station, travel time on the railway networks, and walking time from the destination station to the workplace.
    \item $\alpha_g$ represents the monetary value of time (unit time cost) for income group $g$. This parameter captures how different income groups value their travel time, assumed to increase with income.
    \item $\beta_m$ represents the mode-specific cost components beyond the value of time, expressed per unit of travel time. This encompasses monetary costs (e.g., fuel, tolls, parking fees for private transport; fares for public transit) and non-monetary factors such as discomfort, inconvenience, physical effort (for active travel), or reliability associated with mode $m$.
    
\end{itemize}
The model assumes that individuals have perfect information about the travel times $T_m$ and associated costs represented by $\alpha_g$ and $\beta_m$. It also inherently includes the Independence of Irrelevant Alternatives (IIA) property common to logit models \cite{mcfadden1974}. The model output consists of the mode choice probabilities $p_{gm}$ for each income group and, by extension, for each individual based on their group. These probabilities, combined with the shortest travel paths $\textbf{R}_{m}$, are used as inputs to calculate the mode-specific segregation measure \textit{PSM}, as described in the main text.

\subsection{Parameter calibration}\label{Supplnote4.2}

\noindent\hspace*{1.5em}The model includes several parameters that need to be calibrated: the sensitivity parameter $\delta$, the group-specific value of time parameters $\alpha_g$ (for $g=1, ..., 4$), and the mode-specific cost factors $\beta_m$ (for $m \in \{\text{active, private, railway}\}$). In total, eight parameters constitute the parameter set $\Theta = \{\delta, \alpha_1, \alpha_2, \alpha_3, \alpha_4, \beta_{\text{active}}, \beta_{\text{private}}, \beta_{\text{railway}}\}$. The group-specific value of time parameters $\alpha_g$ are determined prior to the main calibration using the inferred socioeconomic status of the individuals in the dataset. Specifically, the average income level $AvgIncome_g$ is computed for all individuals belonging to each of the four income groups ($g=1, ..., 4$). Based on the economic principle that the value of time generally correlates with income level \cite{Becker1965}, we establish the relative values of $\alpha_g$ by normalizing these average incomes. The highest income group ($g=4$) is chosen as the reference point. The value of time parameter for each group $g$ is then set proportionally to its average income relative to this benchmark:

\begin{equation}
\alpha_g = \frac{AvgIncome_g}{AvgIncome_4} \quad \text{for } g = 1, 2, 3, 4.
\label{eq:supp_alpha_g_determination}
\end{equation}

This approach directly anchors the relative time preferences of different income groups to their empirically observed economic standing, yielding $\alpha_1=0.203, \alpha_2=0.349, \alpha_3=0.596$ and $\alpha_4 = 1$. By pre-determining the $\alpha_g$ values in this manner, we incorporate realistic socioeconomic differentiation into the model structure. Consequently, the main calibration process focuses on estimating the remaining four parameters: the overall sensitivity to cost $\delta$, and the mode-specific cost factors $\beta_{\text{active}}, \beta_{\text{private}}, \beta_{\text{railway}}$. The objective of this calibration is to minimize the discrepancy between model-predicted mode-specific segregation (\textit{PSM}) values and empirically observed \textit{PSM} values derived from mobile phone data during workday morning peak hours (9:00-10:00 AM).

To perform this calibration, we define an objective function $L(\Theta')$ that quantifies the goodness-of-fit between the model predictions and the empirical observations. We use the sum of squared errors (SSE) across all relevant spatial units $s$ for each mode $m$ during the target time frame $t$ (9:00-10:00 AM):

\begin{equation}
L(\delta, \beta_{\text{active}}, \beta_{\text{private}}, \beta_{\text{railway}}) = \sum_{m \in \mathcal{M}} \sum_{s \in S_m} \left( \widehat{PSM}_{s,t,m}(\Theta') - PSM_{s,t,m}^{\text{obs}} \right)^2,
\label{eq:supp_objective_func_reduced}
\end{equation}

\noindent\hspace*{0em}where $\widehat{PSM}_{s,t,m}(\Theta')$ is the model-predicted segregation value for spatial unit $s$, time $t$, and mode $m$, calculated using the fixed $\alpha_g$ values (Eq. \ref{eq:supp_alpha_g_determination}) and the candidate parameter set $\Theta' = \{\delta, \beta_{\text{active}}, \beta_{\text{private}}, \beta_{\text{railway}}\}$. $PSM_{s,t,m}^{\text{obs}}$ is the corresponding empirically observed segregation value, and $S_m$ is the set of all spatial units relevant for mode $m$.

The calibration involves finding the parameter set $\Theta'^* = \{\delta^*, \beta_{\text{active}}^*, \beta_{\text{private}}^*, \beta_{\text{railway}}^*\}$ that minimizes the objective function $L(\Theta')$. We employ a Grid Search approach, a discrete optimization method, to perform this minimization. For each parameter in $\Theta'$, a plausible range of values is defined based on theoretical considerations and preliminary analysis. For $\delta$, the range [0.0001, 0.01]; for $\beta_g$, the range [0, 10] is explored. Each range is then discretized using a fixed step size (0.0001 for $\delta$; 0.01 for $\beta_g$), creating a multi-dimensional candidate parameter combinations.

The model simulation process is executed for each combination of parameter values in the grid. Within this process, we first filter the individuals whose most frequent commute time, as observed in the dataset, falls within the target morning peak hour (9:00-10:00 AM). Then, for this selected subset of individuals and for the current parameter combination being tested, we calculate their perceived travel costs $\mathcal{C}_{gm}$, their mode choice probabilities $p_{gm}$, and subsequently the predicted segregation measures $\widehat{PSM}_{s,t,m}(\Theta')$. The objective function $L(\Theta')$ is then evaluated for this parameter combination using these predicted \textit{PSM} values. This procedure is repeated. The parameter combination yielding the minimum value of the objective function $L(\Theta')$ is selected as the optimal calibrated parameter set $\Theta'^*$. The resulting calibrated parameter values are: $\delta^* = 3\times10^{-4}$, $\beta_{\text{active}}^* = 0.22$, $\beta_{\text{private}}^* = 2.1$, and $\beta_{\text{railway}}^* = 0.07$.

To assess the model's performance with these calibrated parameters, we compare the model-predicted \textit{PSM} values with the empirically observed \textit{PSM} values for each spatial unit across the three modes (active, private, railway). A strong positive correlation, quantified by Pearson correlation coefficients (active: \textit{r} = 0.9327***; private: \textit{r} = 0.9613***; railway: \textit{r} = 0.9551***; see Supplementary Fig. \ref{FigS23}), demonstrates a good fit of the model to the observed segregation patterns.

\begin{figure}[h]
\centering
\includegraphics[width=1\textwidth]{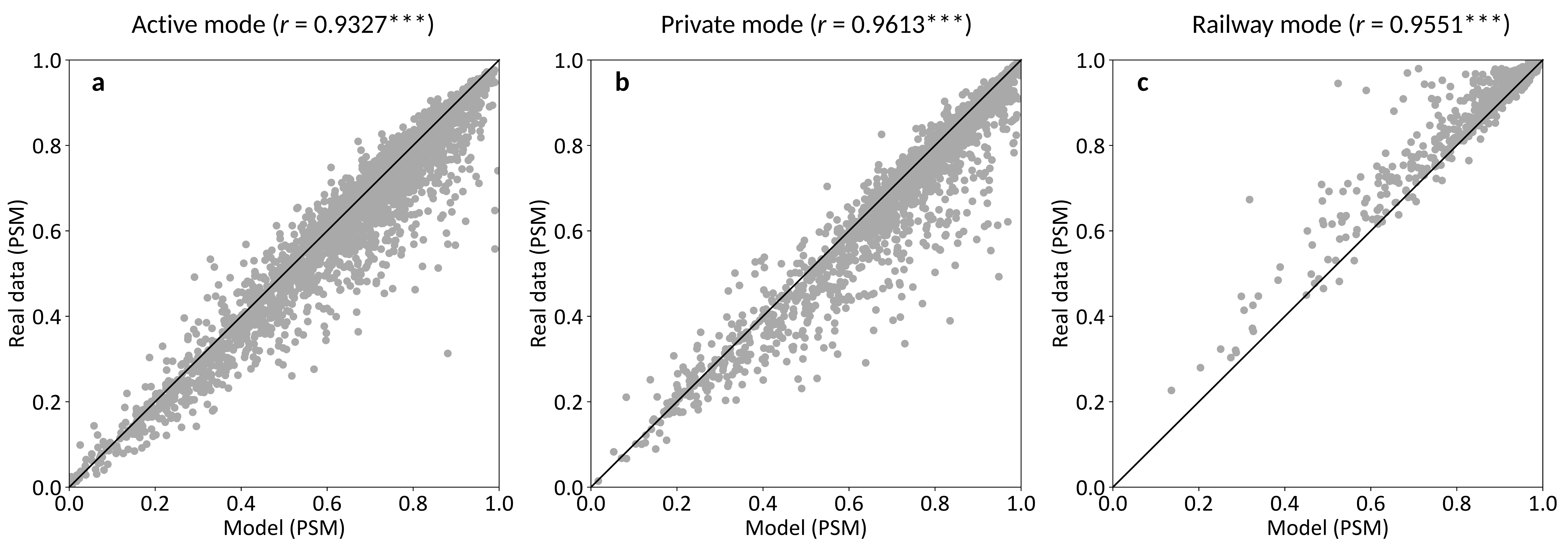}
\caption[Comparison of model-predicted and observed \textit{PSM} values.]{\textbf{Comparison of model-predicted and observed \textit{PSM} values.} Scatter plots showing the relationship between predicted \textit{PSM} (using calibrated parameters) and observed \textit{PSM} for spatial units during the 9:00-10:00 AM peak hour, separated by travel modes: (\textbf{a}) Active, (\textbf{b}) Private, (\textbf{c}) Railway. Pearson correlation coefficients (r) are indicated on each panel (***$p < 0.001$).}\label{FigS23}\end{figure}

\subsection{Simulation for private car use control policies}\label{Supplnote4.3}

\noindent\hspace*{1.5em}A central challenge in urban transport planning lies in designing policies that achieve collective goals, such as reducing congestion or emissions, without disproportionately burdening specific population groups or exacerbating existing social inequalities \cite{Lucas2012TP}. The distributional effects of transport policies -- how their costs and benefits are spread across different socioeconomic groups and geographic areas -- are therefore a critical consideration, particularly when evaluating strategies aimed at altering travel behavior \cite{Pereira2017TR}. This study seeks to investigate the distributional effects by expanding the agent-based mobility model to simulate the impacts of urban transport policies on urban segregation and social equity. Among the various policy levers available, measures aimed at controlling or reducing private car usage are prominent tools used globally to manage urban mobility challenges. These policies, commonly taking forms such as congestion pricing zones \cite{Leape2006,Eliasson2009}, increased fuel taxes \cite{Goodwin2004,Parry2007}, or stricter parking regulations \cite{Marsden2006}, are primarily designed to discourage driving by making it more expensive or less convenient. The primary goal is typically to mitigate negative externalities associated with high vehicle density, including traffic congestion, air and noise pollution, and inefficient use of urban space, thereby encouraging shifts towards public transport, cycling, or walking.

We operationalize these policies within our model framework by systematically increasing the mode-specific cost parameter $\beta_{\text{private}}$. As defined in Eq. \ref{eq:supp_mathcal_C_gm}, this parameter encapsulates the perceived costs associated with private vehicle travel beyond the value of time itself, encompassing factors like fuel expenses, tolls, parking fees, and potentially the perceived inconvenience or stress of driving. Incrementally increasing $\beta_{\text{private}}$ thus serves as a proxy for implementing stricter car control measures that raise the generalized cost of driving, allowing us to explore the sensitivity of travel behavior and segregation patterns to such interventions. To investigate different policy designs, we implement two simulation scenarios. (1) \textbf{Uniform citywide cost increase}: $\beta_{\text{private}}$ is increased uniformly across the entire Beijing metropolitan area, simulating a policy such as a citywide fuel tax or a blanket increase in parking fees that affects all private car users equally, regardless of their location within the city, according to $\beta'_{\text{private}} = \beta_{\text{private}}^* + \Delta\beta_{\text{private}}$. (2) \textbf{Downtown-targeted cost increase}: This scenario mimics geographically specific policies like a downtown congestion charge by increasing $\beta_{\text{private}}$ solely for individuals whose commute either originates from or is destined for a location within the downtown zone, using $\beta'_{\text{private}} = \beta_{\text{private}}^* + \Delta\beta_{\text{private}}$, while costs for other trips starting and ending elsewhere remain at baseline $\beta_{\text{private}}^*$. For both scenarios, we systematically vary the policy-induced cost increment, $\Delta\beta_{\text{private}}$, from 0 (baseline) to 15 in steps of 0.2. For each value of $\Delta\beta_{\text{private}}$ under each scenario, we recalculate the perceived travel costs, update individual mode choice probabilities ($p_{gm}$) using Eq.~\ref{eq:supp_p_gm}, and subsequently compute the resultant mode-specific segregation (\textit{PSM}), changes in mode shares per income group, and the average travel costs per income group.

For the first scenario (detailed in the main text Fig. 4), increasing the citywide cost of driving reduces private car use (disproportionately among lower-income groups, increasing segregation within that mode) and shifts commuters to active and public transport. This leads to complex, non-linear changes in segregation across these alternative modes as the cost increment rises. Overall travel costs increase for all income groups and the burden is distributed unevenly. To illustrate the spatial impact of this policy, we select a specific cost increment $\Delta\beta_{\text{private}}=5$ for detailed comparison against the baseline scenario ($\Delta\beta_{\text{private}}=0$). For this comparison, we compute the \textit{PSM} for each spatial unit $s$ under both the baseline and the policy scenario, across all three travel modes (active, private, railway). We then calculate the change ratio in \textit{PSM} for each spatial unit and mode. The spatial distribution of these changes, specifically mapping the locations of spatial units where segregation increased and decreased for each mode due to the policy intervention, is visualized in Supplementary Fig.~\ref{FigS24}. The policy increases overall segregation among private car users because the reduction in driving is proportionally larger for high-income groups in suburbs and low-income groups downtown, making the remaining user pool less representative in both areas. Conversely, for active and railway travel, the policy drives proportionally more high-income individuals to these modes citywide. This results in a consistent spatial pattern for both: segregation decreases in the predominantly low-income suburbs due to enhanced income mixing, while segregation increases in the predominantly high-income downtown due to further concentration. Despite this shared spatial dynamic, the net effect differs: overall segregation decreases for active travel, whereas for railway travel at this cost increment, the downtown concentration effect dominates, leading to increased overall segregation (main text Fig. 4e).

For the second scenario (downtown-targeted cost increase), individuals undertaking purely suburban trips face no direct change in driving costs. Compared to the uniform policy at the same cost increment $\Delta\beta_{\text{private}}$, the downtown-targeted scenario notably results in smaller proportional mode shifts among lower-income groups (Supplementary Fig.~\ref{FigS25}): they reduce private car use less and increase active/railway use less than higher-income groups. Consistent with this, lower-income groups also experience a smaller proportional increase in their average travel costs (Supplementary Fig.~\ref{FigS26}). This disparity strongly suggests the policy's impact is less felt by lower-income populations overall, primarily because fewer commute to the restricted downtown zone from their largely suburban homes, thus avoiding direct exposure to the cost increase. Additionally, even for those commuting downtown, lower-income individuals might already rely less on cars due to baseline affordability and potentially face fewer viable or suitable non-car alternatives compared to their higher-income counterparts. Consequently, as shown in Supplementary Fig.~\ref{FigS27}, the downtown-targeted policy leads to a higher overall \textit{PSM} for private car travel citywide, indicating reduced segregation among the remaining users compared to the uniform policy outcome. Conversely, for both active and railway modes, the overall citywide \textit{PSM} decreases under this targeted policy, signaling heightened segregation as the influx of former downtown car commuters, potentially skewed towards higher incomes, concentrates usage within these modes.

In summary, the spatial design of private car control policies fundamentally alters their distributional effects. A uniform citywide cost increase spreads the burden broadly but yields mixed results on segregation—worsening it for private cars while improving it for active travel. Conversely, a downtown-targeted policy concentrates costs on downtown commuters (lessening the impact on low-income groups citywide) but paradoxically reduces segregation for private cars while increasing it for active and railway modes. These contrasting outcomes highlight critical trade-offs between travel costs and social mixing across different mobility layers. Policymakers must carefully weigh these effects: targeted policies like congestion pricing may require specific measures (e.g., revenue recycling, targeted discounts) to mitigate impacts on essential low-income downtown commuters. More broadly, achieving equitable outcomes necessitates integrated strategies that consider both cost burdens and segregation impacts across all modes, potentially combining different policy types to counteract unintended negative social consequences.

\begin{figure}[h]
\centering
\includegraphics[width=1\textwidth]{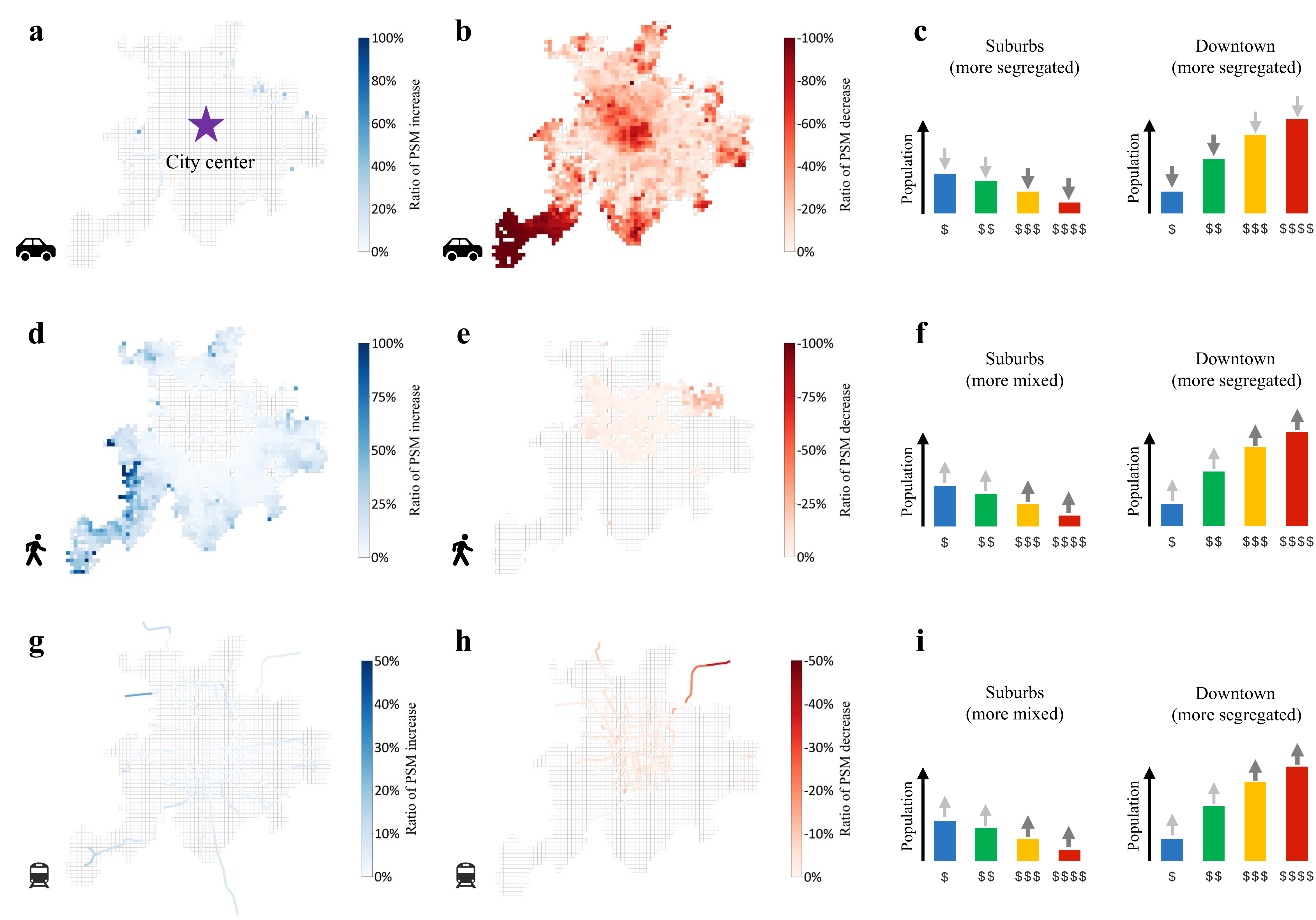}
\caption[Spatial distribution of changes in segregation measure \textit{PSM} for three transport modes under a uniform private car cost increase policy.]{\textbf{Spatial distribution of changes in segregation measure \textit{PSM} for three transport modes under a uniform private car cost increase policy}. The figure compares the policy scenario ($\Delta\beta_{\text{private}}=5$) to the baseline ($\Delta\beta_{\text{private}}=0$) across three transport modes. Rows correspond to modes: \textbf{a--c} Active, \textbf{d--f} Private, \textbf{g--i} Railway. First column of panels (\textbf{a, d, g}) show the spatial units where \textit{PSM} increased (indicating reduced segregation). Second column of panels (\textbf{b, e, h}) show the spatial units where \textit{PSM} decreased (indicating heightened segregation). Third column of panels (\textbf{c, f, i}) represent the illustrative diagrams depicting policy-induced shifts in mode usage by income group for each respective mode. Bars represent the baseline proportion of commuters per income group using the mode. Arrows indicate the net change (increase or decrease) in commuters for that group under the policy ($\Delta\beta_{\text{private}}=5$). Bold arrows denote changes of larger magnitude, summarizing effects relevant to areas of downtown and suburbs.}\label{FigS24}\end{figure}

\begin{figure}[h]
\centering
\includegraphics[width=1\textwidth]{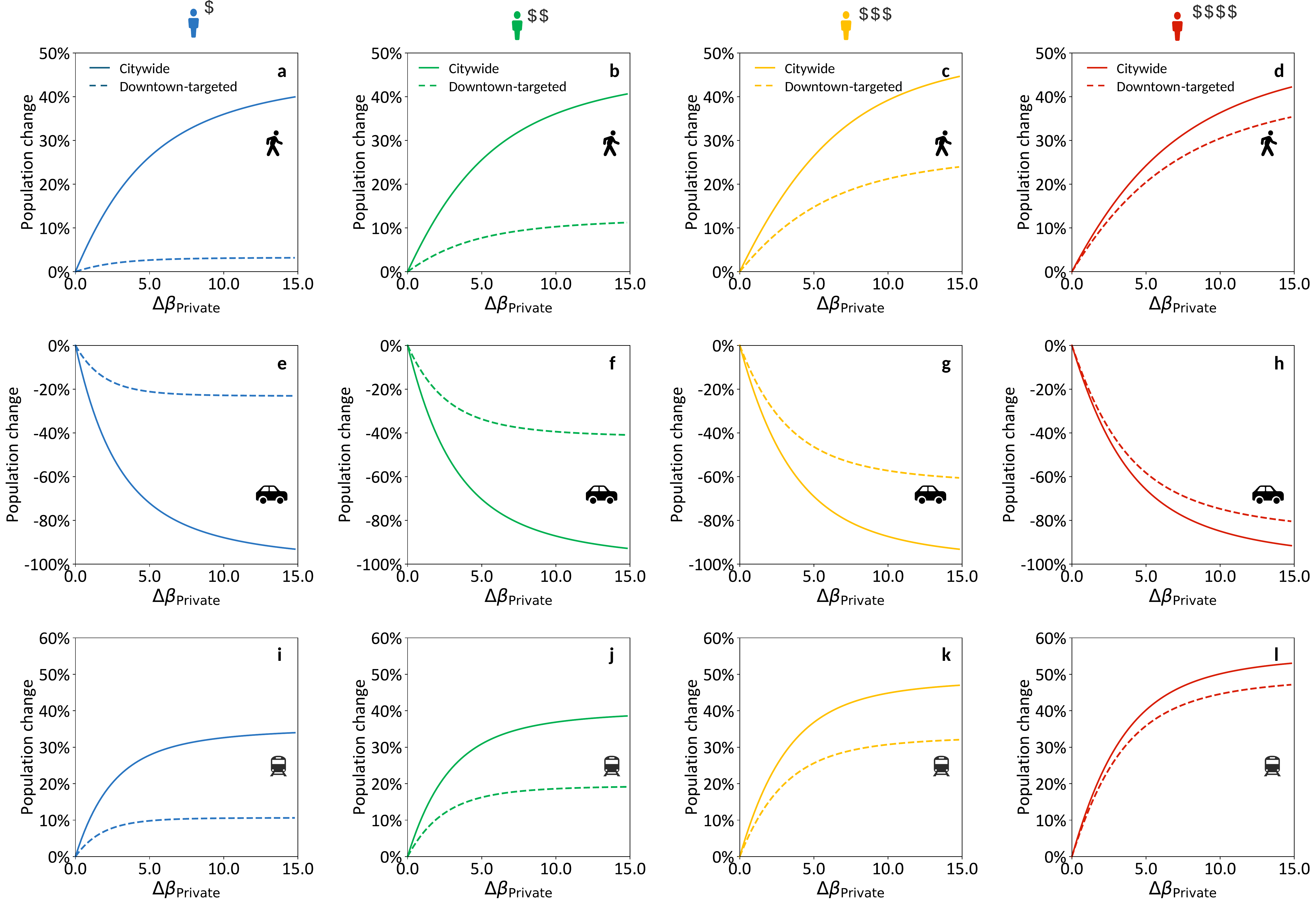}
\caption[Proportional changes in mode usage by income groups under two private car policy scenarios.]{\textbf{Proportional changes in mode usage by income groups under two private car policy scenarios}.}\label{FigS25}\end{figure}

\begin{figure}[h]
\centering
\includegraphics[width=0.8\textwidth]{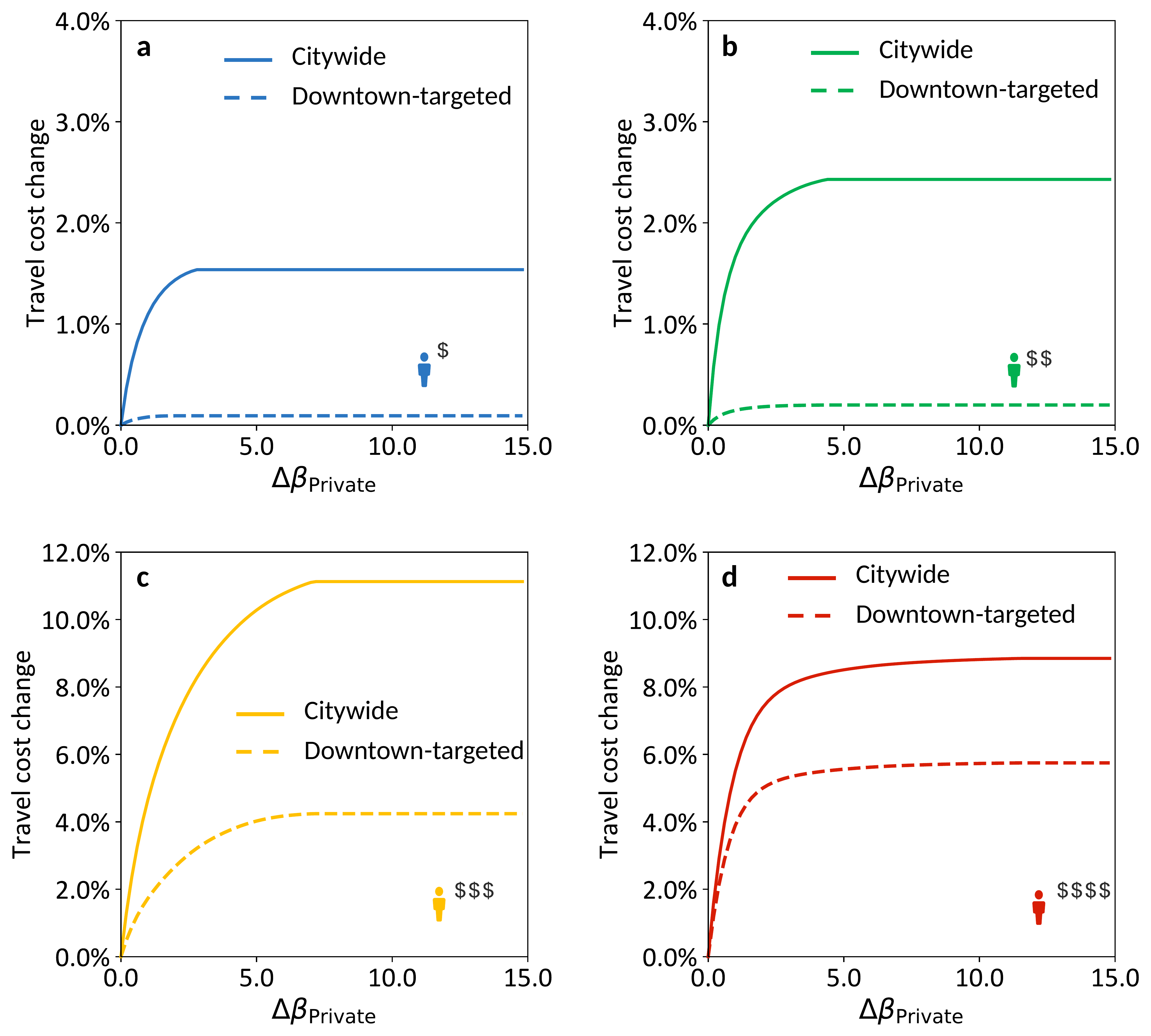}
\caption[Impact of two private car policies on average travel costs by income group.]{\textbf{Impact of two private car policies on average travel costs by income group.} The plot shows the percentage change in average travel cost (relative to the baseline scenario where $\Delta\beta_{\text{private}}=0$) for each of the four income groups. This illustrates the differential cost burden across income groups under the downtown targeted policy, showing a notably smaller proportional cost increase for lower-income groups.}\label{FigS26}\end{figure}

\begin{figure}[h]
\centering
\includegraphics[width=1\textwidth]{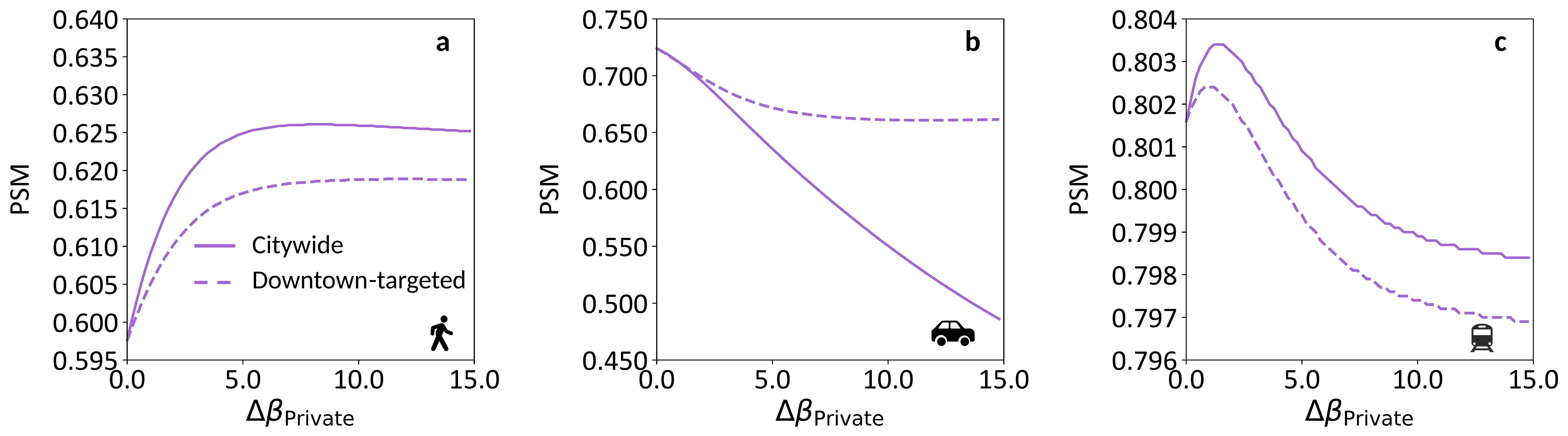}
\caption[Impact of two private car policies on segregation at the city scale by mode.]{\textbf{Impact of two private car policies on segregation at the city scale by mode.} Under the downtown targeted policy, overall segregation decreases for private car users (indicated by a higher \textit{PSM}) while it increases for both active and railway users (indicated by a lower \textit{PSM}) across the city.}\label{FigS27}\end{figure}

\subsection{Simulation for public transport subsidy policies}\label{Supplnote4.4}

\noindent\hspace*{1.5em}Beyond restricting car use, promoting public transport through subsidies is a common strategy for influencing travel behavior and addressing equity concerns \cite{Taylor2002,Pucher2008}. We simulate this by enhancing railway attractiveness, systematically reducing its mode-specific cost parameter $\beta_{\text{railway}}$. This reflects decreased fares or improved service quality, lowering the perceived generalized cost of railway travel. Specifically, we introduce a subsidy factor $\Delta\beta_{\text{public}}$ and modify the railway cost parameter as $\beta'_{\text{railway}} = \beta_{\text{railway}}^* + \Delta\beta_{\text{public}}$. We vary $\Delta\beta_{\text{public}}$ from 0 (baseline) down to -0.07, where the latter value cancels the baseline non-time-related cost factor ($\beta_{\text{railway}}^* = 0.07$), making perceived railway cost solely dependent on travel time value ($\alpha_g T_{\text{railway}}$). For each subsidy level, we recalculate mode choice probabilities ($p_{gm}$), average travel costs per income group, and mode-specific segregation (\textit{PSM}) using the established simulation procedure.

\begin{figure}[h]
\centering
\includegraphics[width=1\textwidth]{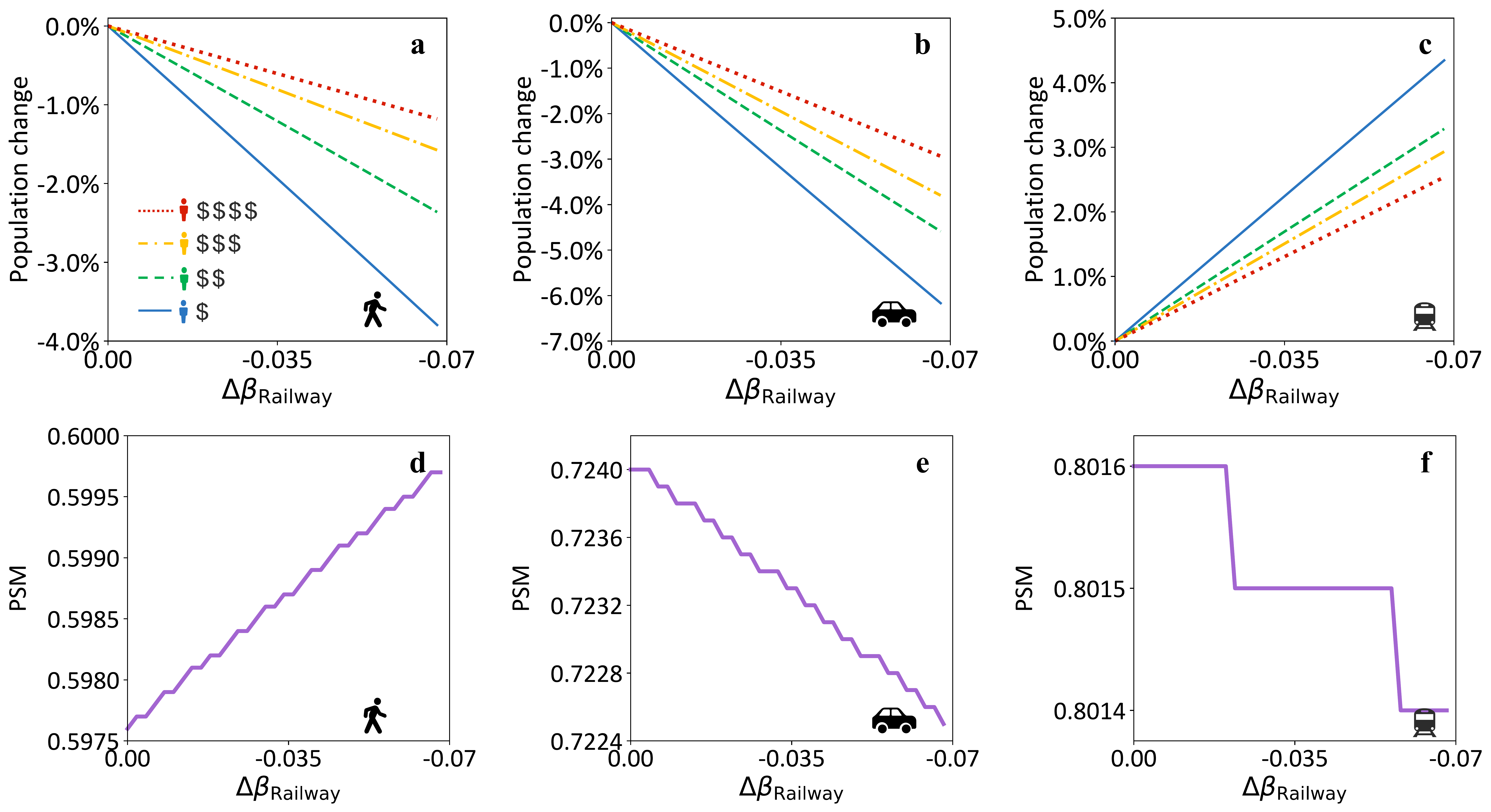}
\caption[Impact of public transport subsidies on mode usage and citywide segregation.]{\textbf{Impact of public transport subsidies on mode usage and citywide segregation.} The figure shows the effects as the railway cost parameter is reduced ($\Delta\beta_{\text{public}}$ varies from 0 to -0.07). \textbf{a--c} Proportional changes in mode usage relative to the baseline for active (panel \textbf{a}), private (panel \textbf{b}), and railway (panel \textbf{c}) modes, separated by income group. Lower-income groups show larger proportional decreases in active/private use and larger increases in railway use. \textbf{d--f} Overall citywide segregation measure (\textit{PSM}) for active (panel \textbf{d}), private (panel \textbf{e}), and railway (panel \textbf{f}) modes. Results indicate reduced segregation (increasing \textit{PSM}) for active travel, but heightened segregation (decreasing \textit{PSM}) for both private and railway travel as subsidies increase.}\label{FigS28}\end{figure}

\begin{figure}[h]
\centering
\includegraphics[width=0.5\textwidth]{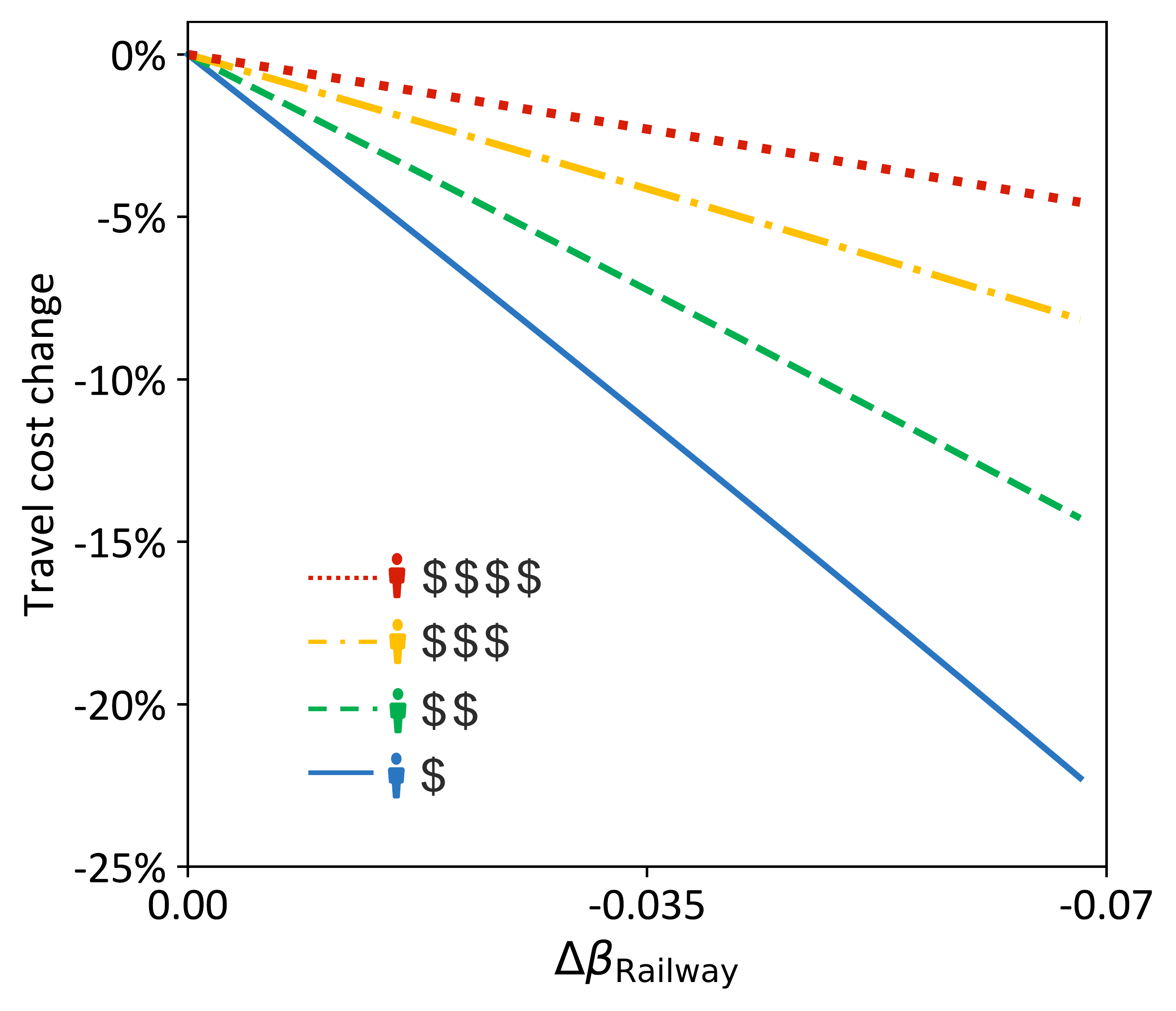}
\caption[Impact of public transport subsidies on average travel costs by income group.]{\textbf{Impact of public transport subsidies on average travel costs by income group.} The plot shows the percentage change in average travel cost (relative to the baseline scenario where $\Delta\beta_{\text{public}}=0$) for each of the four income groups ($\mathcal{G}=1$ to 4) as the railway subsidy increases ($\Delta\beta_{\text{public}}$ varies from 0 to -0.07). The results demonstrate the progressive nature of the subsidy, with lower-income groups experiencing a larger proportional reduction in their travel costs.}\label{FigS29}\end{figure}

The simulation results for increasing public transport subsidies confirm the expected shift towards railway usage, primarily drawing users from active and private modes. Notably, this response is uneven across income groups: lower-income individuals exhibit larger proportional decreases in active and private travel and, correspondingly, larger proportional increases in railway usage as subsidies rise (Supplementary Fig.~\ref{FigS28}a-c). This heightened sensitivity among lower-income groups translates into greater travel cost savings for them, with their average costs decreasing more significantly than higher-income groups (Supplementary Fig.~\ref{FigS29}), highlighting the progressive nature of the subsidy in terms of travel costs. However, the impact on mode-specific segregation presents a more complex picture. The influx of predominantly lower-income users onto the railway network leads to increased concentration and thus heightened segregation within that mode (lower \textit{PSM}, Supplementary Fig.~\ref{FigS28}f). Similarly, as lower- and middle-income individuals shift away from private cars, the remaining pool of drivers becomes more concentrated among higher-income groups, also increasing segregation (lower \textit{PSM}, Supplementary Fig.~\ref{FigS28}e). Conversely, the active travel mode experiences reduced segregation (higher \textit{PSM}, Supplementary Fig.~\ref{FigS28}d), suggesting that the departure of users (disproportionately lower-income) leaves behind a more income-diverse mix of active travelers relative to the baseline. It is crucial to note, however, that the magnitude of these changes in both mode shares and \textit{PSM} is relatively small, even at the maximum simulated subsidy level ($\Delta\beta_{\text{public}}=-0.07$). This muted response likely reflects the already substantial level of public transport support in the baseline scenario, suggesting diminishing returns from further fare-based subsidies alone.

\begin{figure}[h]
\centering
\includegraphics[width=1\textwidth]{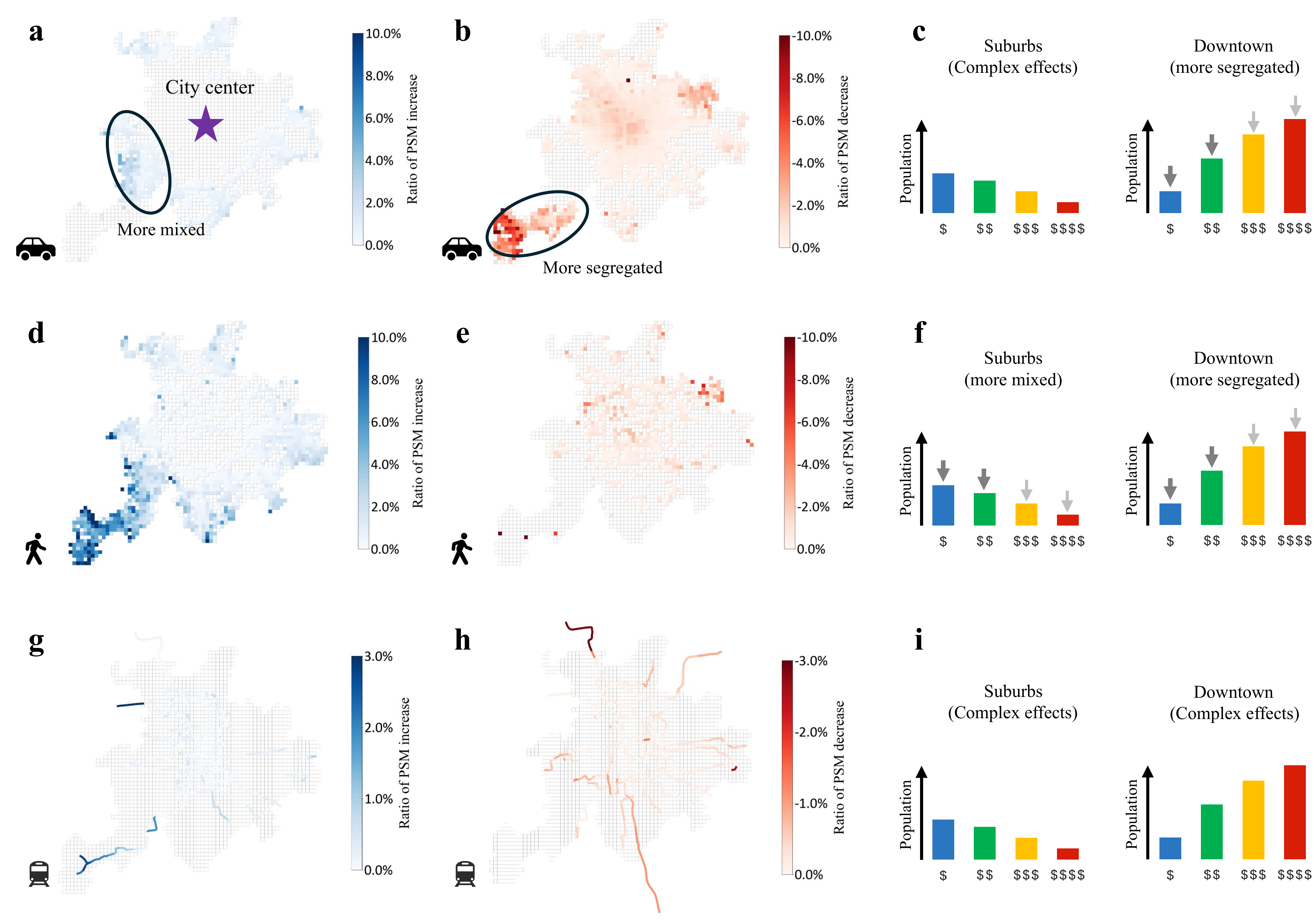}
\caption[Spatial distribution of changes in segregation measure \textit{PSM} for three transport modes under public transport subsidy policies.]{\textbf{Spatial distribution of changes in segregation measure \textit{PSM} for three transport modes under public transport subsidy policies.}  The figure compares the maximum subsidy scenario ($\Delta\beta_{\text{public}}=-0.07$) to the baseline ($\Delta\beta_{\text{public}}=0$) across three transport modes (active, private and railway). First column of panels (\textbf{a, d, g}) show the spatial units where \textit{PSM} increased (indicating reduced segregation). Second column of panels (\textbf{b, e, h}) show the spatial units where \textit{PSM} decreased (indicating heightened segregation). Third column of panels (\textbf{c, f, i}) represent the illustrative diagrams depicting policy-induced shifts in mode usage by income group for each respective mode. Bars represent the baseline proportion of commuters per income group using the mode. Arrows indicate the net change (increase or decrease) in commuters for that group. Bold arrows denote changes of larger magnitude, summarizing effects relevant to areas of downtown and suburbs.}\label{FigS30}\end{figure}

To analyze the spatial impact, we compare the maximum subsidy scenario ($\Delta\beta_{\text{public}}=-0.07$) with the baseline by calculating the change ratio in the mode-specific \textit{PSM} for each spatial unit $s$, analogous to the analysis shown for car policies. The results (Supplementary Fig.~\ref{FigS30}) suggest that for private mode, the policy tends to heighten segregation (lower \textit{PSM}) in downtown areas, as lower-income individuals are more likely to switch to the subsidized public transport, leaving a less diverse group of remaining drivers. Suburban areas show complex effects: enhanced mixing (higher \textit{PSM}) often occurs closer to the city center where railway access might be better, while very remote suburbs can become more segregated (lower \textit{PSM}), as the few lower-income drivers who can switch do so. Regarding active travel, the reduction in usage, particularly among lower-income groups, leads to decreased segregation (higher \textit{PSM}) in the suburbs, suggesting the remaining active travelers form a more evenly distributed mix. Conversely, downtown areas experience increased segregation (lower \textit{PSM}), likely due to the relative concentration of higher-income individuals among those who continue to use active modes there. 
Finally, the railway mode exhibits complex spatial effects on segregation. While the overall citywide trend is towards increased segregation (Supplementary Fig.~\ref{FigS28}f), the specific locations where \textit{PSM} increases or decreases lack a clear regional pattern (e.g., downtown vs. suburb), reflecting the varied geographic origins and destinations of the diverse income groups attracted to the subsidized service across the network.

These findings suggest that while public transport subsidies effectively reduce travel costs, particularly for lower-income groups, they can inadvertently increase segregation within the subsidized public mode and among remaining users of other modes (private cars). The relatively small magnitude of change also implies that further fare reductions might offer diminishing returns if baseline subsidies are already high. Therefore, policy interventions should aim to balance travel costs with social integration goals. Simply deepening existing fare subsidies may not be the most effective approach; instead, strategies could focus on enhancing service quality, network coverage, and reliability across diverse neighborhoods to attract a broader ridership mix, potentially mitigating the observed increase in railway segregation. Furthermore, coordinating public transport enhancements with policies addressing active and private travel could help manage the cross-modal impacts on segregation more effectively, striving for a transportation system that is both affordable and socially inclusive.

\subsection{Simulation for promoting active travel policies}\label{Supplnote4.5}

\noindent\hspace*{1.5em}Next, we explore the distributional impacts of policies designed to promote active travel (walking and cycling). Such policies are increasingly recognized for their potential to yield co-benefits, including improved public health through physical activity, reduced local air and noise pollution, lower greenhouse gas emissions, decreased traffic congestion, and enhanced neighborhood livability. Strategies often involve improving infrastructure (e.g., dedicated bike lanes, wider sidewalks, safer crossings) \cite{Pucher2010}, implementing traffic calming measures \cite{Bunn2003}, or running public awareness campaigns \cite{Cairns2008} to make active modes safer, more convenient, and more appealing alternatives, especially for shorter trips.

\begin{figure}[h]
\centering
\includegraphics[width=1\textwidth]{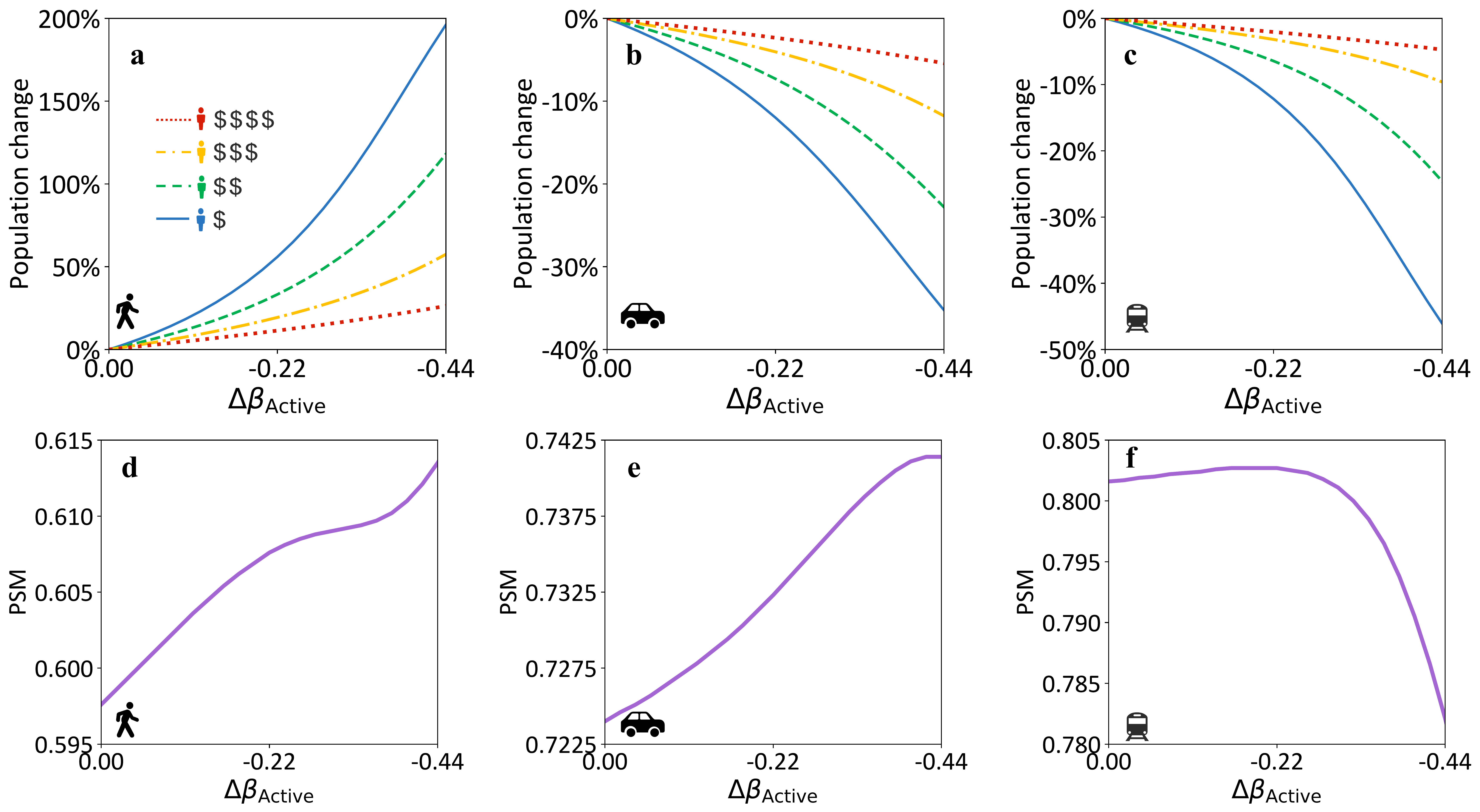}
\caption[Impact of promoting active travel policies on mode usage and citywide segregation.]{\textbf{Impact of promoting active travel policies on mode usage and citywide segregation.} The figure shows the effects as the active travel cost parameter is reduced ($\Delta\beta_{\text{active}}$ varies from 0 to -0.44). \textbf{a--c} Proportional changes in mode usage relative to the baseline for active (panel \textbf{a}), private (panel \textbf{b}), and railway (panel \textbf{c}) modes, separated by income group. \textbf{d--f} Overall citywide segregation measure \textit{PSM} for three modes.}\label{FigS31}\end{figure}

\begin{figure}[h]
\centering
\includegraphics[width=0.5\textwidth]{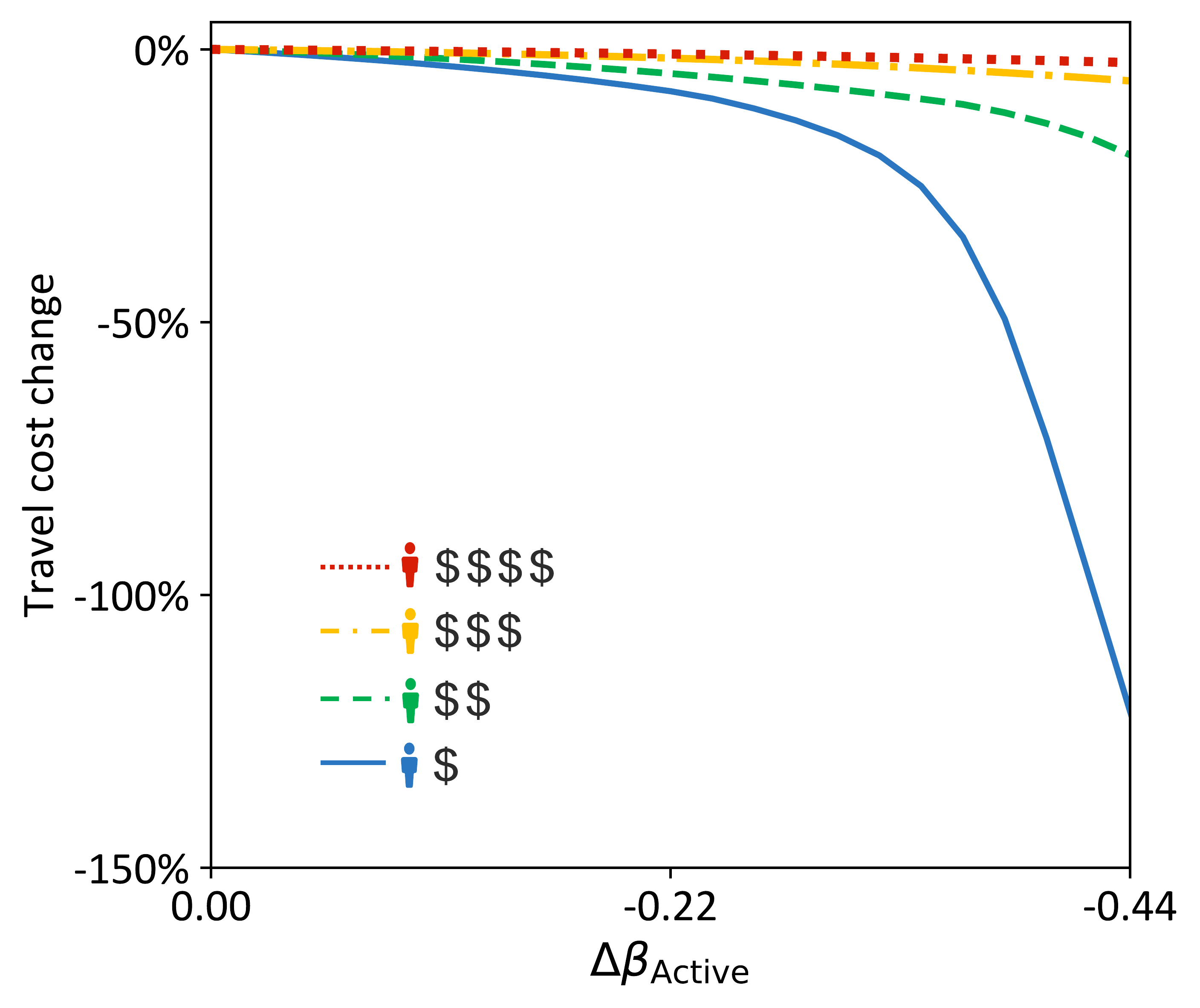}
\caption[Impact of promoting active travel policies on average travel costs by income group.]{\textbf{Impact of promoting active travel policies on average travel costs by income group.} The plot shows the percentage change in average travel cost (relative to the baseline scenario where $\Delta\beta_{\text{active}}=0$) for each of the four income groups ($\mathcal{G}=1$ to 4) as active travel promotion increases ($\Delta\beta_{\text{active}}$ varies from 0 to -0.44).}\label{FigS32} 
\end{figure}

We model these policies by systematically reducing the mode-specific cost parameter associated with active travel, $\beta_{\text{active}}$. This decrease represents a lower perceived generalized cost, reflecting factors like increased safety, convenience, or enjoyment resulting from policy interventions. We introduce a policy variable $\Delta\beta_{\text{active}}$ such that the modified cost parameter becomes $\beta'_{\text{active}} = \beta_{\text{active}}^* + \Delta\beta_{\text{active}}$, where $\beta_{\text{active}}^* = 0.22$ is the calibrated baseline value. We simulate increasing levels of promotion by varying $\Delta\beta_{\text{active}}$ from 0 (baseline) down to -0.44. Notably, when $\Delta\beta_{\text{active}} = -0.22$, the non-time-related cost component $\beta'_{\text{active}}$ becomes zero, implying the perceived cost is driven solely by the value of travel time ($\alpha_g T_{\text{active}}$). Further reductions ($\Delta\beta_{\text{active}} < -0.22$) imply a net perceived benefit beyond time cost (e.g., health benefits outweighing inconvenience). As before, for each policy level, we recalculate segregation measure \textit{PSM} (Supplementary Fig.~\ref{FigS31}d-f), mode usage shifts (Supplementary Fig.~\ref{FigS31}a-c) and average travel costs per income group (Supplementary Fig.~\ref{FigS32}).

\begin{figure}[h]
\centering
\includegraphics[width=1\textwidth]{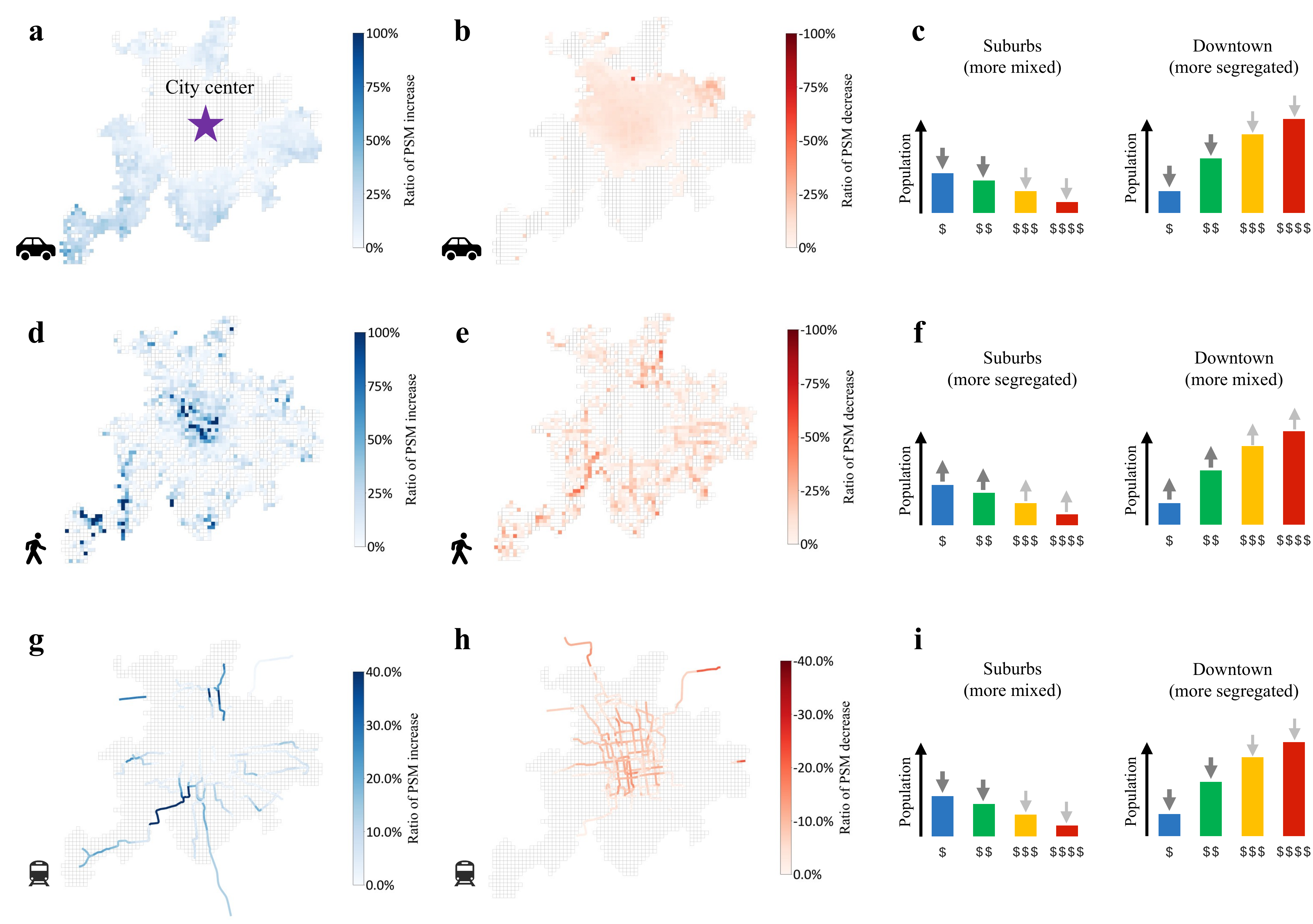}
\caption[Spatial distribution of changes in segregation measure \textit{PSM} for three transport modes under active travel policies.]{\textbf{Spatial distribution of changes in segregation measure \textit{PSM} for three transport modes under active travel policies.}  The figure compares the maximum promotion scenario ($\Delta\beta_{\text{active}}=-0.44$) to the baseline ($\Delta\beta_{\text{active}}=0$) across three transport modes (active, private and railway). First column of panels (\textbf{a, d, g}) show the spatial units where \textit{PSM} increased (indicating reduced segregation). Second column of panels (\textbf{b, e, h}) show the spatial units where \textit{PSM} decreased (indicating heightened segregation). Third column of panels (\textbf{c, f, i}) represent the illustrative diagrams depicting policy-induced shifts in mode usage by income group for each respective mode. Bars represent the baseline proportion of commuters per income group using the mode. Arrows indicate the net change (increase or decrease) in commuters for that group. Bold arrows denote changes of larger magnitude, summarizing effects relevant to areas of downtown and suburbs.}\label{FigS33}\end{figure}

Promoting active travel shows progressive distributional effects regarding costs, disproportionately benefiting lower-income groups who adopt it most readily and experience the largest travel cost savings (Supplementary Fig.~\ref{FigS31}a-c, Supplementary Fig.~\ref{FigS32}). However, the impacts on segregation are multifaceted. While citywide mixing generally increases for active and private modes (railway segregation eventually worsens at high promotion levels, Supplementary Fig.~\ref{FigS31}d-f), spatially the effects diverge significantly between urban core and periphery (Supplementary Fig.~\ref{FigS33}). Active travel promotes income mixing downtown but increases segregation in suburbs. Conversely, private and railway modes become more segregated downtown while mixing improves in suburbs. This highlights a key trade-off: while decreasing travel costs, policies promoting active travel reshape segregation patterns in complex, spatially dependent ways across different mobility layers, enhancing social interaction potential in some contexts (e.g., downtown active travel) while potentially reinforcing concentration or exclusion in others (e.g., suburban active travel or downtown private/railway use). Therefore, policymakers should leverage the social benefits by investing in active travel infrastructure, but carefully consider complementary measures, such as equitable deployment across suburbs and strategies to mitigate heightened segregation in other modes, to ensure fairer social outcomes. Such integrated planning is crucial to maximize the societal benefits of promoting active mobility.

\clearpage
\addcontentsline{toc}{section}{Supplementary references}
\renewcommand{\refname}{Supplementary references}
\bibliography{sn-supplementary}% common bib file